\documentclass[a4paper,11pt]{article}
\pdfoutput=1 

\usepackage{jheppub} 

\usepackage[T1]{fontenc} 
\usepackage[numbers]{natbib}
\usepackage{filecontents}
\usepackage{dynkin-diagrams}
\usepackage{tikz}
\usepackage{placeins}
\usepackage{stackengine}
\usepackage{booktabs}
\tikzset{Rightarrow/.style={double equal sign distance,>={Implies},->},
triple/.style={-,preaction={draw,Rightarrow}},
quadruple/.style={preaction={draw,shorten >=0pt},shorten >=1pt,-,double,double
distance=0.2pt}}
\usetikzlibrary{positioning}
\usepackage{graphicx}
\usepackage{pdflscape}
\usepackage{caption}
\usepackage{subcaption}
\usepackage{comment}
\usepackage{subfiles}
\usepackage{todonotes}

\usetikzlibrary{hobby,intersections} 
\usetikzlibrary{shapes.geometric}

\tikzset{flavour/.style={draw=none,minimum size=0.3mm,fill=white, regular polygon,regular polygon sides=4,draw}}
\tikzset{gaugeBig/.style={inner sep=1mm,draw=none,fill=white,minimum size=2mm,circle, draw}}
\tikzset{bd/.style={circle, draw=black, inner sep=0pt, fill=black, minimum size=2mm}}
\tikzset{wd/.style={circle, draw=black, inner sep=0pt, fill=white, minimum size=2mm}}
\tikzset{Dynkin/.style={circle, draw=black, inner sep=0pt, fill=white, minimum size=2mm}}
\tikzstyle{ligne}=[draw, very thick] 
\tikzstyle{gridline}=[draw, gray] 
\tikzset{gauge/.style={circle, draw,inner sep=2.5pt}}
\tikzset{gaugeo/.style={circle, draw,inner sep=2.5pt,fill=orange}}
\tikzset{gauger/.style={circle, draw,inner sep=2.5pt,fill=red}}
\tikzset{gaugeb/.style={circle, draw,inner sep=2.5pt,fill=blue}}
\tikzset{gaugeg/.style={circle, draw,inner sep=2.5pt,fill=green}}
\tikzset{gaugegoodgreen/.style={circle, draw,inner sep=2.5pt,fill=goodgreen}}
\tikzset{gaugem/.style={circle, draw,inner sep=2.5pt,fill=magenta}}
\tikzset{hasse/.style={circle, fill,inner sep=2pt}}
\tikzset{d2/.style={circle, fill,inner sep=1.3pt}}
\tikzset{shrinky/.style={circle, fill,inner sep=1pt}}
\tikzset{sized/.style={circle, draw, inner sep=1.5pt}}
\tikzset{seven/.style={circle, draw,inner sep=3pt}}

\makeatletter
\DeclareRobustCommand{\rvdots}{%
  \vbox{
    \baselineskip4\p@\lineskiplimit\z@
    \kern-\p@
    \hbox{.}\hbox{.}\hbox{.}
  }}
\makeatother

\newcommand{\Figref}[1]{Figure~\ref{#1}}
\newcommand{\Quiver}[1]{$\mathcal Q_{\ref{#1}}$}
\newcommand{\surm}{\mathrm{SU}}

\newcommand{\urm}{\mathrm{U}}
\newcommand{\sorm}{\mathrm{SO}}
\newcommand{\sprm}{\mathrm{Sp}}
\newcommand{\hs}{\mathrm{HS}}
\newcommand{\hwg}{\mathrm{HWG}}
\newcommand{\pe}{\mathrm{PE}}
\newcommand{\pl}{\mathrm{PL}}
\newcommand{\mHL}{\mathrm{mHL}}

\newcommand{\convexpath}[2]{
  [   
  create hullcoords/.code={
    \global\edef\namelist{#1}
    \foreach [count=\counter] \nodename in \namelist {
      \global\edef\numberofnodes{\counter}
      \coordinate (hullcoord\counter) at (\nodename);
    }
    \coordinate (hullcoord0) at (hullcoord\numberofnodes);
    \pgfmathtruncatemacro\lastnumber{\numberofnodes+1}
    \coordinate (hullcoord\lastnumber) at (hullcoord1);
  },
  create hullcoords
  ]
  ($(hullcoord1)!#2!-90:(hullcoord0)$)
  \foreach [
  evaluate=\currentnode as \previousnode using \currentnode-1,
  evaluate=\currentnode as \nextnode using \currentnode+1
  ] \currentnode in {1,...,\numberofnodes} {
    let \p1 = ($(hullcoord\currentnode) - (hullcoord\previousnode)$),
    \n1 = {atan2(\y1,\x1) + 90},
    \p2 = ($(hullcoord\nextnode) - (hullcoord\currentnode)$),
    \n2 = {atan2(\y2,\x2) + 90},
    \n{delta} = {Mod(\n2-\n1,360) - 360}
    in 
    {arc [start angle=\n1, delta angle=\n{delta}, radius=#2]}
    -- ($(hullcoord\nextnode)!#2!-90:(hullcoord\currentnode)$) 
  }
}

\newcommand{\pal}{\ldots \text{pal}\ldots}
\newcommand{\slice}[1]{\mathcal{S}_{\mathcal N,#1} }

\title{\boldmath Quiver Polymerisation}

\author{Amihay Hanany,}
\author{Rudolph Kalveks,}
\author{and Guhesh Kumaran}

\affiliation{Theoretical Physics Group, The Blackett Laboratory, Imperial College London, Prince Consort Road, SW7 2AZ, UK}

\emailAdd{a.hanany@imperial.ac.uk}
\emailAdd{rudolph.kalveks09@imperial.ac.uk}
\emailAdd{guhesh.kumaran18@imperial.ac.uk}

\abstract{Two new diagrammatic techniques on $3d\;\mathcal N=4$ quiver gauge theories, termed \textit{chain} and \textit{cyclic quiver polymerisation} are introduced. These gauge a diagonal $\surm/\urm(k)$ subgroup of the Coulomb branch global symmetry of a quiver (or pair of quivers) with multiple legs. The action on the Coulomb branch is that of a $\surm/\urm(k)$ hyper-Kähler quotient.
The polymerisation techniques build and generalise known composition methods from class $\mathcal S$. Polymerisation is used to generate a wide range of magnetic quivers from various physical contexts. These include polymerisation constructions for Kronheimer-Nakajima quivers, which generalise the ADHM construction for the moduli space of $k\;\surm(N)$ instantons on $\mathbb C^2$ to A-type singularities. Also a polymerisation construction of the magnetic quiver for the $6d\;\mathcal N=(1,0)$ theory coming from two $\frac{1}{2}$ M5 branes probing an $E_6$ Klein singularity. We find a method of extending magnetic quivers for Class $\mathcal S$ theories to cure the incomplete Higgsing that arises when gluing punctures into the loops associated with higher genus theories. Other novel constructions include a unitary magnetic quiver for the closure of a height four nilpotent orbit of $\sorm(7)$. We explore the relationships between the Coulomb and Higgs branches of quivers under polymerisation.}

\begin{document}
\preprint{Imperial/TP/24/AH/01}
\maketitle
\flushbottom
\section{Introduction}
\label{sec:intro}
Three dimensional gauge theories with 8 supercharges ($3d\;\mathcal N=4$) have moduli spaces of vacua which consist of a Higgs branch and a Coulomb branch, each with a global symmetry. The Higgs branch global symmetry is the flavour symmetry of the theory. The Higgs branch is typically protected from quantum corrections \cite{Argyres:1996eh} so the flavour symmetry is classical and simple to gauge. The D-term and F-term equations are interpreted as real and complex moment maps respectively and thus the action on the Higgs branch is a hyper-Kähler quotient. The Coulomb branch is not protected, receiving perturbative \cite{Seiberg:1996nz} and non-perturbative quantum corrections \cite{Seiberg:1996bs}. In particular, the Coulomb branch global symmetry changes under RG flow; being the maximal torus of the gauge group in the UV (i.e. the topological symmetry) and may be enhanced to some non-Abelian symmetry in the IR. The Coulomb branch global symmetry and the flavour symmetry are dual in the IR under $3d$ mirror symmetry \cite{Intriligator:1996ex} where it exists. The non-perturbative physics at play has made understanding the gauging of Coulomb branch global symmetries difficult.  This is in spite of being dual to flavour symmetries (in the IR) which are simple to gauge.

The focus of this work is on unitary $3d\;\mathcal N=4$ quiver gauge theories. For these theories, the IR Coulomb branch symmetry\footnote{Henceforth Coulomb branch global symmetry refers to this symmetry in the IR and not the UV.} is encoded in the quiver. For example, any \textit{balanced} gauge nodes forming a $G$ Dynkin diagram contributes a factor of $G$ to the Coulomb branch global symmetry \cite{Gaiotto:2008ak}. Systematic ways of reading the Coulomb branch global symmetry for non-Dynkin type quivers have also been found \cite{Gledhill:2021cbe}. This invites the question as to whether there is a way to gauge Coulomb branch global symmetries in quiver theories through combinatorial operations on quivers. Thereby bypassing complications to do with strongly coupled physics using other techniques.

In this paper the gauging of subgroups of the Coulomb branch global symmetry of a unitary $3d\;\mathcal N=4$ quiver theory are further explored from an earlier paper \cite{Hanany:2023tvn}. The earlier paper gave diagrammatic rules for "quotient quiver subtraction" which corresponds to gauging $\surm(k)$ subgroups of the Coulomb branch global symmetry. This $\surm(k)$ subgroup is embedded in a single "leg" of the quiver. Geometrically, the action on the Coulomb branch is a hyper-Kähler quotient by $\surm(k)$. Studies of hyper-Kähler quotients of $3d\;\mathcal N=4$ Coulomb branches were also done in \cite{Bourget:2021qpx,Bourget:2021jwo}. The effect on the Higgs branch is an increase in (quaternionic) dimension by $\textrm{rank}\;\surm(k)=k-1$. The method of gauging Coulomb branch global symmetries as described in \cite{Hanany:2023tvn} was inspired from a study of six-dimensional theories \cite{Hanany:2022itc}.

In this paper, techniques for combining or chaining multiple $3d\;\mathcal N=4$ quivers and/or introducing loops are described. Such a technique has been found in \cite{Benini:2010uu} for the $3d$ mirrors of class $\mathcal S$ theories \cite{Gaiotto:2009we}. From the class $\mathcal S$ perspective, the quiver operation realises a gluing of two "maximal" or "full" punctures. This can be done to connect two different Riemann surfaces or to increase the genus of a given Riemann surface by one using cylinders. The effect is of gauging a diagonal flavour symmetry of the $4d\;\mathcal N=2$ SCFT which is associated to the maximal punctures. The action on the $3d$ mirror is gauging the Coulomb branch global symmetry. In general, the possible (partial) gluings of class $\mathcal S$ theories have been well studied, such as through the "Tinkertoys" programme \cite{Chacaltana:2010ks,Chacaltana:2011ze,Chacaltana:2013oka,Chacaltana:2014jba,Chacaltana:2015bna,Chacaltana:2016shw,Chacaltana:2017boe,Chacaltana:2018vhp} and in \cite{Chacaltana:2012zy,DelZotto:2014hpa,Agarwal:2014rua,Tachikawa:2015bga,Ferlito:2017xdq,Agarwal:2018ejn,Giacomelli:2020jel,Bourget:2020asf,Carta:2021dyx,Baume:2021qho,Etxebarria:2021lmq,Kang:2022zsl,Kim:2023qbx,Lawrie:2024zon}, however, these do not have a realisation as operations on quivers in general. Also within the context of class $\mathcal S$, diagrammatic techniques on $3d$ mirrors of Argyres-Douglas theories corresponding to gauging subgroups of the Coulomb branch global symmetry have been studied in \cite{Xie:2012hs,Giacomelli:2020ryy}.

The techniques described in this work, called "\hyperref[fig:Chain]{chain polymerisation}" and "\hyperref[fig:Cyclic]{cyclic polymerisation}" on quivers, are inspired by and build on the technique established in \cite{Benini:2010uu}. This generalises the gauging of diagonal $\surm/\urm(k)$ subgroups of the Coulomb branch global symmetry of $3d\;\mathcal N=4$ theories. The action on the Coulomb branch can again be understood analytically in terms of hyper-Kähler quotients. The action on the Higgs branch is not precisely known, however the Higgs branch dimension increases by $\textrm{rank}\;\surm/\urm(k)$. For this reason, polymerisation is conjectured to be a $3d$ mirror dual operation to the gauging of flavour symmetries. This duality is summarised in \Figref{fig:MirrorMirror}.

\begin{figure}[h!]
    \centering
    \begin{tikzpicture}
        \node[](Q) at (-4,0){$\mathcal Q$};
        \node[] (Qp) at (-4,-5){$\mathcal Q_P$};
        \node[] (Qd) at (4,0){$\mathcal Q^\vee$};
        \node[] (Qdp) at (4,-5){$\mathcal Q^\vee_P$};
        \node[] (3d) at (0,0.5){$3d$ mirror};

        \node[align=center] (poly) at ($(Q)!0.5!(Qp)$) {Polymer-\\isation};
        \node[align=center] (Flav) at ($(Qd)!0.5!(Qdp)$) {Flavour\\Gauging};

        \draw[->] (Q)--(poly)--(Qp);
        \draw[->] (Qd)--(Flav)--(Qdp);

        \draw [->,out=-30,in=30,looseness=1] (Q) to node[right,align=center]{$\urm/\surm$ HKQ\\on $\mathcal C(\mathcal Q)$} (Qp);
        
        \draw [->,out=-150,in=150,looseness=1] (Q) to node[left,align=center]{$\textrm{dim}\;\mathcal H(\mathcal Q)$\\increases by\\$\textrm{rank}\;\urm/\surm$} (Qp);

        \draw [->,out=-150,in=150,looseness=1] (Qd) to node[left,align=center]{$\urm/\surm$ HKQ\\on $\mathcal H(\mathcal Q^\vee)$} (Qdp);

        \draw [->,out=-30,in=30,looseness=1] (Qd) to node[right,align=center]{$\textrm{dim}\;\mathcal C(\mathcal Q^\vee)$\\increases by\\$\textrm{rank}\;\urm/\surm$} (Qdp);
        \draw[dashed] (0,0.5)--(0,-5.5);
    \end{tikzpicture}
    \caption{Left: Action on Higgs and Coulomb branches of a quiver $\mathcal Q$ under polymerisation (either \hyperref[fig:Chain]{chain} or \hyperref[fig:Cyclic]{cyclic}) producing $\mathcal Q_P$. Right: Image in the "$3d$ mirror" corresponding to action on Coulomb and Higgs branches of $3d$ mirror dual quivers $\mathcal Q^\vee$ and $\mathcal Q^\vee_P$ under flavour gauging.}
    \label{fig:MirrorMirror}
\end{figure}

The quivers that can be subjected to polymerisation need to satisfy certain conditions. Importantly, polymerisation requires at least two quiver "legs" of the form $(1)-\cdots-(k)-\cdots$. Some of these legs have been studied in the Literature before as $T[\surm(k)]$ theories \cite{Cremonesi:2014uva} which are equivalent to "full" $A_{k-1}$ punctures in the language of class $\mathcal S$ theory. The Coulomb branch Hilbert Series of such a leg corresponds to the maximal nilpotent orbit of $A_{k-1}$ (with background magnetic charges), or equivalently to a modified Hall-Littlewood function. The formal equivalence between the polymerisation of quivers containing such legs and hyper-Kähler quotients by $\urm/\surm(k)$ on the Coulomb branch follows from the orthogonality between modified Hall-Littlewood functions \cite{alma99262034401591}.

These two techniques of \hyperref[fig:Chain]{chain} and \hyperref[fig:Cyclic]{cyclic polymerisation} have use beyond the study of $3d\;\mathcal N=4$ theories. This is because they are new additions to the program of \textit{magnetic quivers} \cite{Cabrera:2019izd} which has been successful in the study of moduli spaces for theories with eight supercharges in four, five and six dimensions \cite{Cabrera:2019izd,Bourget:2019aer,Bourget:2019rtl,Cabrera:2019dob,Grimminger:2020dmg,Bourget:2020gzi,Bourget:2020asf,Bourget:2020xdz,Beratto:2020wmn,Closset:2020scj,Akhond:2020vhc,vanBeest:2020kou,Bourget:2020mez,VanBeest:2020kxw,Giacomelli:2020ryy,Akhond:2021knl,Carta:2021whq,Arias-Tamargo:2021ppf,Bourget:2021xex,Gledhill:2021cbe,vanBeest:2021xyt,Carta:2021dyx,Sperling:2021fcf,Nawata:2021nse,Akhond:2022jts,Giacomelli:2022drw,Kang:2022zsl,Hanany:2022itc,Gu:2022dac,Fazzi:2022hal,Bourget:2022tmw,Gledhill:2022hrz,Fazzi:2022yca,Bhardwaj:2023zix,Bourget:2023uhe,Bourget:2023cgs,DelZotto:2023myd,DelZotto:2023nrb,Hanany:2023tvn,Hanany:2023uzn,Lawrie:2023uiu,Bourget:2023dkj,Benvenuti:2023qtv,Mansi:2023faa,Fazzi:2023ulb,Bourget:2024mgn}. The magnetic quiver defines some moduli space of dressed monopole operators which corresponds to the moduli space of the $d=4,5,6$ theory of interest. The success of this program is due to two reasons. Firstly, the monopole formula \cite{Cremonesi:2013lqa} allows for simple computation of the Hilbert series of these moduli spaces. Secondly, there are established and simple diagrammatic techniques, such as \textit{(quotient) quiver subtraction} \cite{Cabrera:2018ann,Hanany:2023tvn}, \textit{decoration} \cite{Bourget:2022ehw,Bourget:2022tmw}, \textit{folding} and \textit{wreathing} \cite{Bourget:2020bxh,Bourget:2021xex}, \textit{collapse} \cite{Hanany:2018vph,Hanany:2018dvd,Hanany:2023uzn}, \textit{multi-lacing} \cite{Hanany:2023uzn} and  \textit{decay} and \textit{fission} \cite{Bourget:2023dkj,Bourget:2024mgn} of magnetic quivers. All of these techniques allow for simple analysis of the moduli spaces of interest. Polymerisation, is the newest addition to the diagrammatic toolbox in the magnetic quiver program. Therefore the techniques of \hyperref[fig:Chain]{chain polymerisation} and \hyperref[fig:Cyclic]{cyclic polymerisation} are not only operations on $3d\;\mathcal N=4$ theories, but may be applied in the study of various physical and mathematical problems through magnetic quivers.

One use of \hyperref[fig:Chain]{chain} and \hyperref[fig:Cyclic]{cyclic polymerisation} on magnetic quivers is to find new and non-trivial relationships between free field theories, nilpotent orbit closures \cite{Hanany:2016gbz,Hanany:2017ooe}, Slodowy slices \cite{Cabrera:2018ldc} and Slodowy intersections \cite{Hanany:2019tji}.

A different application of \hyperref[fig:Chain]{chain polymerisation} is to study the Higgs branch of the worldvolume $6d\;\mathcal N=(1,0)$ gauge theory of two fractional $\frac{1}{2}$ M5 branes \cite{DelZotto:2014hpa} probing a Klein $E_6$ singularity. Magnetic quivers have been instrumental in studying the Higgs branches of six dimensional worldvolume theories of (fractional) M5 branes probing ADE Klein singularities. In particular the Higgs branch has been known to change as the gauge coupling goes from finite to infinite. For example, it may jump in (quaternionic) dimension \cite{Mekareeya:2017sqh} by 29 due to the small $E_8$ instanton transition \cite{Ganor:1996mu} which is required to preserve the anomaly cancellation condition \cite{Green:1984bx,Randjbar-Daemi:1985tdc,Dabholkar:1996zi}. Another way the Higgs branch may change is through the gauging of a discrete global symmetry of the Higgs branch \cite{Hanany:2018vph,Hanany:2018cgo,Cabrera:2019izd} or even through the gauging of a continuous subgroup of the global symmetry \cite{Hanany:2022itc}. Each of these have realisations as diagrammatic operations on magnetic quivers \cite{Hanany:2018uhm,Hanany:2018vph,Hanany:2022itc} respectively. In the dual F-theory description these occur when curves of self-intersection $(-1),(-2),$ and $(-3)$ respectively collapse.

When M5 branes probe $AD$ Klein singularities it is possible to reduce M-theory to Type IIA and read off the electric and magnetic quiver from the brane system \cite{Hanany:2018vph,Cabrera:2019izd,Cabrera:2019dob}. For the case of $E_{6,7,8}$ Klein singularity this is not possible and so  a dual description from F-theory constrains the form of the electric gauge theory \cite{Aspinwall:1998xj,DelZotto:2014hpa}. The case of interest here is one M5 brane on an $E_6$ Klein singularity; there may be fractionation to at most four $\frac{1}{4}$ M5 branes. If pairs of these $\frac{1}{4}$ M5 branes have zero separation, resulting in two separated $\frac{1}{2}$ M5 branes, the gauge theory description is as two rank-1 E-strings coupled to an $\surm(3)$ gauge group with finite coupling. This is constructed from the collapse of a $(-3)$-curve supporting $\surm(3)$ gauge symmetry. Therefore, the Higgs branch of this theory has a realisation through magnetic quivers as $\surm(3)$ \hyperref[fig:Chain]{chain polymerisation} of two affine $E_8$ quivers. Similar techniques were used in \cite{Mansi:2023faa,Lawrie:2024zon} to find magnetic quivers for certain $6d$ $\mathcal N=(1,0)$ theories.

Another application of \hyperref[fig:Cyclic]{cyclic polymerisation} is to give a new construction of the moduli space of $k\;\surm(N)$ instantons on $\mathbb C^2/\mathbb Z_l$. The first construction of instanton solutions for classical gauge groups on $\mathbb C^2$ was done by Atiyah-Drinfeld-Hitchin-Manin (ADHM) \cite{Atiyah:1978ri}. An interpretation of the ADHM solutions were found in Type IIA string theory \cite{Douglas:1995bn} whose low energy effective theory are the ADHM quivers. The moduli space of instantons is the Higgs branch of the ADHM quivers. By definition this is a hyper-Kähler quotient construction on the Higgs branch. A later development by Kronheimer-Nakajima generalised the ADHM construction to find instanton solutions for $k\;\surm(N)$ instantons on $\mathbb C^2/\mathbb Z_l$ \cite{Kronheimer1990}. A result from this work in the context of supersymmetric gauge theories was the Kronheimer-Nakajima quiver which gave a Higgs branch construction for these moduli spaces. The Kronheimer-Nakajima quivers were then interpreted in Type IIA and in Type IIB string theory \cite{Witten:1995gx,deBoer:1996mp,Kapustin:1998fa,Cremonesi:2014xha}. This gave Higgs branch and Coulomb branch constructions for these moduli spaces through \textit{electric} and \textit{magnetic} quivers respectively. However, the Coulomb branch construction coming from magnetic quivers was as a moduli space of dressed monopole operators and not as a hyper-Kähler quotient.

Importantly for this work, the A-type Kronheimer-Nakajima quivers for the moduli space of $k\;\surm(N)$ instantons on $\mathbb C^2/\mathbb Z_l$ consist of gauge nodes forming an affine A-type Dynkin diagram (a "necklace" of gauge nodes) with flavours attached to the gauge nodes. Therefore a natural application of \hyperref[fig:Cyclic]{cyclic polymerisation} is to construct these Kronheimer-Nakajima quivers from quivers without loops. This gives a Coulomb branch hyper-Kähler quotient construction for these moduli spaces which is different to the known Coulomb branch construction. In fact, this construction is dual under $3d$ mirror symmetry to the Higgs branch hyper-Kähler quotient construction of these moduli spaces.

Finally, returning to the original inspiration of polymerisation from the class $\mathcal S$ construction, a long standing problem regarding the agreement of the Hall-Littlewood limit of the superconformal index with the Coulomb branch Hilbert series of the $3d$ mirror is addressed. As first pointed out in \cite{Gadde:2011uv} there is agreement between the two when gluing separate trivial genus Riemann surfaces (except possibly when \textit{twisted} punctures are used \cite{Kang:2022zsl}) but disagreement when increasing the genus of a Riemann surface by one. The source of this disagreement is the incomplete Higgsing observed in the latter case.

From the view of the $3d$ mirror theory, when there is incomplete Higgsing the hyper-Kähler quotients of the Coulomb branches of the $3d$ mirrors are difficult to compute using known Hilbert series methods. This motivates a way to cure the incomplete Higgsing so that the simple Hilbert series techniques are applicable. The incomplete Higgsing may be cured by another new operation on quivers called \hyperref[fig:ClassSCyclicFix]{quiver extension}. Quiver extension takes a star-shaped quiver with at least two maximal legs and adds one more node so that the quiver is not star-shaped anymore. This increases the dimension of the Coulomb branch and changes it in a generally non-trivial way. Then a \hyperref[fig:Cyclic]{cyclic polymerisation} is performed leading to the $3d$ mirror of the class $\mathcal S$ theory. There may be many ways of extending a given star-shaped quiver however under \hyperref[fig:Cyclic]{cyclic polymerisation} all possible extensions lead to the same resulting quiver. The most natural quiver extension dualises the Coulomb branch \cite{Gaiotto:2008ak}, which introduces extra free twisted hypermultiplets that are eaten in the Higgsing. A similar idea where extra free matter is introduced in the gauging of certain $4d\;\mathcal N=2$ SCFTs was employed in \cite{Argyres:2007cn} and realised on the $3d$ mirror quiver in \cite{Hanany:2023tvn}.

\paragraph{Outline} In Section \ref{sec:Theory}, the diagrammatic prescription for \hyperref[fig:Chain]{chain} and \hyperref[fig:Cyclic]{cyclic polymerisation} is presented. For the corresponding Hilbert series analysis, the relevant embeddings into the Coulomb branch global symmetry and Weyl integration are also discussed. The properties of modified Hall-Littlewood functions are discussed to provide an analytical account of the equivalence between polymerisation and hyper-Kähler quotients over $T[\surm(k)]$ theories. Then, a discussion of the connection of polymerisation to class $\mathcal S$ theories is presented. Finally, a brief overview of the Kronheimer-Nakajima quiver and the moduli space of $k\;\surm(N)$ instantons on $\mathbb C^2/\mathbb Z_l$ is presented.

In Section \ref{sec:Chain}, the method of \hyperref[fig:Chain]{chain polymerisation} is demonstrated. Examples are drawn upon magnetic quivers for moduli spaces of free fields and nilpotent orbits. Some of the resulting quivers take the form of Dynkin diagrams for Kac-Moody algebras.


In Section \ref{sec:instantonconstruc}, the method of 
\hyperref[fig:Cyclic]{cyclic polymerisation} is applied to construct the Kronheimer-Nakajima quivers for the moduli spaces of $k\;\surm(N)$ instantons on $\mathbb C^2/\mathbb Z_l$. Firstly, for specific choices of $k,\;N,$ and $l$, then generically.

In Section \ref{sec:3dmirror}, the commutation between polymerisation and $3d$ mirror symmetry is examined using examples presented earlier.

In Section \ref{sec:ADEOrb}, \hyperref[fig:Cyclic]{cyclic polymerisation} is applied to quivers which are of finite and affine $ADE$ type. A new magnetic quiver for a height 4 nilpotent orbit of $\sorm(7)$ is found as well as magnetic quivers for slices in exceptional nilcones.

In Section \ref{sec:cyclicproduct}, the combined techniques of  \hyperref[fig:ClassSCyclicFix]{quiver extension} and \hyperref[fig:Cyclic]{cyclic polymerisation} are applied to $3d$ mirrors of class $\mathcal S$ theories.

In Section \ref{sec:conc}, conclusions are drawn and future work is discussed.

\paragraph{Note added.} While this work was in its final stages we were made aware of \cite{Dancer:2024lra} which uses similar constructions.

\section{Chain and Cyclic Polymerisation}
\label{sec:Theory}
\subsection{Diagrammatic methods}
\label{subsec:diagrams}
The diagrammatic techniques of \hyperref[fig:Chain]{chain polymerisation} and \hyperref[fig:Cyclic]{cyclic polymerisation} are shown schematically in \Figref{fig:Chain} and \Figref{fig:Cyclic} respectively. For simplicity all the quivers are taken to be unframed. These techniques realise the gauging of a subgroup of the Coulomb branch global symmetry of a $3d\;\mathcal N=4$ quiver theory. The action on the Coulomb branch is a hyper-Kähler quotient, as will be shown. The geometric action on the Higgs branch is not systematically known, but the Higgs branch dimension increases by the rank of the subgroup of the Coulomb branch global symmetry being gauged, as will be shown. Details of the specific embeddings of the Coulomb branch global symmetry being gauged under polymerisation are discussed in Section \ref{subsec:weyl}.

\begin{figure}[h!]
    \centering
    \begin{tikzpicture}
    \node (a) at (0,0) {
    $\begin{tikzpicture}
    \node[gauge, label=below:$1$] (1l) []{};
    \node[gauge, label=below:$2$] (2l) [right=of 1l]{};
    \node[] (cdotsl) [right=of 2l]{$\cdots$};
    \node[gauge, label=below:$k$] (kl) [right=of cdotsl]{};
    \node[gauge] (Q1) [right=of kl]{$Q_1$};
    \node[gauge] (Q2) [right=of Q1]{$Q_2$};
    \node[gauge, label=below:$k$] (kr) [right=of Q2]{};
    \node[] (cdotsr) [right=of kr]{$\cdots$};
    \node[gauge, label=below:$2$] (2r) [right=of cdotsr]{};
    \node[gauge, label=below:$1$] (1r) [right=of 2r]{};

    \draw[-] (1l)--(2l)--(cdotsl)--(kl)(kr)--(cdotsr)--(2r)--(1r);
    \draw[-] (kr)--(cdotsr)--(2r)--(1r);
    \draw[green,very thick] (kl)--(Q1);
    \draw[orange,very thick] (Q2)--(kr);

    \draw (kl) to [out=135, in=45,looseness=8]  node[pos=0.5,above]{$g_1$} (kl);
    \draw (kr) to [out=135, in=45,looseness=8]  node[pos=0.5,above]{$g_2$} (kr);
    \end{tikzpicture}$};

    \node (b) at (0,-5){$\begin{tikzpicture}
        \node[gauge, label=above:$k$] (k) []{};
        \node[gauge] (Q1) [above left=of k]{$Q_1$};
        \node[gauge] (Q2) [above right=of k]{$Q_2$};

        \draw[green,very thick] (k)--(Q1);
        \draw[orange,very thick] (Q2)--(k);

        \draw (k) to [out=225, in=-45,looseness=8]  node[pos=0.5,below]{$g_1+g_2$} (k);
    \end{tikzpicture}$};

    \draw[->] (a)--(b);
    \end{tikzpicture}
    \caption{Schematic showing the 
    \hyperref[fig:Chain]{chain polymerisation} of two unframed quivers (assumed to be "good" or "ugly") with a node of $\urm(k)$ on a leg to give a single unframed quiver.}
    \label{fig:Chain}
\end{figure}

The first technique is \hyperref[fig:Chain]{chain polymerisation}, shown in \Figref{fig:Chain}, which gauges an $\surm(k)$ subgroup of the Coulomb branch global symmetry. \hyperref[fig:Chain]{Chain polymerisation} requires two "good" or "ugly" quivers (in the sense of \cite{Gaiotto:2008ak}) $\mathcal Q_1$ and $\mathcal Q_2$, each containing a leg of $(1)-(2)-\cdots-(k)-$. Each quiver also contains a background quiver, $Q_1$ and $Q_2$ respectively. These background quivers can be quite general and may have a multiplicity of links to the $\urm(k)$ nodes. The background quivers may even be disconnected as long as each part is attached to the node of $\urm(k)$. The gauge node of $\urm(k)$ in each leg may also carry any number of adjoint hypermultiplet loops. The polymerisation action on the quivers superimposes the $\urm(k)$ nodes and amputates the other nodes of the two legs. The resulting unframed quiver $\mathcal Q_3$ contains a single $\urm(k)$ node connected to the background quivers $Q_1$ and $Q_2$, along with any adjoint hypermultiplet loops on each of the $\urm(k)$ nodes. The action on the Coulomb branch is an $\surm(k)$ hyper-Kähler quotient of the product theory. The rank of the gauge group reduces by $k^2-1$ which is consistent with the reduction in Coulomb branch dimension after polymerisation being $\textrm{dim}\;\surm(k)=k^2-1$. From counting Higgs branch dimensions, it is clear that the polymerised theory $Q_3$ has a Higgs branch that is greater in dimension by $\textrm{rank}\;\surm(k)=k-1$ compared to the unpolymerised product theory $\mathcal Q_1\times \mathcal Q_2$. The only contribution to the difference in Higgs branch dimension comes from the amputation of $(1)-\cdots-(k-1)-$:
\begin{align}
    \mathrm{dim}\left[\mathcal H(\mathcal Q_3)\right]-\left(\mathrm{dim}\left[\mathcal H(\mathcal Q_1)\right]+\mathrm{dim}\left[\mathcal H(\mathcal Q_2)\right]\right)&=\left(0-k^2+1\right)-2\left(\sum_{i=1}^{k-1}i(i+1)-\sum_{i=1}^{k}i^2+1\right)\nonumber\\&=k-1\nonumber\\&=\mathrm{rank}\;\surm(k).
\end{align}


In special cases where both of the starting quivers are $3d$ mirrors of class $\mathcal S$ theories, \hyperref[fig:Chain]{chain polymerisation} at the central node reproduces the action of gluing two Riemann surfaces via maximal punctures.

\begin{figure}[h!]
    \centering
    \begin{tikzpicture}
\node (a) at (0,0){$\begin{tikzpicture}
    \node[gauge, label=below:$1$] (1l) []{};
    \node[gauge, label=below:$2$] (2l) [right=of 1l]{};
    \node[] (cdotsl) [right=of 2l]{$\cdots$};
    \node[gauge, label=below:$k$] (kl) [right=of cdotsl]{};
    \node[gauge] (Q) [right=of kl]{$Q$};
    \node[gauge, label=below:$k$] (kr) [right=of Q]{};
    \node[] (cdotsr) [right=of kr]{$\cdots$};
    \node[gauge, label=below:$2$] (2r) [right=of cdotsr]{};
    \node[gauge, label=below:$1$] (1r) [right=of 2r]{};

    \draw[-] (1l)--(2l)--(cdotsl)--(kl)(kr)--(cdotsr)--(2r)--(1r);
    \draw[-] (kr)--(cdotsr)--(2r)--(1r);
    \draw[green,very thick] (kl)--(Q);
    \draw[orange,very thick] (Q)--(kr);
    \draw (kl) to [out=135, in=45,looseness=8]  node[pos=0.5,above]{$g_1$} (kl);
    \draw (kr) to [out=135, in=45,looseness=8]  node[pos=0.5,above]{$g_2$} (kr);
    \end{tikzpicture}$};

    \node (b) at (0,-5){$\begin{tikzpicture}
    \node[gauge] (Qb) []{$Q$};
    \node[gauge, label=above:$k$] (kb) [below=of Qb]{};

   \draw (kb) to [out=-135, in=-45,looseness=8]  node[pos=0.5,below]{$g_1+g_2$} (kb);
   \draw [orange,very thick] (Qb) to [out=-60,in=60,looseness=1] (kb);
    \draw[green,very thick] (kb) to [out=120,in=-120, looseness=1] (Qb);
    \end{tikzpicture}$};

    \draw[->] (a)--(b);
        
    \end{tikzpicture}
    \caption{Schematic showing the \hyperref[fig:Cyclic]{cyclic polymerisation} of some generic quiver (assumed good or ugly) with two external legs of $(1)-\cdots-(k)-$.}
    \label{fig:Cyclic}
\end{figure}

\hyperref[fig:Cyclic]{Cyclic polymerisation}, shown in \Figref{fig:Cyclic}, is similar to \hyperref[fig:Chain]{chain polymerisation} in that two distinct legs are involved. However, in this case, the background quiver $Q$ is common, so that the polymerisation acts on a single connected quiver $\mathcal Q$ (although the links from $Q$ to the two $\urm(k)$ nodes may differ). As before, each of the $\urm(k)$ nodes may have adjoint hypermultiplet loops. The polymerisation action on the quivers superimposes the $\urm(k)$ nodes in each leg and removes the other nodes in these legs. The net effect this time is a gauging of a $\urm(k)$ subgroup of the Coulomb branch global symmetry, which corresponds to a $\urm(k)$ hyper-Kähler quotient on the Coulomb branch. This is consistent with the rank of the gauge group being reduced by $\textrm{dim}\;\urm(k)=k^2$. The polymerised theory $\mathcal Q'$ has a Higgs branch that is greater in dimension by $\textrm{rank}\;\urm(k)=k$ compared to the unpolymerised theory $\mathcal Q$. Again, the difference in the Higgs branch dimension comes from the amputated legs:
\begin{align}
    \mathrm{dim}\left[\mathcal H(\mathcal Q')\right]-\mathrm{dim}\left[\mathcal H(\mathcal Q)\right]&=\left(0-k^2+1\right)-\left(2\sum_{i=1}^{k-1}i(i+1)-2\sum_{i=1}^{k}i^2+1\right)\nonumber\\&=k\nonumber\\&=\mathrm{rank}\;\urm(k).
\end{align}

In the special case of $\urm(1)$ \hyperref[fig:Cyclic]{cyclic polymerisation} the $\urm(1)$ hyper-Kähler quotient of the Higgs branch of the polymerised theory is the Higgs branch of the unpolymerised theory. This special case exists because $\textrm{dim}\;\urm(1)=\textrm{rank}\;\urm(1)=1$. There is no such special case for \hyperref[fig:Chain]{chain polymerisation} as there is no non-trivial choice of $k$ for which $\textrm{dim}\;\surm(k)=\textrm{rank}\;\surm(k)$. The reason for the different $\surm/\urm(k)$ hyper-Kähler quotient outcomes between \hyperref[fig:Chain]{chain polymerisation} and \hyperref[fig:Cyclic]{cyclic polymerisation} is explained in Section \ref{subsec:weyl}.

Polymerisation is a different type of quiver subtraction to that presented in \cite{Hanany:2023tvn}: in the case of polymerisation, the nodes being subtracted form complete and distinct legs, and do not constitute a single continuous sub-chain. 

\subsection{Weyl integration for polymerisation}
\label{subsec:weyl}
The theoretical justification underpinning \hyperref[fig:Chain]{chain} and \hyperref[fig:Cyclic]{cyclic polymerisation} is outlined in Section \ref{subsec:analytic}. For pedagogical reasons, it is also helpful to check the results for specific cases from the hyper-Kähler quotient. This can be done using the Hilbert series for the moduli space \cite{Benvenuti:2006qr,Feng:2007ur,Cremonesi:2013lqa} and Weyl integration in the case where there is \textit{complete} Higgsing. Complete Higgsing on the Coulomb branch results in a moduli space which is a symplectic singularity. Incomplete Higgsing on the Coulomb branch results in a moduli space which is typically not a symplectic singularity.

The hyper-Kähler quotient is easily computed using Weyl integration on the Hilbert series, provided complete Higgsing occurs. There is always complete Higgsing on the Coulomb branch from polymerisation but not on the Higgs branch (through the reverse action of polymerisation) except for the case of $\urm(1)$ \hyperref[fig:Cyclic]{cyclic polymerisation}. The Weyl integration formula for complete Higgsing is: 

\begin{equation}
    \hs\left[\mathcal M///G\right](t;x_1,\cdots,x_{r})=\oint_{G}d\mu_G\frac{ \hs\left[\mathcal M\right](t;x_1,\cdots,x_r,y_1,\cdots,y_{\mathrm{rank}(G)})}{\pe[\chi_{\mathrm{Adj}}^G(y_1,\cdots,y_{\mathrm{rank}(G)})t^2]}.
\label{eq:weyl}
\end{equation}

The Weyl integration formula \eqref{eq:weyl} proceeds as follows. Take the Hilbert series of some moduli space $\mathcal M$ with global symmetry $G_{\mathcal M}$ and branch the global symmetry to $G_{\mathcal M}\supset G\times G'$. $G$ is the group for which the hyper-Kähler quotient is to be performed and $G'$, of rank $r$, is the commutant with $G$ in $G_{\mathcal M}$. The $\pe$ gives symmetrisations of the adjoint representation of $G$ graded by $t^2$. The quotient of the moduli space by these terms is taken and then integrated over $G$. The overall effect is to project out the $G$ singlets that remain in the Hilbert series $\hs\left[\mathcal M\right]$ after a hyper-Kähler quotient.

 The F-flat space is a complete intersection and so the reduction in the (quaternionic) dimension of a moduli space under the hyper-Kähler quotient of semi-simple Lie group $G$ is $\mathrm{dim}\,G$.

The choice of embedding of $G$ into $G_{\mathcal M}$, the global symmetry of $\mathcal M$, is extremely important. In \cite{Hanany:2023tvn}, the embedding used to gauge $\surm(k)$ subgroups of the Coulomb branch global symmetry was a direct ``Dynkin'' type embedding, based on the Dynkin diagram of a single sub-chain within the quiver. For polymerisation a \textit{diagonal} subgroup of the global symmetry is required.

Both \hyperref[fig:Chain]{chain polymerisation} and \hyperref[fig:Cyclic]{cyclic polymerisation} involve two legs of $(1)-(2)-\cdots-(k)-$, which from consideration of balance \cite{Gledhill:2021cbe} contributes at least a factor of $\surm(k)\times\urm(1)$ to the Coulomb branch global symmetry of the theory. There may be enhancements of this symmetry depending on the specific examples. \hyperref[fig:Chain]{Chain} and \hyperref[fig:Cyclic]{cyclic polymerisation} require explicitly breaking the symmetry of each leg to $\surm(k)\times\urm(1)$, and the diagonal nature of the embedding requires that each leg has a conjugate embedding w.r.t this $\surm(k)\times\urm(1)$ symmetry. The fugacity map is explicitly shown for \hyperref[fig:Chain]{chain} and \hyperref[fig:Cyclic]{cyclic polymerisation} in \Figref{fig:ChainEmbed} and \Figref{fig:CyclicEmbed} respectively. For cyclic polymerisation the $\urm(1)\subset \urm(k)$ Weyl integration is performed w.r.t the $\urm(1)$ with fugacity $p$.

\begin{figure}[h!]
    \centering
    \begin{tikzpicture}
    \node[gauge, label=below:$\frac{y_1^2}{y_2}$] (1l) []{};
    \node[gauge, label=below:$\frac{y_2^2}{y_1 y_3}$] (2l) [right=of 1l]{};
    \node[] (cdotsl) [right=of 2l]{$\cdots$};
    \node[gauge, label=below:$\frac{y_{k-1}^2}{y_{k-2}}$] (km1l) [right=of cdotsl]{};
    \node[gauge, label=below:$\frac{1}{y_{k-1}{\bf q}_1}$] (kl) [right=of km1l]{};
    \node[gauge] (Q1) [right=of kl]{$Q_1$};
    \node[gauge] (Q2) [right=of Q1]{$Q_2$};
    \node[gauge, label=below:$\frac{1}{y_{1}{\bf q}_2}$] (kr) [right=of Q2]{};
    \node[gauge, label=below:$\frac{y_1^2}{y_2}$] (km1r) [right=of kr]{};
    \node[] (cdotsr) [right=of km1r]{$\cdots$};
    \node[gauge, label=below:$\frac{y_{k-2}^2}{y_{k-1}y_{k-3}}$] (2r) [right=of cdotsr]{};
    \node[gauge, label=below:$\frac{y_{k-1}^2}{y_{k-2}}$] (1r) [right=of 2r]{};
    
    \draw[-] (1l)--(2l)--(cdotsl)--(km1l)--(kl) (kr)--(km1r)--(cdotsr)--(2r)--(1r);
    \draw[green,very thick] (kl)--(Q1);
    \draw[orange,very thick] (Q2)--(kr);

    \draw (kl) to [out=135, in=45,looseness=8]  node[pos=0.5,above]{$g_1$} (kl);
    \draw (kr) to [out=135, in=45,looseness=8]  node[pos=0.5,above]{$g_2$} (kr);
    
    \end{tikzpicture}
    \caption{The assignment of fugacities in each quiver for the diagonal $\surm(k)$ hyper-Kähler quotient on the Coulomb branch from \hyperref[fig:Chain]{chain polymerisation}. The $y_i$ are $\surm(k)$ CSA fugacities and ${\bf q}_{1,2}$ refers to any other fugacities associated to links to the quivers $Q_{1,2}$ respectively.}
    \label{fig:ChainEmbed}
\end{figure}

It is time to address why \hyperref[fig:Chain]{chain polymerisation} and \hyperref[fig:Cyclic]{cyclic polymerisation} gauges respectively a $\surm/\urm(k)$ subgroup of the Coulomb branch global symmetry. Firstly, consider \hyperref[fig:Chain]{chain polymerisation}. \hyperref[fig:Chain]{Chain polymerisation} starts with two unframed quivers and polymerises them to give a single unframed quiver. Each of these starting unframed quivers has an overall $\urm(1)$ which may mix with the $\urm(1)$ in their respective broken legs. In doing \hyperref[fig:Chain]{chain polymerisation} one overall $\urm(1)$ is lost since two quivers are chained into one. Hence the quaternionic dimensional reduction from the polymerisation is $\mathrm{dim}\,\surm(k)=k^2-1$ and the net effect on the Coulomb branch is that of an $\surm(k)$ hyper-Kähler quotient.

\begin{figure}[h!]
    \centering
    \begin{tikzpicture}
         \node[gauge, label=below:$\frac{y_1^2}{y_2}$] (1l) []{};
    \node[gauge, label=below:$\frac{y_2^2}{y_1 y_3}$] (2l) [right=of 1l]{};
    \node[] (cdotsl) [right=of 2l]{$\cdots$};
    \node[gauge, label=below:$\frac{y_{k-1}^2}{y_{k-2}}$] (km1l) [right=of cdotsl]{};
    \node[gauge, label=below:$\frac{qp}{y_{k-1}{\bf q}_1}$] (kl) [right=of km1l]{};
    \node[gauge] (Q) [right=of kl]{$Q$};
    \node[gauge, label=below:$\frac{q}{py_1 {\bf q}_2}$] (kr) [right=of Q]{};
    \node[gauge, label=below:$\frac{y_1^2}{y_2}$] (km1r) [right=of kr]{};
    \node[] (cdotsr) [right=of km1r]{$\cdots$};
    \node[gauge, label=below:$\frac{y_{k-2}^2}{y_{k-1}y_{k-3}}$] (2r) [right=of cdotsr]{};
    \node[gauge, label=below:$\frac{y_{k-1}^2}{y_{k-2}}$] (1r) [right=of 2r]{};
   
    \draw[-] (1l)--(2l)--(cdotsl)--(km1l)--(kl) (kr)--(km1r)--(cdotsr)--(2r)--(1r);
    \draw[green,very thick] (kl)--(Q);
    \draw[orange,very thick] (Q)--(kr);
    \draw (kl) to [out=135, in=45,looseness=8]  node[pos=0.5,above]{$g_1$} (kl);
    \draw (kr) to [out=135, in=45,looseness=8]  node[pos=0.5,above]{$g_2$} (kr);
    \end{tikzpicture}
    \caption{The assignment of fugacities of each leg for the diagonal $\urm(k)$ hyper-Kähler quotient on the Coulomb branch from \hyperref[fig:Cyclic]{cyclic polymerisation}. The $y_i$ are $\surm(k)$ CSA fugacities, the $p$ and $q$ are $\urm(1)$ fugacities, and  ${\bf q}_{1,2}$ refer to any fugacities associated to links to the quiver $Q$.}
    \label{fig:CyclicEmbed}
\end{figure}

The difference with \hyperref[fig:Cyclic]{cyclic polymerisation} is that the operation happens on a single given quiver. Thus the overall $\urm(1)$ mixes equally, if at all, amongst the two $\urm(1)$ from breaking the Coulomb branch global symmetry in each leg. Hence \hyperref[fig:Cyclic]{cyclic polymerisation} involves gauging a diagonal $\urm(k)$ subgroup of the Coulomb branch global symmetry.

In some cases the global symmetry of the resulting moduli space enhances from that expected from the embedding specified in \Figref{fig:ChainEmbed} and \Figref{fig:CyclicEmbed}. In such cases the details of the embedding to the enhanced group are shown instead.

\subsection{Comparison to Flavour Gauging}
It is instructive to compare the gauging of Coulomb branch global symmetries with the gauging of flavour symmetries. Suppose we start with a theory $\mathcal T_1$ and gauge (a subgroup of) the flavour symmetry to produce a new theory $\mathcal T_2$. It is well known that the gauging of a (subgroup of the) flavour symmetry, $G_F$, of a gauge theory decreases the (quaternionic) dimension of the Higgs branch by $\textrm{dim}\,G_F$ through a hyper-Kähler quotient action due to the additional F-terms that arise. This hyper-Kähler quotient involves complete Higgsing so can be checked using Weyl integration as in Section \ref{subsec:weyl}. For $3d\;\mathcal N=4$ theories, the introduction of this additional gauge group introduces opens new directions in the Coulomb branch thereby increasing its dimension by $\textrm{rank}\,G_F$ and enhancing the Coulomb branch global symmetry by at least a factor of $\urm(1)$. If the gauged node is balanced then there may be non-trivial enhancements. The operation on the Coulomb branch is conjectured to be a hyper-Kähler quotient $G_F$ on $\mathcal T_2$ to $\mathcal T_1$. This often involves \textit{incomplete} Higgsing since  $\textrm{rank}\,G_F<\textrm{dim}\,G_F$ for many simple Lie groups. Therefore the Weyl integration techniques described in Section \ref{subsec:weyl} cannot be used to check in most cases. The best that can be done is a check at the level of the dimension.

As stated before, both \hyperref[fig:Chain]{chain} and \hyperref[fig:Cyclic]{cyclic polymerisation} gauge the Coulomb branch global symmetry of a $3d\;\mathcal N=4$ theory with the action on the Coulomb branch being a hyper-Kähler quotient. This involves complete Higgsing so the Weyl integration techniques can be used. This is the main check of the claim that \hyperref[fig:Chain]{chain} and \hyperref[fig:Cyclic]{cyclic polymerisation} correspond to gauging the Coulomb branch global symmetry of a theory. The Higgs branch dimension of the theory after polymerisation goes up by $\textrm{rank}\;\surm/\urm(k)$ respectively. Additionally, in the case of \hyperref[fig:Cyclic]{cyclic polymerisation} the creation of a loop of gauge nodes also enhances the Higgs branch global symmetry by at least a factor of $\urm(1)$. The reason is that the loop of gauge nodes in the quiver corresponds to a new invariant which could be rotated by some new flavour symmetry which is at least a $\urm(1)$ factor. Note that the action on the moduli spaces from polymerisation is precisely dual to the effect on the moduli spaces from gauging flavour symmetries. Again, this duality is summarised in \Figref{fig:MirrorMirror}.




The precise analogue of \hyperref[fig:Chain]{chain} and \hyperref[fig:Cyclic]{cyclic polymerisation} with flavour gauging can be stated, although it is exactly what one expects. \hyperref[fig:Chain]{Chain polymerisation} is the Coulomb branch global symmetry analogue of gauging two separate quiver gauge theories by a diagonal flavour symmetry. \hyperref[fig:Cyclic]{Cyclic polymerisation} is the Coulomb branch analogue of gauging a diagonal flavour symmetry in one quiver gauge theory \cite{Cremonesi:2014kwa, Cremonesi:2014vla}.

The $\surm(k)$ quotient quiver subtraction in \cite{Hanany:2023tvn} also realises gauging of subgroups of the Coulomb branch global symmetry of a theory in a similar fashion. 

It should be noted that the diagrammatic procedure of \hyperref[fig:Cyclic]{cyclic polymerisation} for $\urm(1)$ hyper-Kähler quotient encompasses the well-known ungauging of a $\urm(1)$ gauge node in a framed quiver, which also produces a $\urm(1)$ hyper-Kähler quotient on the Coulomb branch. The focus of this work is on illustrating the new technique of \hyperref[fig:Cyclic]{cyclic polymerisation} for non-trivial cases, so $\urm(k)$ hyper-Kähler quotients will be presented mainly for $k\geq 2$.

Readers simply interested in examples and applications of the \hyperref[fig:Chain]{chain} and \hyperref[fig:Cyclic]{cyclic polymerisation} techniques may skip the following sub-sections and proceed to Section \ref{sec:Chain}.

\subsection{Analytic methods}
\label{subsec:analytic}
Now that the diagrammatic techniques of \hyperref[fig:Chain]{chain} and \hyperref[fig:Cyclic]{cyclic polymerisation} have been presented, and the relevant embeddings for Weyl integration have been discussed, it remains to justify the polymerisation techniques. Ultimately, the polymerisation techniques follow from the orthogonality of modified Hall-Littlewood functions. The formulation is quite general and will be presented for a group $G$ where $G$ is $\surm(k)/\urm(k)$ for \hyperref[fig:Chain]{chain} and \hyperref[fig:Cyclic]{cyclic polymerisation} respectively.

Although the focus in this work is on unitary quivers, the formulation of the arguments easily generalises to orthosymplectic quivers as well.

\subsubsection{Justifying Chain Polymerisation --  Unitary Case}
\label{subsubsec:ChainUnitary}
The starting point is the pair of quivers $\mathcal Q_1$ and $\mathcal Q_2$ shown in \Figref{fig:ChainEmbed}. Each quiver in the pair contains a leg which takes the form of $(1)-\cdots-(k)-$. Each leg may have $g_1\;(\text{resp.}\;g_2)$ adjoint hypermultiplet loops on the $\urm(k)$ node. The leg is also attached to some background quiver $Q_1\;(\text{resp.}\;Q_2)$. 

The quiver $\mathcal Q_1$ can be thought of as the theory that results from gauging a diagonal $G=\surm(k)$ flavour symmetry of a $T[G]^{(1^k)}$ theory, $Q_1$, and $g_1$ free $G$ adjoint hypermultiplets. The Coulomb branch Hilbert series may be computed by gluing together the monopole formula contributions as:
\begin{equation}
\hs_{\mathcal C}\left[{\mathcal Q_1}\right](t, {\bf y},{\bf q}_1) = \sum\limits_{\lambda} {\underbrace {P^{G}_{\lambda}(t){~}{t^{ 2\Delta (\lambda)}}}_{\text{central node}}} 
{~}
\underbrace {{\left( {t^{- 2\Delta ( \lambda)}} \right)}^{g_1}}_{\text{adjoint loops}}
{~}
\underbrace 
{ \hs_{\mathcal C}\left[{Q_1}\right](t,{\bf q}_1,\lambda)}_{Q_1}{~}\underbrace{\hs_{\mathcal C}\left[{T[G]^{(1^k)}}\right](t,{\bf y},\lambda)}_{\text{leg}}.
\label{eq:GluedQ1}
\end{equation}
The notation is as follows: $\lambda$ defines the partition of magnetic charges for the node of $G$ according to the Dirac quantisation condition, $P^{G}_\lambda(t)$ is the dressing function, $\Delta(\lambda)$ refers to the contribution to the conformal dimension from matter that transforms in the adjoint representation of $G$ \cite{Bourget:2020xdz}. The Hilbert series may depend on some fugacities, ${\bf y}=\{y_1,\cdots,y_{\mathrm{rank}\;G}\}$ and $\bf{q_1}$ for the leg and for the background quiver $Q_1$, respectively. These fugacities are defined in \Figref{fig:ChainEmbed}.

The Coulomb branch Hilbert series for $\mathcal Q_2$ is found by simply exchanging $1\leftrightarrow 2$ and ${\bf y}\leftrightarrow{\bf y}^*$.

The form of the Hilbert series as shown in \eqref{eq:GluedQ1} is the typical form for a glued Hilbert series. However, it will be convenient to use a different but equivalent form of Hilbert series.

Firstly, the Hilbert series for the $T[G]^{\rho}$ theory in the presence of magnetic charges for $G$ (specified by partitions $\lambda$) are related to the Hilbert series for charged Slodowy slices \cite{Cabrera:2018ldc,Bourget:2020xdz} in the following way:

\begin{equation} 
{{\hs_{\mathcal C}}}\left[ {T[G]^{\rho}} \right](t,{\bf y},\lambda)
=\frac{\hs\left[\slice{\rho}^{G}\right](t,{\bf y},\lambda)}{\hs\left[\slice{{\cal N}}^{G}\right](t,\lambda)}
= {t^{-\Delta \left( \lambda  \right)}}{~}\hs\left[\slice{\rho}^{G}\right](t,{\bf y},\lambda),
\label{eq:theory2}
\end{equation}
where the following relationship is used:

\begin{equation}
\hs\left[\slice{{\cal N}}^{G}\right](t,\lambda) = {t^{\Delta (\lambda)}}.
\label{eq:theory3}
\end{equation}

Secondly, the partitions $\lambda$, which are partitions for the magnetic charges of $G$, are in one-to-one correspondence with irreps of $G$. Therefore $\lambda$ may be replaced by a collection of Dynkin labels $[\bf n]$ for $G$, with the sum now being over Dynkin labels of $G$ instead. This change of $\lambda\rightarrow [\bf n]$ will be mostly implicit. However it is useful for pedagogical purposes to show how the $\Delta(\lambda)$ become explicit functions of the $[\bf n]$ for the specific case of $G=\surm(k)$.  Let $\bf A$ be the Cartan matrix of $\surm(k)$, then $ {\Delta(\lambda [{\bf n}]) \equiv \Delta [{\bf n}]  ={- [{\bf n}]}\cdot{{\bf{A}}^{{ - 1}}}\cdot{\bf{2}}}$. Alternatively, drawing on the Characteristics and weight maps $\omega$ of nilpotent orbits, $\Delta[\bf n]$ can be found simply from the weight map of the regular orbit as, ${{\Delta [\bf n]}=- [\bf n] \cdot \omega ({\text{reg.}})}$ \cite{Hanany:2016gbz,Hanany:2017ooe}.

After making these two changes, one further observation that can be made is the equivalence of $\hs\left[\slice{(1^k)}^{G}\right](t,{\bf y},[{\bf n}])=\mHL^{G}_{[\bf n]}(t,{\bf y})$ where $\mHL$ is a \textit{modified Hall-Littlewood} function. A full account of $\mHL$ and the notations used herein are found in \cite{Hanany:2015hxa}. This finally brings \eqref{eq:GluedQ1} to the following form:

\begin{equation}
  \hs_{\mathcal C}\left[{\mathcal Q_1}\right](t, {\bf y},{\bf q}_1) = \sum\limits_{[{\bf n}]} P^{G}_{[{\bf n}]}(t){~}{t^{\Delta [{\bf n}](1-2g_1)}}
{~}
 \hs_{\mathcal C}\left[{Q_1}\right](t,{\bf q}_1,[{\bf n}]){~}\mHL^{G}_{[{\bf n}]}(t,{\bf y}).\label{eq:GluedQ1Two}
\end{equation}

\hyperref[fig:Chain]{Chain polymerisation} of $\mathcal Q_1$ and $\mathcal Q_2$ produces a new quiver $\mathcal Q_3$. The action on the Coulomb branch is a hyper-Kähler quotient by the diagonal $G$ in each leg of $\mathcal Q_1$ and $\mathcal Q_2$; this gives the Coulomb branch of $\mathcal Q_3$. A key assumption is that the hyper-Kähler quotient involves complete Higgsing of the Coulomb branch global symmetry. With this assumption, the hyper-Kähler quotient is evaluated using the Weyl integration formula \eqref{eq:weyl}:

\begin{equation}
    \hs_{\mathcal C}\left[{\mathcal Q_3}\right](t,{\bf q}_1,{\bf q}_2)=\oint_{G}d\mu^{G}({\bf y})\pe\left[-\chi^{G}_{\text{adj}}({\bf y})t^2\right]\hs_{\mathcal C}\left[{\mathcal Q_1}\right](t,{\bf y},{\bf q}_1)\hs_{\mathcal C}\left[{\mathcal Q_2}\right](t,{\bf y^*},{\bf q}_2).\label{eq:Q12Weyl}
\end{equation}
Recall the key orthogonality relations between the $\mHL$ functions:

\begin{equation}
\begin{aligned}
\oint\limits_{G} {d\mu _{\mHL}^{G}}({\bf y})
{~}\mHL_{[\bf m]}^{G}({{{\bf y}}^*},t)
{~}\mHL_{[\bf n]}^{G}({{\bf y}},t) = \frac{{{\delta _{\bf [ m ][ n ]}}}}{{{{\left( {1 - {t^2}} \right)}^{\mathrm{rank}\;G}}P_{[\bf n]}^{G}(t)}},
\label{eq:mHLOrthog}
\end{aligned}
\end{equation}
where ${d\mu _{\mHL}^{G}}$ is the modified Hall-Littlewood Haar measure,

\begin{equation}
\begin{aligned}
{d\mu _{\mHL}^{G}}({\bf y})\equiv{d\mu^{G}}({\bf y}){~}{\pe \left[\left( \mathrm{rank}\;G - \chi _{\mathrm{\text{adj}}}^{G}({\bf y})\right){t^2} \right]}.
\label{eq:mHLMeasure}
\end{aligned}
\end{equation}
Substituting \eqref{eq:mHLOrthog} into \eqref{eq:mHLMeasure} gives the convenient result:

\begin{equation}
\begin{aligned}
\oint\limits_{G} {d\mu ^{G}}
{~}{\pe \left[ {- \chi _{\mathrm{adjoint}}^{G}{t^2}} \right]}
{~}\mHL_{[\bf m]}^{G}({{\bf{y}}^*},t)
{~}\mHL_{[\bf n]}^{G}({\bf{y}},t) = \frac{{{\delta _{\bf [ m ][ n ]}}}}
{{P_{[\bf n]}^{G}(t)}}.
\label{eq:Haar2mHLMeasure}
\end{aligned}
\end{equation}

Now, the explicit form of the Coulomb branch Hilbert series for $\mathcal Q_1$ \eqref{eq:GluedQ1Two}, and similarly for $\mathcal Q_2$, may be substituted into \eqref{eq:Q12Weyl}. Upon performing the integration and using the orthogonality property of the modified Hall-Littlewood functions \eqref{eq:Haar2mHLMeasure}, the Coulomb branch Hilbert series for $\mathcal Q_3$ is computed as \begin{equation}
     \hs_{\mathcal C}\left[{\mathcal Q_3}\right](t,{\bf q}_1,{\bf q}_2)=\sum\limits_{[{\bf n}]} P^{G}_{[{\bf n}]}(t){~}{t^{2\Delta [{\bf n}](1-(g_1+g_2))}}
{~}
 \hs_{\mathcal C}\left[{Q_1}\right](t,{\bf q}_1,[{\bf n}]){~}\hs_{\mathcal C}\left[{Q_2}\right](t,{\bf q}_2,[{\bf n}]).\label{eq:Q12Polymerised}
\end{equation}It is clear that the Coulomb branch Hilbert series of $\mathcal Q_3$ shown in \eqref{eq:Q12Polymerised} is that of the Coulomb branch of the resulting quiver in \Figref{fig:Chain}.
\subsubsection{Justifying Cyclic Polymerisation -- Unitary Case}
\label{subsubsec:CyclicUnitary}
Much of the above analysis carries over to the case of \hyperref[fig:Cyclic]{cyclic polymerisation}, except with some minor differences.

The starting point is the quiver $\mathcal Q$ shown in \Figref{fig:CyclicEmbed}. This quiver contains two legs of the form $(1)-\cdots-(k)-$ with $g_1$ and $g_2$ adjoint hypermultiplets on each $\urm(k)$. These legs are connected, perhaps in different ways, to some background quiver $Q$. 

The Coulomb branch Hilbert series of $\mathcal Q$ may be computed by gluing the two legs to the background quiver $Q$. The embedding given in \Figref{fig:CyclicEmbed} requires that the symmetry in each leg is broken to $\surm(k)\times\urm(1)$. Each leg carries a diagonal $G=\urm(k)$ symmetry with the remaining $\urm(1)$ dependence being absorbed into the background quiver $Q$. The background quiver $Q$ has fugacities $\bf q$, which include this remaining $\urm(1)$ from the legs. Each leg carries conjugate $G$ fugacities as before. Therefore, the Coulomb branch Hilbert series of $\mathcal Q$ is computed as: \begin{align}
\hs_{\mathcal C}\left[{\mathcal Q}\right](t, {\bf y},{\bf q}) &= \sum\limits_{\lambda,\lambda'}  P^{G}_{\lambda}(t){~}P^{G}_{\lambda'}(t){~}t^{ 2\Delta (\lambda)\left(1-g_1\right)}{~}t^{ 2\Delta (\lambda')\left(1-g_2\right)}\nonumber\\&\times\hs_{\mathcal C}\left[{Q}\right](t,{\bf q},\lambda,\lambda'){~}\hs_{\mathcal C}\left[{T[G]^{(1^k)}}\right](t,{\bf y},\lambda)\hs_{\mathcal C}\left[{T[G]^{(1^k)}}\right](t,{\bf y^*},\lambda').
\end{align}

The notation is similar, but as there are two gluings involved this time, there is another set of partitions $\lambda'$ for $G$ magnetic charges. The sum over partitions $\lambda$ and $\lambda'$ may be changed to a sum over Dynkin labels $[\bf n]$ and $[\bf m]$ of $G$ to get to the following, but equivalent, Hilbert series: \begin{align}
    \hs_{\mathcal C}\left[{\mathcal Q}\right](t, {\bf y},{\bf q},q) &= \sum\limits_{[{\bf n}],[{\bf m}]}  P^{G}_{[{\bf n}]}(t){~}P^{G}_{[{\bf m}]}(t){~}t^{\Delta[{\bf n}]\left(1-2g_1\right)}{~}t^{\Delta[{\bf m}]\left(1-2g_2\right)}\nonumber\\&\times\hs_{\mathcal C}\left[{Q}\right](t,{\bf q},[{\bf n}],[{\bf m}]){~}\mHL^{G}_{[{\bf n}]}(t,{\bf y}){~}\mHL^{G}_{[{\bf m}]}(t,{\bf y^*}).
\end{align}

\hyperref[fig:Cyclic]{Cyclic polymerisation} takes the quiver $\mathcal Q$ and produces a new quiver $\mathcal Q'$. The hyper-Kähler quotient action on the Hilbert series is again computed by Weyl integration by $G$ with fugacities $\bf y$. The assumption that there is complete Higgsing is made again. The Coulomb branch Hilbert series of $\mathcal Q'$ is: 

\begin{equation}
     \hs_{\mathcal C}\left[{\mathcal Q'}\right](t,{\bf q})=\oint_{G}d\mu^{G}({\bf y})\pe\left[-\chi^{G}_{\text{adj}}({\bf y})t^2\right]\hs_{\mathcal C}\left[{\mathcal Q}\right](t, {\bf y},{\bf q}).
\end{equation}
The integral over $G$ proceeds in a similar fashion to the case of \hyperref[fig:Chain]{chain polymerisation}. The final result is: \begin{equation}
    \hs_{\mathcal C}\left[{\mathcal Q'}\right](t,{\bf q})=\sum_{[\bf n]}P^{G}_{[\bf n]}(t){~}t^{2\Delta[{\bf n}]\left(1-(g_1+g_2)\right)}{~}\hs_{\mathcal C}[Q](t,{\bf q},[{\bf n}],[{\bf n}]).
\end{equation}
This is the correct form for the Coulomb branch Hilbert series of $\mathcal Q'$ as expected from polymerisation and the monopole formula.

\subsection{Connections to class $\mathcal S$}
\label{subsec:ClassS}
The \hyperref[fig:Chain]{chain} and \hyperref[fig:Cyclic]{cyclic polymerisation} techniques are related to those presented in \cite{Benini:2010uu} and \cite{Chacaltana:2010ks} for star-shaped quivers which are $3d$ mirrors of class $\mathcal S$ theories \cite{Gaiotto:2009we} with \textit{regular} punctures. For this reason a brief account of class $\mathcal S$ theories and their $3d$ mirrors and their gluing is presented \cite{Benini:2010uu}. There is discussion of the incomplete Higgsing in the class $\mathcal S$ theories which are not visible in the $3d$ mirror theories. Finally, the concept of \hyperref[fig:ClassSCyclicFix]{quiver extensions} is introduced as a way to cure the incomplete Higgsing.
\subsubsection{Class $\mathcal S$ theories and their $3d$ mirrors}
\begin{figure}[h!]
    \begin{subfigure}{\textwidth}
    \centering
        \begin{tikzpicture}
        
       \draw[] (0,0+8) circle (3);

        \node[label=left:$\rho_1$] (p1) at (-1.5,1.5+8) {$\times$};
        \node (vdots) at (-1.5,0+8) {$\vdots$};
        \node[label=left:$\rho_n$] (pn) at (-1.5,-1.5+8) {$\times$};
        
         \draw (-0.75,1.5+8) to[bend left] (0.75,1.5+8);
        \draw (-0.95,1.6+8) to[bend right] (0.95,1.6+8);

        \node[label=right:$g\;\mathrm{times}$] (vdots2) at (0,0+8) {$\vdots$};

        \draw (-0.75,1.5-3+8) to[bend left] (0.75,1.5-3+8);
        \draw (-0.95,1.6-3+8) to[bend right] (0.95,1.6-3+8);
        \end{tikzpicture}
        \subcaption{}
        \label{fig:ClassSRS}
    \end{subfigure}
       \begin{subfigure}{\textwidth}
       \centering
    \begin{tikzpicture}
    
        \node[gauge, label=above:$k$] (k) []{};
    
        \node[] (p1) [above left=of k]{$T_{\rho_1}$};
        \node[] (cdotst) [above=of k]{$\cdots$};
        \node[] (pn) [above right=of k]{$T_{\rho_n}$};

        \draw[-] (p1)--(k)--(pn);
         \draw (k) to [out=225, in=-45,looseness=8]  node[pos=0.5,below]{$g$} (k);
    \end{tikzpicture}
    \subcaption{}
    \label{fig:ClassSQuiv}
    \end{subfigure}
    \caption{In \Figref{fig:ClassSRS} is a punctured Riemann surface $\mathcal C_{n,g}$ of genus $g$ with $n$ punctures, $\rho_i$, of type $A_{k-1}$. In \Figref{fig:ClassSQuiv} is the corresponding $3d$ mirror which consists of legs which are the $T^{\rho_i}[\surm(k)]$ glued at the common flavour symmetry. There are $g$ adjoint on the central $\urm(k)$ gauge node corresponding to the genus of the Riemann surface.}
    \label{fig:ClassS3dMirror}
\end{figure}

A class $\mathcal S$ theory is a $4d\;\mathcal N=2$ SCFT specified by a genus $g$ Riemann surface $\mathcal C_{n,g}$ with $n$ punctures labelled by nilpotent orbits of some $ADE$ Lie algebra $\mathfrak{g}$. This $4d\;\mathcal N=2$ SCFT comes from compactification of a $6d\;\mathcal N=(2,0)$ theory specified by the same $ADE$ algebra $\mathfrak{g}$ on $\mathcal C_{n,g}$ with a topological twist.

For the examples herein, the algebra is $A_{k-1}$ with partitions of $k$ labelling the punctures. In terms of notation, $\rho_i,\,i \in \{1,\cdots,n\}$ denotes a puncture and also the corresponding partition of $k$ labelling that puncture. The $3d$ mirror theory for a class $\mathcal S$ theory is a star-shaped quiver with each leg corresponding to a $T^{\rho_i}[\surm(k)]$ quiver \cite{Cremonesi:2014uva}, which is also a magnetic quiver for the Slodowy slice labelled by the same partition \cite{Cabrera:2018ldc}. The common $\surm(k)$ flavour symmetry is gauged leading to a star-shaped quiver with $\urm(k)$ gauge group at the centre of the star once the overall $\urm(1)$ is included. Additionally, $g$ adjoint hypermultiplets are added to the central $\urm(k)$ gauge node due to the genus $g$ of $\mathcal C_{n,g}$. This is illustrated schematically in \Figref{fig:ClassS3dMirror}.
\FloatBarrier
\subsubsection{Gluing Maximal Punctures}
\begin{figure}[h!]
  \centering
     \begin{subfigure}{\textwidth}
    \centering
    \resizebox{!}{0.43\paperheight}{\begin{tikzpicture}
    \node (a) at (0,0){$\begin{tikzpicture}

        \draw[] (0,0) circle (3);

        \node[label=left:$\rho_1$] (p1) at (-1.5,1.5) {$\times$};
        \node (vdots) at (-1.5,0) {$\vdots$};
        \node[label=below:$\rho_{n_1-1}$] (pn) at (-1.5,-1.5) {$\times$};
        
         \draw (-0.75,1.5) to[bend left] (0.75,1.5);
        \draw (-0.95,1.6) to[bend right] (0.95,1.6);

        \node[label=right:$g_1\;\mathrm{times}$] (vdots2) at (0,0) {$\vdots$};

        \draw (-0.75,1.5-3) to[bend left] (0.75,1.5-3);
        \draw (-0.95,1.6-3) to[bend right] (0.95,1.6-3);
        \node[label=below:$(1^k)$] (k1) at (2.5,0){$\times$};
        
        \draw[] (8,0) circle (3);

        \node[label=below:$(1^k)$] (k2) at (8-2.5,0){$\times$};

         \draw (-0.75+8,1.5) to[bend left] (0.75+8,1.5);
        \draw (-0.95+8,1.6) to[bend right] (0.95+8,1.6);

        \node[label=left:$g_2\;\mathrm{times}$] (vdots2) at (0+8,0) {$\vdots$};

        \draw (-0.75+8,1.5-3) to[bend left] (0.75+8,1.5-3);
        \draw (-0.95+8,1.6-3) to[bend right] (0.95+8,1.6-3);

        \node[label=right:$\rho'_1$] (pp1) at (8+1.5,1.5) {$\times$};
        \node (vdots) at (8+1.5,0) {$\vdots$};
        \node[label=below:$\rho'_{n_2-1}$] (ppm) at (8+1.5,-1.5) {$\times$};
    \end{tikzpicture}$};
    \node (b) at (0,-7){$\begin{tikzpicture}

        \node[label=left:$\rho_1$] (p1) at (-1.5,1.5) {$\times$};
        \node (vdots) at (-1.5,0) {$\vdots$};
        \node[label=below:$\rho_{n_1-1}$] (pn) at (-1.5,-1.5) {$\times$};
        
         \draw (-0.75,1.5) to[bend left] (0.75,1.5);
        \draw (-0.95,1.6) to[bend right] (0.95,1.6);

        \node[label=right:$g_1\;\mathrm{times}$] (vdots2) at (0,0) {$\vdots$};

        \draw (-0.75,1.5-3) to[bend left] (0.75,1.5-3);
        \draw (-0.95,1.6-3) to[bend right] (0.95,1.6-3);

         \draw (-0.75+8,1.5) to[bend left] (0.75+8,1.5);
        \draw (-0.95+8,1.6) to[bend right] (0.95+8,1.6);

        \node[label=left:$g_2\;\mathrm{times}$] (vdots2) at (0+8,0) {$\vdots$};

        \draw (-0.75+8,1.5-3) to[bend left] (0.75+8,1.5-3);
        \draw (-0.95+8,1.6-3) to[bend right] (0.95+8,1.6-3);

        \node[label=right:$\rho'_1$] (pp1) at (8+1.5,1.5) {$\times$};
        \node (vdots) at (8+1.5,0) {$\vdots$};
        \node[label=below:$\rho'_{n_2-1}$] (ppm) at (8+1.5,-1.5) {$\times$};

        \draw ({sqrt(9-0.04)-0.3},{0.2}) to ({sqrt(9-0.04)+2+0.3},{0.2});

        \draw (2,0.3) to[out=-20,in=180] (2.69332590942,0.2);

        \draw (2.2,-0.3) to[out=20,in=180] (2.89332590942,-0.2);

         \draw ({sqrt(9-0.04)},{0.2})  arc  (+12/pi:360-12/pi:3);
        
        \draw ({sqrt(9-0.04)-0.3+0.2},{-0.2}) to ({sqrt(9-0.04)+2+0.3-0.2},{-0.2});

        \draw (6,0.3) to[out=-160,in=0] (5.29332590942,0.2);

        \draw (5.8,-0.3) to[out=160,in=0] (5.09332590942,-0.2);

        \draw ({sqrt(9-0.04)+2},{-0.2})  arc  (180+12/pi:540-12/pi:3);

    \end{tikzpicture}$};

    \draw[->] (a)--(b);
        \end{tikzpicture}}
    \subcaption{}
    \label{fig:ClassSChainRS}
    \end{subfigure}
    \centering
    \begin{subfigure}{\textwidth}
    \centering
    \begin{tikzpicture}
    \node (a) at (0,0) {$\begin{tikzpicture}
        \node[gauge, label=above:$k$] (kl) []{};
        \node[] (cdotsl) [right=of kl]{$\cdots$};
        \node[gauge, label=below:$2$] (2l) [right=of cdotsl]{};
        \node[gauge, label=below:$1$] (1l) [right=of 2l]{};
        \node[] (p1) [above left=of kl] {$T_{\rho_1}$};
        \node[] (cdotstl) [above=of kl] {$\cdots$};
        \node[] (pn) [above right=of kl] {$T_{\rho_{n_1-1}}$};

        \draw[-] (1l)--(2l)--(cdotsl)--(kl)--(p1) (kl)--(pn);

        \draw (kl) to [out=225, in=-45,looseness=8]  node[pos=0.5,below]{$g_1$} (kl);

        \node[gauge, label=below:$1$] (1r) [right=of 1l]{};
        \node[gauge, label=below:$2$] (2r) [right=of 1r]{};
        \node[] (cdotsr) [right=of 2r]{$\cdots$};
        \node[gauge, label=above:$k$] (kr) [right=of cdotsr]{};
        \node[] (pp1) [above left=of kr] {$T_{\rho'_1}$};
        \node[] (cdotstt) [above=of kr] {$\cdots$};
        \node[] (ppm) [above right=of kr] {$T_{\rho'_{n_2-1}}$};

        \draw[-] (1r)--(2r)--(cdotsr)--(kr)--(pp1) (kr)--(ppm);

        \draw (kr) to [out=225, in=-45,looseness=8]  node[pos=0.5,below]{$g_2$} (kr);

    \end{tikzpicture}$};
     \node (b) at (0,-5){$\begin{tikzpicture}
        \node[gauge, label=above:$k$] (k) []{};
        \node[] (p1) [above left=of k]{$T_{\rho_1}$};
        \node[] (pp1) [below left=of k]{$T_{\rho'_1}$};
        \node[] (pn) [above right=of k]{$T_{\rho_{n_1-1}}$};
        \node[] (ppm) [below right=of k]{$T_{\rho'_{n_2-1}}$};
        \node[] (cdotst) [above=of k]{$\cdots$};
        \node[] (cdotsb) [below=of k]{$\cdots$};

        \draw[-] (p1)--(k)--(pn) (pp1)--(k)--(ppm);

        \draw (k) to [out=225, in=-45,looseness=8]  node[pos=0.5,below]{$g_1+g_2$} (k);
    \end{tikzpicture}$};
    \draw[->] (a)--(b);
    \end{tikzpicture}
    \subcaption{}
    \label{fig:ClassSChainQuiv}
    \end{subfigure}
    \caption{In (\Figref{fig:ClassSChainRS}) is the gluing of maximal punctures on the Riemann surfaces. In (\Figref{fig:ClassSChainQuiv}) is the the perspective on the $3d$ mirror quiver, this is the same as \hyperref[fig:Chain]{chain polymerisation} at the node of $\urm(k)$.}
    \label{fig:ClassSChain}
\end{figure}

In class $\mathcal S$ theories with $A_{k-1}$ Lie algebra, two punctured Riemann surfaces $\mathcal C_{n_1,g_1}$ and $\mathcal C_{n_2,g_2}$, with at least one maximal puncture each, may be joined by a cylinder to form a new Riemann surface $\mathcal C_{n_1+n_2-2,g_1+g_2}$. Without loss of generality let the $n_1^{\text{th}}$ and $n_2^{\text{th}}$ punctures be the maximal punctures to be joined. The effect on the $4d\;\mathcal N=2$ theory is of gauging the diagonal $\surm(k)$ flavour symmetry. The action on the $3d$ mirror quiver is the gauging of a diagonal $\surm(k)$ Coulomb branch global symmetry as discussed in \cite{Benini:2010uu} which is the same as \hyperref[fig:Chain]{chain polymerisation} at the central node. This is summarised in \Figref{fig:ClassSChain}.

The \hyperref[fig:Chain]{chain polymerisation} introduced in this work generalises this action on $3d\;\mathcal N=4$ quivers which are not necessarily star-shaped or simply laced. Additionally the subgroup of the Coulomb branch global symmetry being gauged is not restricted to just $\surm(k)$ but any $\surm(k')$ for $k'\leq k$, given a suitable choice of nodes \cite{Ferlito:2017xdq}.
\begin{figure}[h!]

    \begin{subfigure}{\textwidth}
    \centering
        \begin{tikzpicture}
        
       \draw[] (0,0+8) circle (3);

        \node[label=left:$\rho_1$] (p1) at (-1.5,1.5+8) {$\times$};
        \node (vdots) at (-1.5,0+8) {$\vdots$};
        \node[label=left:$\rho_n$] (pn) at (-1.5,-1.5+8) {$\times$};
        
         \draw (-0.75,1.5+8) to[bend left] (0.75,1.5+8);
        \draw (-0.95,1.6+8) to[bend right] (0.95,1.6+8);

        \node[label=right:$g\;\mathrm{times}$] (vdots2) at (0,0+8) {$\vdots$};

        \draw (-0.75,1.5-3+8) to[bend left] (0.75,1.5-3+8);
        \draw (-0.95,1.6-3+8) to[bend right] (0.95,1.6-3+8);
        \node[label=below:$(1^k)$] (k1) at (1.5,1.5+8){$\times$};
        \node[label=below:$(1^k)$] (k2) at (1.5,-1.5+8){$\times$};
        
        \node[label=left:$\rho_1$] (P1) at (-1.5,1.5) {$\times$};
        \node (Vdots) at (-1.5,0) {$\vdots$};
        \node[label=left:$\rho_n$] (Pn) at (-1.5,-1.5) {$\times$};
        
         \draw (-0.75,1.5) to[bend left] (0.75,1.5);
        \draw (-0.95,1.6) to[bend right] (0.95,1.6);

        \node[label=right:$g\;\mathrm{times}$] (Vdots2) at (0,0) {$\vdots$};

        \draw (-0.75,1.5-3) to[bend left] (0.75,1.5-3);
        \draw (-0.95,1.6-3) to[bend right] (0.95,1.6-3);
        
        \draw ({sqrt(6.11)},{1.7}) to 
        ({3},{1.7});

        \draw ({sqrt(7.31)},{1.3}) to ({3},{1.3});

        \draw ({sqrt(6.11)},{-1.7}) to 
        ({3},{-1.7});

        \draw ({sqrt(7.31)},{-1.3}) to ({3},{-1.3});

        \draw ({3},{1.7}) to [out=0,in=0,looseness=1.7] ({3},{-1.7});
        \draw ({3},{1.3}) to [out=0,in=0,looseness=1.7] ({3},{-1.3});

         \draw ({1.5},{1.8}) to[out=-20,in=180] (2.47184141886,1.7);

         \draw ({1.5},{1.2}) to[out=20,in=180] (2.70370116692,1.3);

         \draw ({1.5},{-1.8}) to[out=20,in=180] (2.47184141886,-1.7);

         \draw ({1.5},{-1.2}) to[out=-20,in=180] (2.70370116692,-1.3);

         \draw (2.47184141886,1.7)  arc  ({asin(1.7/3)}:{360-asin(1.7/3)}:3);

         \draw (2.70370116692,-1.3)  arc  ({-asin(1.3/3)}:{asin(1.3/3)}:3);

         \draw[->] (0,4.5)--(0,3.5);
        \end{tikzpicture}
        \subcaption{}
        \label{fig:ClassSCyclicRS}
    \end{subfigure}
       \begin{subfigure}{\textwidth}
    \begin{tikzpicture}
        \node[gauge, label=below:$1$] (1l) []{};
        \node[gauge, label=below:$2$] (2l) [right=of 1l]{};
        \node[] (cdotsl) [right=of 2l]{$\cdots$};
        \node[gauge, label=above:$k$] (k) [right=of cdotsl]{};
        \node[] (cdotsr) [right=of k]{$\cdots$};
        \node[gauge, label=below:$2$] (2r) [right=of cdotsr]{};
        \node[gauge, label=below:$1$] (1r) [right=of 2r]{};

        \node[] (p1) [above left=of k]{$T_{\rho_1}$};
        \node[] (cdotst) [above=of k]{$\cdots$};
        \node[] (pn) [above right=of k]{$T_{\rho_n}$};

        \draw[-] (1l)--(2l)--(cdotsl)--(k)--(cdotsr)--(2r)--(1r);
        \draw[-] (p1)--(k)--(pn);
         \draw (k) to [out=225, in=-45,looseness=8]  node[pos=0.5,below]{$g$} (k);
        
        \node[] (Cdotst) [below=of k] {$\cdots$};
        \node[gauge, label=above:$k$] (K) [below=of Cdotst]{};
        \node[] (P1) [above left=of K]{$T_{\rho_1}$};
        \node[] (Pn) [above right=of K]{$T_{\rho_n}$};

        \draw[-] (P1)--(K)--(Pn);
        \draw (K) to [out=225, in=-45,looseness=8] node[pos=0.5,below]{$g+1$} (K);

        \node[] (topghost) [left=of 1l]{};
        \node[] (bottomghost) [left=of K]{};
        \draw [->] (topghost) to [out=-150,in=150,looseness=1] (bottomghost);
    \end{tikzpicture}
    \subcaption{}
    \label{fig:ClassSCyclicQuiv}
    \end{subfigure}
    \caption{In \Figref{fig:ClassSCyclicRS} is the gluing of two maximal punctures on the same Riemann surface. In \Figref{fig:ClassSCyclicQuiv} is the corresponding action on the $3d$ mirror quivers, this differs from \hyperref[fig:Cyclic]{cyclic polymerisation}.}
    \label{fig:ClassSCyclic}
\end{figure}

Also in class $\mathcal S$, a Riemann surface $\mathcal C_{n+2,g}$ with at least two maximal $\surm(k)$ punctures may also be glued to itself with a cylinder. The resulting Riemann surface is $\mathcal C_{n,g+1}$. The action on the $3d$ mirror quiver consists of the amputation of these maximal external legs and the addition of an adjoint hypermultiplet at the central node of $\surm(k)$. This adjoint hypermultiplet corresponds to a unit increase in the genus of the Riemann surface. This is summarised in \Figref{fig:ClassSCyclic}. The \hyperref[fig:Cyclic]{cyclic polymerisation} introduced in this work is not the same as the action in \Figref{fig:ClassSCyclic}. This is because cyclic polymerisation gives complete Higgsing of the Coulomb branch global symmetry which is not the case for the action in \Figref{fig:ClassSCyclic} as will be reviewed in Section \ref{sec:quiverextensions}.

The \hyperref[fig:Cyclic]{cyclic polymerisation} introduced in this work gauges a $\surm(k')\subset\surm(k)$ subgroup of the symmetry of the puncture. 

It is important to reiterate that although for the $4d\;\mathcal N=2$ SCFT gauging $\surm(k')\subset \surm(k)$ subgroups of the flavour symmetry in general lead to IR free theories\footnote{We thank Gabi Zafrir for discussion about this point.}, the focus of this study is on $3d\;\mathcal N=4$ theories.

\FloatBarrier
\subsubsection{Quiver Extensions and Cyclic Polymerisation}
\label{sec:quiverextensions}
When self-gluing a Riemann surface, as shown in \Figref{fig:ClassSCyclic}, there is an increase in the genus of the Riemann surface by one. This increase in the genus means that the Cartan subalgebra of the flavour symmetry being gauged cannot receive a mass. Therefore there is incomplete Higgsing in this case as there remain $\textrm{rank}\;\surm(k)=k-1$ massless vectormultiplets.

The action on the $3d$ mirror quivers in \Figref{fig:ClassSCyclic} differs from \hyperref[fig:Cyclic]{cyclic polymerisation}. However, it is possible to engineer a new quiver theory that is amenable to \hyperref[fig:Cyclic]{cyclic polymerisation} to recover the desired class $\mathcal S$ theory, while respecting complete Higgsing. The way to engineer such a quiver is to take the star-shaped quiver for the class $\mathcal S$ theory defined through $\mathcal C_{n+2,g}$, with at least two maximal punctures, and to \hyperref[fig:ClassSCyclicFix]{extend} one leg of the quiver by a further node of $\urm(k)$.

The legs corresponding to the punctures are split between the two nodes of $\urm(k)$ in any way and similarly the adjoint loops. The engineered moduli spaces are generally different in each case. Nevertheless, under $\urm(k)$ \hyperref[fig:Cyclic]{cyclic polymerisation} the same resulting quiver is found. This resulting quiver is also the one which is expected from purely class $\mathcal S$ arguments. This is summarised in \Figref{fig:ClassSCyclicFix}.

\begin{figure}[h!]
    \centering
    \begin{tikzpicture}
    \node (a) at (0,0){$\begin{tikzpicture}
        \node[gauge, label=below:$1$] (1l) []{};
        \node[gauge, label=below:$2$] (2l) [right=of 1l]{};
        \node[] (cdotsl) [right=of 2l]{$\cdots$};
        \node[gauge, label=above:$k$] (k) [right=of cdotsl]{};
        \node[] (cdotsr) [right=of k]{$\cdots$};
        \node[gauge, label=below:$2$] (2r) [right=of cdotsr]{};
        \node[gauge, label=below:$1$] (1r) [right=of 2r]{};

        \node[] (p1) [above left=of k]{$T_{\rho_1}$};
        \node[] (cdotst) [above=of k]{$\cdots$};
        \node[] (pn) [above right=of k]{$T_{\rho_n}$};

        \draw[-] (1l)--(2l)--(cdotsl)--(k)--(cdotsr)--(2r)--(1r);
        \draw[-] (p1)--(k)--(pn);

        \draw (k) to [out=225, in=-45,looseness=8]  node[pos=0.5,below]{$g$} (k);
    \end{tikzpicture}$};

    \node (b) at (0,-5){$\begin{tikzpicture}
         \node[gauge, label=below:$1$] (1l) []{};
        \node[gauge, label=below:$2$] (2l) [right=of 1l]{};
        \node[] (cdotsl) [right=of 2l]{$\cdots$};
        \node[gauge, label=below left:$k$] (kl) [right=of cdotsl]{};
        \node[gauge, label=below right:$k$] (kr) [right=of kl]{};
        \node[] (cdotsr) [right=of kr]{$\cdots$};
        \node[gauge, label=below:$2$] (2r) [right=of cdotsr]{};
        \node[gauge, label=below:$1$] (1r) [right=of 2r]{};

        \node[] (p1) [above left=of kl]{$T_{\rho_1}$};
        \node[] (pj) [above =of kl]{$T_{\rho_j}$};

        \node[] (pj1) [above=of kr]{$T_{\rho_{j+1}}$};
        \node[] (pn) [above right=of kr] {$T_{\rho_n}$};

        \draw[-] (1l)--(2l)--(cdotsl)--(kl)--(kr)--(cdotsr)--(2r)--(1r) (p1)--(kl)--(pj) (pj1)--(kr)--(pn);

        \draw (kl) to [out=225, in=-45,looseness=8]  node[pos=0.5,below]{$g'$} (kl);
        \draw (kr) to [out=225, in=-45,looseness=8]  node[pos=0.5,below]{$g-g'$} (kr);

    \end{tikzpicture}$};

    \node (c) at (0,-10){$\begin{tikzpicture}
         \node[] (Cdotst) {$\cdots$};
        \node[gauge, label=above:$k$] (K) [below=of Cdotst]{};
        \node[] (P1) [above left=of K]{$T_{\rho_1}$};
        \node[] (Pn) [above right=of K]{$T_{\rho_n}$};

        \draw[-] (P1)--(K)--(Pn);
        \draw (K) to [out=225, in=-45,looseness=8] node[pos=0.5,below]{$g+1$} (K);
    \end{tikzpicture}$};
    \draw[->] (a)--(b) node[midway,left] {$\text{Extension}$};
    \draw[->] (b)--(c) node[midway,left]{$\text{\hyperref[fig:Cyclic]{Cyclic polymerisation}}$};
    \end{tikzpicture}
    \caption{The extension of a star-shaped quiver to a form which is amenable to $\urm(k)$ \hyperref[fig:Cyclic]{cyclic polymerisation}. Then performing the $\urm(k)$ \hyperref[fig:Cyclic]{cyclic polymerisation}.}
    \label{fig:ClassSCyclicFix}
\end{figure}

Among the different possibilities of \hyperref[fig:ClassSCyclicFix]{quiver extensions}, there is one which stands out. This is the case where all of the extra quiver legs, not participating in \hyperref[fig:Cyclic]{cyclic polymersiation}, and all of the adjoint hypermultiplets are on the same $\urm(k)$ and none on the other. This "naked" $\urm(k)$ is now part of a leg that goes as $\urm(1)-\cdots-\urm(k)-\urm(k)-$, with the penultimate $\urm(k)$ having balance $-1$ and hence being "ugly" in the sense of \cite{Gaiotto:2008ak}. Following the \textit{dualisation} algorithm in \cite{Gaiotto:2008ak}, it is simple to see that the Coulomb branch is the Coulomb branch of the original theory with $k$ free twisted hypermultiplets. Thus obtaining the general result on the Coulomb branch,
\begin{equation}
    \left(\mathcal M_\mathcal C\left[\mathcal C_{n+2,g}\right]\times \mathbb H^k\right)///\urm(k)=\mathcal M_\mathcal C\left[\mathcal C_{n,g+1}\right].
\end{equation}

Physically, the introduction of free twisted hypermultiplets is a natural solution to solve the incomplete Higgsing problem as the massless twisted vector multiplets, coming from the cylinder, can now "eat" and completely Higgs the theory. Note that there were initially $k-1$ free massless twisted vectors left from the gauging of $\surm(k)$ from class $\mathcal S$ arguments; with \hyperref[fig:Cyclic]{cyclic polymerisation} there are an additional $k$ twisted hypermultiplets but the gauging is by $\urm(k)$. This is still consistent in terms of Coulomb branch dimensions (of the mirror $3d\;\mathcal N=4$ theory). 

The idea of adding free matter fields in the gauging of $4d\;\mathcal N=2$ SCFTs has been seen in \cite{Argyres:2007cn}, which was realised using $3d\;\mathcal N=4$ quivers in \cite{Hanany:2023tvn}.

Examples of \hyperref[fig:ClassSCyclicFix]{quiver extension} and \hyperref[fig:Cyclic]{cyclic polymerisation} are set out in Section \ref{sec:cyclicproduct}.

\FloatBarrier
\subsection{Kronheimer-Nakajima Quivers and Moduli Spaces of Instantons}
\label{sec:intantonreview}
\begin{figure}[h!]
    \centering
    \begin{tikzpicture}
        \node[gauge, label=left:$k_1$] (K1){};
        \node[] (ghostl) [below left=of K1]{};
        \node[] (ghostr) [below right=of K1]{};
        \node[gauge, label=below left:$k_2$] (K2)[below left=of ghostl]{};
        \node[gauge, label=below right:$k_N$] (KN)[below right=of ghostr]{};
        \node[] (cdotsres) at ($(K2)!0.5!(KN)$) {$\cdots$};

        \node[flavour,label=left:$l_1$] (F1) [above=of K1]{};
        \node[flavour,label=left:$l_2$] (F2) [above=of K2]{};
        \node[flavour,label=left:$l_N$] (FN) [above=of KN]{};

        \draw[-] (K1)--(K2)--(cdotsres)--(KN)--(K1);
        \draw[-] (F1)--(K1) (F2)--(K2) (FN)--(KN);
    
    \end{tikzpicture}
    \caption{Most general A-type Kronheimer-Nakajima quiver. The Higgs branch is the moduli space of instantons $\mathcal M^{\mathbb C^2/\mathbb Z_N;\surm(l)}_{(k_1,\cdots,k_N),(l_1,\cdots,l_N)}$, where $l=\sum_{i=1}^{N}l_i$. The Coulomb branch is the moduli space of instantons $\mathcal M^{\mathbb C^2/\mathbb Z_l;\surm(N)}_{(k'_1,\cdots,k'_l),(l'_1,\cdots,l'_l)}$, where the $k'_i$ and $l'_i$ are linear functions of the $k_i$ and $l_i$ respectively.}
    \label{fig:KNFrame}
\end{figure}
Another application of the technique of \hyperref[fig:Cyclic]{cyclic polymerisation} is in the construction of A-type Kronheimer-Nakajima quivers \cite{Kronheimer1990}, as will be presented in Section \ref{sec:instantonconstruc}. Both the Coulomb and Higgs branches of these quivers are moduli spaces of $\surm$ instantons on A-type singularity. The most general A-type Kronheimer-Nakajima quiver is shown in \Figref{fig:KNFrame}.

The quiver data encodes the geometric data of the instanton bundle. For example, the framing determines the monodromy of gauge fields at infinity and the balance determines the monodromy of gauge fields at the origin \cite{Witten:2009xu,Cherkis:2009hpw}. Both of these have interpretations in the Type IIB brane system as linking numbers of NS5 branes and D5 branes respectively. Further comments on brane realisations will be made in Section \ref{subsubsec:BraneKN}.

Examples of such geometric data are the first and second Chern classes, the monodromies of the gauge fields at infinity and the rank of the instanton bundle. Often this dictionary between quiver data and geometric data is cumbersome. In special cases of the Kronheimer-Nakajima quiver the dictionary is straight forward as will be illustrated in Section \ref{subsubsec:SpecialKN}.

\subsubsection{Brane Realisations}
\label{subsubsec:BraneKN}

Instantons can be thought of as solitonic codimension 4 objects. In string theory instantons are realised in Type IIA string theory as D2 branes dissolved in D6 branes \cite{Douglas:1995bn} which may also include orientifold planes. The low energy effective theory are the ADHM quivers whose Higgs branch are the moduli space of instantons of type $\surm,\,\sorm$, and $\sprm$ on $\mathbb C^2$. 

Application of T-duality and then S-duality to the Type IIA brane system gives a Type IIB brane system whose Coulomb phase corresponds to the same moduli space of instantons. The low energy effective field theories for moduli spaces of $\surm,\,\sorm$, and $\sprm$ instantons on $\mathbb C^2$ are extended Dynkin quivers as found in \cite{Intriligator:1996ex,deBoer:1996mp,Kapustin:1998fa,Hanany:1999sj,Cremonesi:2014xha}. For exceptional gauge groups, which do not have any perturbative open string formulations, the moduli space of instantons of these gauge symmetries on $\mathbb C^2$ are also extended Dynkin diagrams \cite{Intriligator:1996ex,Cremonesi:2014xha}.

Importantly, the Kronheimer-Nakajima quivers have constructions in Type IIB string theory as a system of NS5 and D5 branes with D3 branes on $S^1$ \cite{Hanany:1996ie}. Here, the action of S-duality on this brane system exchanges NS5 and D5 branes.  The effect on the quiver is that the number of gauge nodes and the total number of flavours swaps. 

\subsubsection{Special Cases}
\label{subsubsec:SpecialKN}



\begin{figure}[h!]
    \centering
    \begin{tikzpicture}
        \node[gauge, label=left:$k$] (K1){};
        \node[] (ghostl) [below left=of K1]{};
        \node[] (ghostr) [below right=of K1]{};
        \node[gauge, label=below left:$k$] (K2)[below left=of ghostl]{};
        \node[gauge, label=below right:$k$] (KN)[below right=of ghostr]{};
        \node[] (cdotsres) at ($(K2)!0.5!(KN)$) {$\cdots$};

        \node[flavour,label=left:$l_1$] (F1) [above=of K1]{};
        \node[flavour,label=left:$l_2$] (F2) [above=of K2]{};
        \node[flavour,label=left:$l_N$] (FN) [above=of KN]{};

        \draw[-] (K1)--(K2)--(cdotsres)--(KN)--(K1);
        \draw[-] (F1)--(K1) (F2)--(K2) (FN)--(KN);

         \draw [decorate, decoration = {brace, raise=10pt, amplitude=5pt}] (KN) --  (K2) node[pos=0.5,below=15pt,black]{$N-1$};
    \end{tikzpicture}
    \caption{The Kronheimer-Nakajima quiver with all gauge nodes the same. The Coulomb branch is the moduli space of $k\;\surm(N)$ instantons on $\mathbb C^2/\mathbb Z_{l}$ where $l=\sum_{i=1}^N l_i$.}
    \label{fig:KNFrameSame}
\end{figure}

A particularly special case of the Kronheimer-Nakajima quiver \Quiver{fig:KNFrame} is to set all of the gauge nodes to be the same $\urm(k)$, resulting in \Quiver{fig:KNFrameSame}. In particular, the balance of each gauge node is given by its framing. This also corresponds to the vanishing of the first Chern class $c_1$ of the instanton bundle. The second Chern class is $c_2=k$ which is the instanton number. Another simplification is in finding the residual gauge group broken by the monodromies of the gauge fields.

In the case where all $l_i>0$, the monodromy of the gauge fields breaks the flavour symmetry $\surm(l)
\rightarrow \mathrm{S}\left(\prod_{i=1}^{N}\urm(l_i)\right)$. The instantons live on $\mathbb C^2/\mathbb Z_N$ and the instanton number is $k$. Thus $\mathcal H(\text{\Quiver{fig:KNFrameSame}})=\mathcal M^{\mathbb C^2/\mathbb Z_N}_{k,\mathrm{S}\left(\prod_{i=1}^{N}\urm(l_i)\right)}$, where a simplification of the notation used in \Figref{fig:KNFrame} is now possible. The Coulomb branch is easily found from application of S-duality to the brane system for \Quiver{fig:KNFrameSame}, which results in $l\;\urm(k)$ gauge nodes in a necklace with framing $(0^{l_1-1},1,0^{l_2-1},1\cdots,0^{l_N-1},1)$. The Coulomb branch of \Quiver{fig:KNFrameSame} is $\mathcal C\left(\text{\Quiver{fig:KNFrameSame}}\right)=\mathcal M_{k,\urm(1)^{N-1}}^{\mathbb C^2/\mathbb Z_l}$.

\begin{figure}[h!]
    \centering
    \begin{tikzpicture}
        \node[gauge, label=left:$k$] (K1){};
        \node[] (ghostl) [below left=of K1]{};
        \node[] (ghostr) [below right=of K1]{};
        \node[gauge, label=below left:$k$] (K2)[below left=of ghostl]{};
        \node[gauge, label=below right:$k$] (KN)[below right=of ghostr]{};
        \node[] (cdotsres) at ($(K2)!0.5!(KN)$) {$\cdots$};

        \node[flavour,label=left:$l$] (F1) [above=of K1]{};

        \draw[-] (K1)--(K2)--(cdotsres)--(KN)--(K1);
        \draw[-] (F1)--(K1);

        \draw [decorate, decoration = {brace, raise=10pt, amplitude=5pt}] (KN) --  (K2) node[pos=0.5,below=15pt,black]{$N-1$};
    \end{tikzpicture}
    \caption{The Kronheimer-Nakajima quiver with all gauge nodes the same. The Coulomb branch is the moduli space of $k\;\surm(N)$ instantons on $\mathbb C^2/\mathbb Z_{l}$ where $l=\sum_{i=1}^N l_i$.}
    \label{fig:KNFrameSame1}
\end{figure}

A yet further simplification can be made if only one of the $\urm(k)$ gauge node has flavours, say $l$ flavours, and the others do not. In this case the full $\surm(l)$ symmetry is preserved and the Higgs branch is $\mathcal H(\text{\Quiver{fig:KNFrameSame1}})=\mathcal M_{k,\surm(l)}^{\mathbb C^2/\mathbb Z_N}$. In this case, application of S-duality simply swaps the role of NS5 and D5 branes and hence the Coulomb branch is  $\mathcal C(\text{\Quiver{fig:KNFrameSame1}})=\mathcal M_{k,\surm(N)}^{\mathbb C^2/\mathbb Z_l}$, which simply swaps $N\leftrightarrow l$.

The Kronheimer-Nakajima quivers give Higgs and Coulomb branch constructions of the moduli space of $\surm$ instantons on A-type singularities. The Higgs branch construction is as a hyper-Kähler quotient. However, it is important to emphasise that the Coulomb branch construction here is as a moduli space of dressed monopole operators of the theory \Quiver{fig:KNFrame}. This is different to the Coulomb construction of the same moduli spaces, in Section \ref{sec:instantonconstruc}, which is a hyper-Kähler quotient.

\section{Chain Polymerisation Examples}
\label{sec:Chain}
The trivial case of \hyperref[fig:Chain]{chain polymerisation} arises when two star shaped quivers are combined on their central nodes. The Literature already contains many examples \cite{Chacaltana:2010ks,Gadde:2011uv}. A brief theoretical review is presented in Section \ref{subsec:ClassS} and illustrated in \Figref{fig:ClassSChain}, and a simple example is presented in Appendix \ref{sec:triskelion}.

Here, the focus is on quivers that are not necessarily star-shaped, and on cases where the nodes being superimposed under the polymerisation are not generally central nodes. This in turn leads to the chaining of quivers. The examples presented below yield magnetic quivers for nilpotent orbit closures or slices in nilcones. Some of the quivers that are constructed take the form of Dynkin diagrams for some Kac-Moody algebras. This includes the construction for unitary quivers with B and D type endings at two ends that were first studied from brane systems with orientifold planes \cite{Hanany:2001iy}. The string theory construction of \cite{Hanany:2001iy} is reviewed in Appendix \ref{sec:orientifolds}. These quivers take the form of affine or twisted affine Dynkin diagrams.

Quivers that take the form of finite Dynkin diagrams are labelled by the corresponding algebra and rank. Examples of these quivers, which are used in this work are in Table \ref{tab:FiniteQuivers}. Following the notation of \cite{KacInfiniteDimensional}, affine Dynkin diagrams are denoted with a superscript $(1)$ and twisted affine Dynkin diagrams are denoted with a superscript $(2)$. The quivers of this type which are used in this work are given in Table \ref{tab:AffineQuivers}. Additionally, Dynkin diagrams which correspond to generalised Cartan matrices which admit \textit{additive} functions\footnote{This means that the cokernel of the Cartan matrix is non-trivial, i.e. one is able to define non-trivial Coxeter labels for this Cartan matrix.} \cite{alma9955948401201591} are denoted with a preceding $\frac{1}{2}$ in the label also following the notation of \cite{KacInfiniteDimensional}. This type of Dynkin diagram involves loops on the nodes. The quivers of this type which are used in this work are given in Table \ref{tab:AdditiveQuivers}. The gauge nodes for quivers that take the form of affine, twisted affine, or "additive" Dynkin diagrams are specified by the dual Coxeter labels unless otherwise stated.

\begin{table}[h!]
    \centering
    \begin{tabular}{cccc}
    \toprule
    Label & Quiver & Coulomb Branch & Higgs Branch\\\midrule
        $\begin{array}{c}
	A_n\\n\geq1
	\end{array}$ & $\raisebox{-.5\height}{\begin{tikzpicture}
            \node[gauge, label=below:$1$](l)[]{};
            \node[] (cdots) [right=of l]{$\cdots$};
            \node[gauge, label=below:$1$] (r) [right=of cdots]{};
            \draw[-] (l)--(cdots)--(r);
             \draw [decorate, decoration = {brace, raise=10pt, amplitude=5pt}] (r) --  (l) node[pos=0.5,below=15pt,black]{$n$};
             \end{tikzpicture}}$&$\mathbb H^{n-1}$&Trivial\\
       $\begin{array}{c}
	B_n\\n\geq2
	\end{array}$ &$\raisebox{-.5\height}{\begin{tikzpicture}
            \node[gauge, label=below:$1$] (l)[]{};
            \node[gauge, label=below:$2$] (l2) [right=of l]{};
            \node[] (cdots) [right=of l2]{$\cdots$};
            \node[gauge, label=below:$2$] (r) [right=of cdots]{};
            \node[gauge, label=below:$1$] (rr) [right=of r]{};
            \draw[-] (l)--(l2)--(cdots)--(r);
            \draw [line width=1pt, double distance=3pt,
             arrows = {-Latex[length=0pt 2 0]}] (r) -- (rr);
             \draw [decorate, decoration = {brace, raise=15pt, amplitude=5pt}] (rr) --  (l) node[pos=0.5,below=20pt,black]{$n$};
        \end{tikzpicture}}$&$\mathbb H^{2n-3}$&$-$\\
        $\begin{array}{c}
	D_n\\n\geq2
	\end{array}$ &$\raisebox{-.5\height}{\begin{tikzpicture}
        \node[gauge, label=below:$1$] (1lt) []{};
        \node[gauge, label=below:$2$] (1l2) [right=of 1lt]{};
        \node[] (cdotst) [right=of 1l2]{$\cdots$};
        \node[gauge, label=below:$2$] (2rt) [right=of cdotst]{};
        \node[gauge, label=below:$1$] (1rt) [right=of 2rt]{};
        \node[gauge, label=right:$1$] (1ft) [above=of 2rt]{};

        \draw[-] (1lt)--(1l2)--(cdotst)--(2rt)--(1rt) (1ft)--(2rt);

        \draw [decorate, 
    decoration = {brace,
        raise=15pt,
        amplitude=5pt}] (1rt) --  (1lt) node[pos=0.5,below=20pt,black]{$n-1$};
        \end{tikzpicture}}$&$\mathbb H^{2n-4}$&Trivial\\
       $\begin{array}{c}
	E_6\\
	\end{array}$ &$\raisebox{-.5\height}{\begin{tikzpicture}
            \node[gauge, label=below:$1$] (1l) []{};
    \node[gauge, label=below:$2$] (2l) [right=of 1l]{};
    \node[gauge, label=below:$3$] (3) [right=of 2l]{};
    \node[gauge, label=below:$2$] (2r) [right=of 3]{};
    \node[gauge, label=below:$1$] (1r) [right=of 2r]{};
    \node[gauge, label=right:$2$] (2t) [above=of 3]{};

    \draw[-] (1l)--(2l)--(3)--(2r)--(1r);
    \draw[-] (2t)--(3);
        \end{tikzpicture}}$&$\mathbb H^{10}$&Trivial\\\bottomrule
    \end{tabular}
    \caption{Some quivers which take the form of finite Dynkin diagrams. The gauge nodes are specified by the dual Coxeter labels. The Coulomb branch and Higgs branch are identified where possible.}
    \label{tab:FiniteQuivers}
\end{table}

\begin{table}[h!]
    \centering
    \begin{tabular}{cccc}
    \toprule
    Label & Quiver & Coulomb Branch & Higgs Branch\\\midrule
    $\begin{array}{c}A_n^{(1)}\\n\geq1\end{array}$&$\raisebox{-.5\height}{\begin{tikzpicture}
            \node[gauge, label=below:$1$](l)[]{};
            \node[] (cdots) [right=of l]{$\cdots$};
            \node[gauge, label=below:$1$] (r) [right=of cdots]{};
            \node[gauge, label=above:$1$] (t) [above=of cdots]{};
            \draw[-] (t)--(l)--(cdots)--(r)--(t);
             \draw [decorate, decoration = {brace, raise=10pt, amplitude=5pt}] (r) --  (l) node[pos=0.5,below=15pt,black]{$n$};
             \end{tikzpicture}}$&$\overline{min. A_n}$&$A_n$\\
        $\begin{array}{c}B_n^{(1)}\\n\geq2\end{array}$&$\raisebox{-.5\height}{\begin{tikzpicture}
             \node[gauge, label=below:$1$] (l)[]{};
            \node[gauge, label=below:$2$] (l2) [right=of l]{};
            \node[] (cdots) [right=of l2]{$\cdots$};
            \node[gauge, label=below:$2$] (r) [right=of cdots]{};
            \node[gauge, label=below:$1$] (rr) [right=of r]{};
            \node[gauge, label=left:$1$] (t) [above=of l2]{};
            \draw[-] (l)--(l2)--(cdots)--(r) (t)--(l2);
           \draw [line width=1pt, double distance=3pt,
             arrows = {-Latex[length=0pt 2 0]}] (r) -- (rr);
             \draw [decorate, decoration = {brace, raise=15pt, amplitude=5pt}] (rr) --  (l) node[pos=0.5,below=20pt,black]{$n$};
        \end{tikzpicture}}$&$\overline{min. B_n}$&$-$\\
        $\begin{array}{c}D_n^{(1)}\\n\geq3\end{array}$&$\raisebox{-.5\height}{\begin{tikzpicture}
        \node[gauge, label=below:$1$] (1lt) []{};
        \node[gauge, label=below:$2$] (1l2) [right=of 1lt]{};
        \node[] (cdotst) [right=of 1l2]{$\cdots$};
        \node[gauge, label=below:$2$] (2rt) [right=of cdotst]{};
        \node[gauge, label=below:$1$] (1rt) [right=of 2rt]{};
        \node[gauge, label=right:$1$] (1ft) [above=of 2rt]{};
        \node[gauge, label=left:$1$] (1t) [above=of 1l2]{};

        \draw[-] (1lt)--(1l2)--(cdotst)--(2rt)--(1rt) (1ft)--(2rt) (1t)--(1l2);

        \draw [decorate, 
    decoration = {brace,
        raise=15pt,
        amplitude=5pt}] (1rt) --  (1lt) node[pos=0.5,below=20pt,black]{$n-1$};
        \end{tikzpicture}}$&$\overline{min. D_n}$&$D_n$\\
        $E_6^{(1)}$&$\raisebox{-.5\height}{\begin{tikzpicture}
            \node[gauge, label=below:$1$] (1l) []{};
    \node[gauge, label=below:$2$] (2l) [right=of 1l]{};
    \node[gauge, label=below:$3$] (3) [right=of 2l]{};
    \node[gauge, label=below:$2$] (2r) [right=of 3]{};
    \node[gauge, label=below:$1$] (1r) [right=of 2r]{};
    \node[gauge, label=right:$2$] (2t) [above=of 3]{};
    \node[gauge, label=right:$1$] (1t) [above=of 2t]{};

    \draw[-] (1l)--(2l)--(3)--(2r)--(1r);
    \draw[-] (1t)--(2t)--(3);
        \end{tikzpicture}}$&$\overline{min. E_6}$&$E_6$\\
        $E_7^{(1)}$&$\raisebox{-.5\height}{\begin{tikzpicture}
             \node[gauge, label=below:$1$] (1l) []{};
    \node[gauge, label=below:$2$] (2l) [right=of 1l]{};
    \node[gauge, label=below:$3$] (3l) [right=of 2l]{};
    \node[gauge, label=below:$4$] (4) [right=of 3l]{};
    \node[gauge, label=below:$3$] (3r) [right=of 4]{};
    \node[gauge, label=below:$2$] (2r) [right=of 3r]{};
    \node[gauge, label=below:$1$] (1r) [right=of 2r]{};
    \node[gauge, label=right:$2$] (2t) [above=of 4]{};

    \draw[-] (1l)--(2l)--(3l)--(4)--(3r)--(2r)--(1r);
    \draw[-] (4)--(2t);
        \end{tikzpicture}}$&$\overline{min. E_7}$&$E_7$\\$E_8^{(1)}$&$\raisebox{-0.5\height}{\begin{tikzpicture}
            \node[gauge, label=below:$1$] (1l) []{};
            \node[gauge, label=below:$2$] (2l)[right=of 1l]{};
            \node[gauge, label=below:$3$] (3l) [right=of 2l]{};
            \node[gauge, label=below:$4$] (4l) [right=of 3l]{};
            \node[gauge, label=below:$5$] (5l) [right=of 4l]{};
            \node[gauge, label=below:$6$] (6l) [right=of 5l]{};
            \node[gauge, label=below:$4$] (4r) [right=of 6l]{};
            \node[gauge, label=below:$2$] (2r) [right=of 4r]{};
            \node[gauge, label=right:$3$] (3t) [above=of 6l]{};

            \draw[-] (1l)--(2l)--(3l)--(4l)--(5l)--(6l)--(4r)--(2r) (6l)--(3t);
        \end{tikzpicture}}$&$\overline{min. E_8}$&$E_8$\\
        $\begin{array}{c}D_{n}^{(2)}\\n\geq 3\end{array}$&$\raisebox{-.5\height}{\begin{tikzpicture}
            \node[gauge, label=below:$2$] (2rb) []{};
        \node[] (cdotsrb) [left=of 2rb] {$\cdots$};
        \node[gauge, label=below:$2$] (2mb) [left=of cdotsrb]{};
        \node[] (cdotslb) [left=of 2mb]{$\cdots$};
        \node[gauge, label=below:$2$] (2lb) [left=of cdotslb]{};
        \node[gauge, label=below:$1$] (1lb) [left=of 2lb]{};
        \node[gauge, label=below:$1$] (1rb) [right=of 2rb]{};

        \draw[-] (1lb)--(2lb)--(cdotslb)--(2mb)--(cdotsrb)--(2rb)--(1rb);

        \draw [line width=1pt, double distance=3pt,
             arrows = {-Latex[length=0pt 2 0]}] (2rb) -- (1rb);
         \draw [line width=1pt, double distance=3pt,
             arrows = {-Latex[length=0pt 2 0]}] (2lb) -- (1lb);

        \draw [decorate, 
    decoration = {brace,
        raise=15pt,
        amplitude=5pt}] (1rb) --  (1lb) node[pos=0.5,below=20pt,black]{$n$};
        \end{tikzpicture}}$&$\overline{min. D_n}$ & $-$\\\bottomrule
    \end{tabular}
    \caption{Some quivers which take the form of affine or twisted affine Dynkin diagrams. The gauge nodes are specified by the dual Coxeter labels. The Coulomb branch and Higgs branch are identified where possible.}
    \label{tab:AffineQuivers}
\end{table}

\begin{table}[h!]
    \centering
    \begin{tabular}{cccc}
    \toprule
    Label & Quiver & Coulomb Branch & Higgs Branch\\\midrule
    
    $\begin{array}{c}\frac{1}{2}D_{2n+1}^{(1)}\\n\geq 3\end{array}$&$\raisebox{-.5\height}{\begin{tikzpicture}
            \node[gauge, label=below:$2$] (2mb) []{};
        \node[] (cdotsrb) [left=of 2mb]{$\cdots$};
        \node[gauge, label=below:$2$] (2rb) [left=of cdotsrb]{};
        \node[gauge, label=below:$1$] (1rb) [left=of 2rb]{};
        \node[gauge, label=left:$1$] (1tb) [above=of 2rb]{};

        \draw[-] (2mb)--(cdotsrb)--(2rb)--(1rb) (2rb)--(1tb);
         \draw (2mb) to [out=-45, in=45,looseness=8] (2mb);
        \draw [decorate, 
    decoration = {brace,
        raise=15pt,
        amplitude=5pt}] (2mb) --  (1rb) node[pos=0.5,below=20pt,black]{$n$};
        \end{tikzpicture}}$& $\overline{n. min. B_n}$& $\mathcal S^{C_n}_{\mathcal N,(2n-2,1^2)}\times \mathbb H$ \\
        $\begin{array}{c}\frac{1}{2}A_{2n-1}^{(1)}\\n\geq1\end{array}$&$\raisebox{-.5\height}{\begin{tikzpicture}
            \node[gauge, label=below:$2$] (2mb) []{};
        \node[] (cdotsrb) [right=of 2mb]{$\cdots$};
        \node[gauge, label=below:$2$] (2rb) [right=of cdotsrb]{};
        \node[] (cdotslb) [left=of 2mb]{$\cdots$};
        \node[gauge, label=below:$2$] (2lb) [left=of cdotslb]{};

        \draw[-] (2lb)--(cdotslb)--(2mb)--(cdotsrb)--(2rb);

         \draw (2lb) to [out=135, in=225,looseness=8] (2lb);

         \draw (2rb) to [out=-45, in=45,looseness=8] (2rb);

        \draw [decorate, 
    decoration = {brace,
        raise=15pt,
        amplitude=5pt}] (2rb) --  (2lb) node[pos=0.5,below=20pt,black]{$n$};
        \end{tikzpicture}}$& $\mathcal S^{C_{2n}}_{(4,2^{2n-2}),(2^{2n})}$ & $\mathbb H^2\times \mathcal S^{C_{2n}}_{((2n)^2),((2n-1)^2,1^2)}$ \\
        $\begin{array}{c}\frac{1}{2}D^{(2)}_{2n+1}\\n\geq2\end{array}$ &$\raisebox{-.5\height}{\begin{tikzpicture}
             \node[gauge, label=below:$2$] (2l) []{};
        \node[] (cdots) [right=of 2l]{$\cdots$};
        \node[gauge, label=below:$2$] (2r) [right=of cdots]{};
        \node[gauge, label=below:$1$] (1) [right=of 2r]{};
        \draw[-] (2l)--(cdots)--(2r);        
         \draw [line width=1pt, double distance=3pt,
             arrows = {-Latex[length=0pt 2 0]}] (2r) -- (1);
         \draw (2l) to [out=135, in=225,looseness=8] (2l);
        \draw [decorate, 
    decoration = {brace,
        raise=15pt,
        amplitude=5pt}] (1) --  (2l) node[pos=0.5,below=20pt,black]{$n$};
        \end{tikzpicture}}$& $\overline{\mathcal O}^{D_n}_{(3,1^{2n-3})}$ & $-$ \\ \bottomrule
    \end{tabular}
     \caption{Some quivers which take the form of generalised Dynkin diagrams for Cartan matrices that admit additive functions. The gauge nodes are specified by the dual Coxeter labels, except for $\frac{1}{2}A^{(1)}_{2n-1}$ where twice the dual Coxeter labels are used. The Coulomb branch and Higgs branch are identified where possible.}
    \label{tab:AdditiveQuivers}
\end{table}

\subsection{Affine $D^{(1)}_{k+p}$}
\begin{figure}[h!]
    \centering
    \begin{tikzpicture}
        \node[gauge, label=below:$1$] (1lt) []{};
        \node[gauge, label=below:$2$] (2lt) [right=of 1lt]{};
        \node[] (cdotst) [right=of 2lt]{$\cdots$};
        \node[gauge, label=below:$2$] (2rt) [right=of cdotst]{};
        \node[gauge, label=below:$1$] (1rt) [right=of 2rt]{};
        \node[gauge, label=left:$1$] (1ft) [above=of 2lt]{};

        \draw[-] (1lt)--(2lt)--(cdotst)--(2rt)--(1rt) (1ft)--(2lt);

        \draw [decorate, 
    decoration = {brace,
        raise=15pt,
        amplitude=5pt}] (1rt) --  (1lt) node[pos=0.5,below=20pt,black]{$k+1$};

        \node[] (times) [below=of 2rt]{$\times$};

        \node[gauge, label=below:$2$] (2l) [below=of times]{};
        \node[gauge, label=below:$1$] (1l) [left=of 2l]{};
        \node[] (cdots) [right=of 2l]{$\cdots$};
        \node[gauge, label=below:$2$] (2r) [right=of cdots]{};
        \node[gauge, label=below:$1$] (1r) [right=of 2r]{};
        \node[gauge, label=right:$1$] (1f) [above=of 2r]{};

        \draw[-] (1l)--(2l)--(cdots)--(2r)--(1r) (2r)--(1f);

        \draw [decorate, 
    decoration = {brace,
        raise=15pt,
        amplitude=5pt}] (1r) --  (1l) node[pos=0.5,below=20pt,black]{$p+1$};

        \node[gauge, label=right:$1$] (1frb) [below=of 2r]{};
        \node[gauge, label=below:$2$] (2rb) [below=of 1frb]{};
        \node[] (cdotsrb) [left=of 2rb] {$\cdots$};
        \node[gauge, label=below:$2$] (2mb) [left=of cdotsrb]{};
        \node[] (cdotslb) [left=of 2mb]{$\cdots$};
        \node[gauge, label=below:$2$] (2lb) [left=of cdotslb]{};
        \node[gauge, label=below:$1$] (1lb) [left=of 2lb]{};
        \node[gauge, label=below:$1$] (1rb) [right=of 2rb]{};
        \node[gauge, label=left:$1$] (1flb) [above=of 2lb]{};

        \draw[-] (1lb)--(2lb)--(cdotslb)--(2mb)--(cdotsrb)--(2rb)--(1rb) (1flb)--(2lb) (1frb)--(2rb);

        \draw [decorate, 
    decoration = {brace,
        raise=15pt,
        amplitude=5pt}] (1rb) --  (1lb) node[pos=0.5,below=20pt,black]{$k+p-1$};
        
    \node[] (topghost) [below=of 1lt]{};
        \node[] (bottomghost) [left=of 1lb]{};
        \draw [->] (topghost) to [out=-150,in=150,looseness=1] (bottomghost);
        
    \end{tikzpicture}
    \caption{$\surm(2)$ \hyperref[fig:Chain]{chain polymerisation} of the finite $D_{k+2}$ and $D_{p+2}$ quivers to produce the affine $D^{(1)}_{k+p}$ quiver.}
    \label{fig:nminDkpm3}
\end{figure}

A magnetic quiver for the product space of $\mathbb H^{2k}\times \mathbb H^{2p}$ may be given as a product of the finite $D_{k+2}$ and $D_{p+2}$ quivers. An $\surm(2)$ \hyperref[fig:Chain]{chain polymerisation}, shown in \Figref{fig:nminDkpm3}, produces the affine $D^{(1)}_{k+p}$ quiver. The resulting action on the Coulomb branch is:
\begin{equation}
    \mathbb H^{2k}\times \mathbb H^{2p}///\surm(2)=\overline{min. D_{k+p}}.
\end{equation} 

A further trivial $\urm(1)$ hyper-Kähler quotient may be taken using \hyperref[fig:Cyclic]{cyclic polymerisation} (discussed later) to obtain either $\overline{n.min. A_{k+p-1}}$ or $\overline{n.min. D_{k+p-1}}$, depending on whether the hyper-Kähler quotient is taken over opposing or adjacent $\urm(1)$ nodes.

Turning to the action of $\surm(2)$ \hyperref[fig:Chain]{chain polymerisation} on the Higgs branches of these theories; the $D_{k+2}$ and $D_{p+2}$ quivers have trivial Higgs branches and the Higgs branch of the $D^{(1)}_{k+p}$ quiver is the Kleinian $D_{k+p}$ singularity which is of dimension 1. There is an increase in the Higgs branch dimension by $\textrm{rank}\;\surm(2)=1$ as expected.

\subsection{Affine $B^{(1)}_{k+p-1}$}

\begin{figure}[h!]
    \centering
    \begin{tikzpicture}
        \node[gauge, label=below:$1$] (1lt) []{};
        \node[gauge, label=below:$2$] (2lt) [right=of 1lt]{};
        \node[] (cdotst) [right=of 2lt]{$\cdots$};
        \node[gauge, label=below:$2$] (2rt) [right=of cdotst]{};
        \node[gauge, label=below:$1$] (1rt) [right=of 2rt]{};
        \node[gauge, label=left:$1$] (1ft) [above=of 2lt]{};

        \draw[-] (1lt)--(2lt)--(cdotst)--(2rt)--(1rt) (1ft)--(2lt);

        \draw [decorate, 
    decoration = {brace,
        raise=15pt,
        amplitude=5pt}] (1rt) --  (1lt) node[pos=0.5,below=20pt,black]{$k+1$};

        \node[] (times) [below=of 2rt]{$\times$};

        \node[gauge, label=below:$2$] (2l) [below=of times]{};
        \node[gauge, label=below:$1$] (1l) [left=of 2l]{};
        \node[] (cdots) [right=of 2l]{$\cdots$};
        \node[gauge, label=below:$2$] (2r) [right=of cdots]{};
        \node[gauge, label=below:$1$] (1r) [right=of 2r]{};

        \draw[-] (1l)--(2l)--(cdots)--(2r);
          \draw [line width=1pt, double distance=3pt,
             arrows = {-Latex[length=0pt 2 0]}] (2r) -- (1r);

        \draw [decorate, 
    decoration = {brace,
        raise=15pt,
        amplitude=5pt}] (1r) --  (1l) node[pos=0.5,below=20pt,black]{$p+1$};

        \node[](2rghost)[below=of 2r]{};
        
        \node[gauge, label=below:$2$] (2rb) [below=of 2rghost]{};
        \node[] (cdotsrb) [left=of 2rb] {$\cdots$};
        \node[gauge, label=below:$2$] (2mb) [left=of cdotsrb]{};
        \node[] (cdotslb) [left=of 2mb]{$\cdots$};
        \node[gauge, label=below:$2$] (2lb) [left=of cdotslb]{};
        \node[gauge, label=below:$1$] (1lb) [left=of 2lb]{};
        \node[gauge, label=below:$1$] (1rb) [right=of 2rb]{};
        \node[gauge, label=left:$1$] (1flb) [above=of 2lb]{};

        \draw[-] (1lb)--(2lb)--(cdotslb)--(2mb)--(cdotsrb)--(2rb)--(1rb) (1flb)--(2lb);

        \draw [line width=1pt, double distance=3pt,
             arrows = {-Latex[length=0pt 2 0]}] (2rb) -- (1rb);

        \draw [decorate, 
    decoration = {brace,
        raise=15pt,
        amplitude=5pt}] (1rb) --  (1lb) node[pos=0.5,below=20pt,black]{$k+p-1$};
        
    \node[] (topghost) [below=of 1lt]{};
        \node[] (bottomghost) [left=of 1lb]{};
        \draw [->] (topghost) to [out=-150,in=150,looseness=1] (bottomghost);
        
    \end{tikzpicture}
    \caption{$\surm(2)$ \hyperref[fig:Chain]{chain polymerisation} of the finite $D_{k+2}$ and $B_{p+1}$ quivers to produce the affine $B^{(1)}_{k+p-1}$ quiver.}
    \label{fig:nminBkpm3}
\end{figure}
The product moduli space $\mathbb H^{2k}\times \mathbb H^{2p-1}$ has a Coulomb branch construction as the product of the finite $D_{k+2}$ and $B_{p+1}$ quivers. The $\surm(2)$ \hyperref[fig:Chain]{chain polymerisation} is performed in \Figref{fig:nminBkpm3} to produce the affine $B^{(1)}_{k+p-1}$ quiver. 

The action on the Coulomb branch is:
\begin{equation}
    \mathbb H^{2k}\times \mathbb H^{2p-1}///\surm(2)=\overline{min. B_{k+p-1}}.
\end{equation}

No statement about the Higgs branch can be made as it is unknown how to compute Higgs branches of non-simply laced quivers.

\subsection{Twisted affine $D^{(2)}_{k+p-1}$}
\label{subsec:TwistedAffD}
\begin{figure}[h!]
    \centering
    \begin{tikzpicture}
        \node[gauge, label=below:$1$] (1lt) []{};
        \node[gauge, label=below:$2$] (2lt) [right=of 1lt]{};
        \node[] (cdotst) [right=of 2lt]{$\cdots$};
        \node[gauge, label=below:$2$] (2rt) [right=of cdotst]{};
        \node[gauge, label=below:$1$] (1rt) [right=of 2rt]{};

        \draw[-] (2lt)--(cdotst)--(2rt)--(1rt);
        \draw [line width=1pt, double distance=3pt,
             arrows = {-Latex[length=0pt 2 0]}] (2lt) -- (1lt);

        \draw [decorate, 
    decoration = {brace,
        raise=15pt,
        amplitude=5pt}] (1rt) --  (1lt) node[pos=0.5,below=20pt,black]{$k+1$};

        \node[] (times) [below=of 2rt]{$\times$};

        \node[gauge, label=below:$2$] (2l) [below=of times]{};
        \node[gauge, label=below:$1$] (1l) [left=of 2l]{};
        \node[] (cdots) [right=of 2l]{$\cdots$};
        \node[gauge, label=below:$2$] (2r) [right=of cdots]{};
        \node[gauge, label=below:$1$] (1r) [right=of 2r]{};

        \draw[-] (1l)--(2l)--(cdots)--(2r);
          \draw [line width=1pt, double distance=3pt,
             arrows = {-Latex[length=0pt 2 0]}] (2r) -- (1r);

        \draw [decorate, 
    decoration = {brace,
        raise=15pt,
        amplitude=5pt}] (1r) --  (1l) node[pos=0.5,below=20pt,black]{$p+1$};

        \node[](2rghost)[below=of 2r]{};
        
        \node[gauge, label=below:$2$] (2rb) [below=of 2rghost]{};
        \node[] (cdotsrb) [left=of 2rb] {$\cdots$};
        \node[gauge, label=below:$2$] (2mb) [left=of cdotsrb]{};
        \node[] (cdotslb) [left=of 2mb]{$\cdots$};
        \node[gauge, label=below:$2$] (2lb) [left=of cdotslb]{};
        \node[gauge, label=below:$1$] (1lb) [left=of 2lb]{};
        \node[gauge, label=below:$1$] (1rb) [right=of 2rb]{};

        \draw[-] (1lb)--(2lb)--(cdotslb)--(2mb)--(cdotsrb)--(2rb)--(1rb);

        \draw [line width=1pt, double distance=3pt,
             arrows = {-Latex[length=0pt 2 0]}] (2rb) -- (1rb);
         \draw [line width=1pt, double distance=3pt,
             arrows = {-Latex[length=0pt 2 0]}] (2lb) -- (1lb);

        \draw [decorate, 
    decoration = {brace,
        raise=15pt,
        amplitude=5pt}] (1rb) --  (1lb) node[pos=0.5,below=20pt,black]{$k+p-1$};
        
    \node[] (topghost) [below=of 1lt]{};
        \node[] (bottomghost) [left=of 1lb]{};
        \draw [->] (topghost) to [out=-150,in=150,looseness=1] (bottomghost);
        
    \end{tikzpicture}
    \caption{$\surm(2)$ \hyperref[fig:Chain]{chain polymerisation} of the finite $B_{k+1}$ and $B_{p+1}$ quivers to produce the twisted affine $D^{(2)}_{k+p-1}$ quiver.}
    \label{fig:minDkTwo}
\end{figure}
The product moduli space $\mathbb H^{2k-1}\times \mathbb H^{2p-1}$ has a Coulomb branch construction as the product of the finite $B_{k+1}$ and $B_{p+1}$ quivers. The $\surm(2)$ \hyperref[fig:Chain]{chain polymerisation} is performed in \Figref{fig:minDkTwo} to produce the twisted affine $D^{(2)}_{k+p-1}$ quiver. 

The action on the Coulomb branch is:
\begin{equation}
    \mathbb H^{2k-1}\times \mathbb H^{2p-1}///\surm(2)=\overline{min. D_{k+p-1}}.
\end{equation}

No statement about the Higgs branch can be made as it is unknown how to compute Higgs branches of non-simply laced quivers.

\subsection{$\mathcal Q_M(\overline{n.min. B_{k+p-1}})$}

\begin{figure}[h!]
    \centering
    \begin{tikzpicture}
        \node[gauge, label=below:$2$](2l) []{};
        \node[] (cdots) [right=of 2l]{$\cdots$};
        \node[gauge, label=below: $2$] (2r) [right=of cdots] {};
        \node[gauge, label=below:$1$] (1) [right=of 2r]{};
        \node[gauge, label=below:$1$] (1f) [left=of 2l]{};

        \draw[-] (1f)--(2l)--(cdots)--(2r)--(1);
        \draw[double,double distance=3pt,line width=0.4pt] (2l)--(1f);

        \draw [decorate, 
    decoration = {brace,
        raise=15pt,
        amplitude=5pt}] (1) --  (1f) node[pos=0.5,below=20pt,black]{$k+2$};
    \end{tikzpicture}
    \caption{Magnetic quiver \Quiver{fig:A1H2km2} for $\mathbb C^2/\mathbb Z_2\times \mathbb H^{2k}$.}
    \label{fig:A1H2km2}
\end{figure}

\begin{figure}[h!]
    \centering
    \begin{tikzpicture}
        \node[gauge, label=below:$2$](2lt) []{};
        \node[] (cdotst) [right=of 2lt]{$\cdots$};
        \node[gauge, label=below: $2$] (2rt) [right=of cdots] {};
        \node[gauge, label=below:$1$] (1t) [right=of 2r]{};
        \node[gauge, label=below:$1$] (1ft) [left=of 2l]{};

        \draw[-] (1ft)--(2lt)--(cdotst)--(2rt)--(1t);
        \draw[double,double distance=3pt,line width=0.4pt] (2l)--(1f);

        \draw [decorate, 
    decoration = {brace,
        raise=15pt,
        amplitude=5pt}] (1t) --  (1ft) node[pos=0.5,below=20pt,black]{$k+2$};

        \node[] (times) [below=of 2rt]{$\times$};

        \node[gauge, label=below:$2$] (2l) [below=of times]{};
        \node[gauge, label=below:$1$] (1l) [left=of 2l]{};
        \node[] (cdots) [right=of 2l]{$\cdots$};
        \node[gauge, label=below:$2$] (2r) [right=of cdots]{};
        \node[gauge, label=below:$1$] (1r) [right=of 2r]{};

        \draw[-] (1l)--(2l)--(cdots)--(2r);
          \draw [line width=1pt, double distance=3pt,
             arrows = {-Latex[length=0pt 2 0]}] (2r) -- (1r);

        \draw [decorate, 
    decoration = {brace,
        raise=15pt,
        amplitude=5pt}] (1r) --  (1l) node[pos=0.5,below=20pt,black]{$p+1$};

        \node[] (ghost) [below=of 2l]{};
        
        \node[gauge, label=below:$2$] (2mb) [below=of ghost]{};
        \node[] (cdotsrb) [right=of 2mb]{$\cdots$};
        \node[gauge, label=below:$2$] (2rb) [right=of cdotsrb]{};
        \node[gauge, label=below:$1$] (1rb) [right=of 2rb]{};
        \node[] (cdotslb) [left=of 2mb]{$\cdots$};
        \node[gauge, label=below:$2$] (2lb) [left=of cdotslb]{};
        \node[gauge, label=below:$1$] (1lb) [left=of 2lb]{};

        \draw[-] (1lb)--(2lb)--(cdotslb)--(2mb)--(cdotsrb)--(2rb)--(1rb);

        \draw [line width=1pt, double distance=3pt,
             arrows = {-Latex[length=0pt 2 0]}] (2rb) -- (1rb);

           \draw[double,double distance=3pt,line width=0.4pt] (1lb)--(2lb);

        \draw [decorate, 
    decoration = {brace,
        raise=15pt,
        amplitude=5pt}] (1rb) --  (1lb) node[pos=0.5,below=20pt,black]{$k+p$};
        
        \node[] (topghost) [below=of 1lt]{};
        \node[] (bottomghost) [left=of 1lb]{};
        \draw [->] (topghost) to [out=-150,in=150,looseness=1] (bottomghost);
        
    \end{tikzpicture}
    \caption{$\surm(2)$ \hyperref[fig:Chain]{chain polymerisation} of \Quiver{fig:A1H2km2} with the finite $B_{p+1}$ quiver to give  \Quiver{fig:nminBkpm3One}.}
    \label{fig:nminBkpm3One}
\end{figure}
The Coulomb branch of \Quiver{fig:A1H2km2} is $\mathbb C^2/\mathbb Z_2\times \mathbb H^{2k}$, which can be seen from explicit computation of the Hilbert series or argued from its Hasse diagram. The product may be taken with the $B_{p+1}$ quiver, whose Coulomb branch is $\mathbb H^{2p-1}$. The $\surm(2)$ \hyperref[fig:Chain]{chain polymerisation} of these theories is shown in \Figref{fig:nminBkpm3One}.

The resulting quiver \Quiver{fig:nminBkpm3One} is an unframed magnetic quiver for $\overline{n.min. B_{k+p-1}}$. The conclusion on the Coulomb branch is that:
\begin{equation}
    \left(\mathbb C^2/\mathbb Z_2\times \mathbb H^{2k}\right)\times \mathbb H^{2p-1}///\surm(2)=\overline{n.min. B_{k+p-1}}.
\end{equation}
Again, no statement about the Higgs branch can be made as the quivers involved have non-simply laced edges.

Similarly, using the finite $D_{p+2}$ quiver instead of the finite $B_{p+1}$ quiver gives a magnetic quiver for $\overline{n.min. D_{k+p}}$. However in this case it can be confirmed that there is an increase in the Higgs branch dimension by 1.

\subsection{Additive $\frac{1}{2}D^{(1)}_{2k+2p-1}$}
\begin{figure}[h!]
    \centering
    \begin{tikzpicture}
        \node[gauge, label=below:$2$](2l) []{};
        \node[] (cdots) [right=of 2l]{$\cdots$};
        \node[gauge, label=below: $2$] (2r) [right=of cdots] {};
        \node[gauge, label=below:$1$] (1) [right=of 2r]{};

        \draw[-] (2l)--(cdots)--(2r)--(1);
        \draw (2l) to [out=135, in=225,looseness=8] (2l);

        \draw [decorate, 
    decoration = {brace,
        raise=15pt,
        amplitude=5pt}] (1) --  (2l) node[pos=0.5,below=20pt,black]{$k+1$};
    \end{tikzpicture}
    \caption{Unframed magnetic quiver \Quiver{fig:H2km2Z2} for $\mathbb C^2/\mathbb Z_2\times \mathbb H^{2k-1}$.}
    \label{fig:H2km2Z2}
\end{figure}
\begin{figure}[h!]
    \centering
    \begin{tikzpicture}
        \node[gauge, label=below:$2$](2lt) []{};
        \node[] (cdotst) [right=of 2lt]{$\cdots$};
        \node[gauge, label=below: $2$] (2rt) [right=of cdotst] {};
        \node[gauge, label=below:$1$] (1t) [right=of 2rt]{};

        \draw[-] (2lt)--(cdotst)--(2rt)--(1t);
        \draw (2lt) to [out=135, in=225,looseness=8] (2lt);

        \draw [decorate, 
    decoration = {brace,
        raise=15pt,
        amplitude=5pt}] (1t) --  (2lt) node[pos=0.5,below=20pt,black]{$k+1$};

        \node[] (times) [below=of 2rt]{$\times$};

        \node[gauge, label=below:$2$] (2l) [below=of times]{};
        \node[gauge, label=below:$1$] (1l) [left=of 2l]{};
        \node[] (cdots) [right=of 2l]{$\cdots$};
        \node[gauge, label=below:$2$] (2r) [right=of cdots]{};
        \node[gauge, label=below:$1$] (1r) [right=of 2r]{};
        \node[gauge, label=right:$1$] (1t) [above=of 2r]{};

        \draw[-] (1l)--(2l)--(cdots)--(2r)--(1r) (2r)--(1t);

        \draw [decorate, 
    decoration = {brace,
        raise=15pt,
        amplitude=5pt}] (1r) --  (1l) node[pos=0.5,below=20pt,black]{$p+1$};

        \node[] (ghost) [below=of 2l]{};
        
        \node[gauge, label=below:$2$] (2mb) [below=of ghost]{};
        \node[] (cdotsrb) [right=of 2mb]{$\cdots$};
        \node[gauge, label=below:$2$] (2rb) [right=of cdotsrb]{};
        \node[gauge, label=below:$1$] (1rb) [right=of 2rb]{};
        \node[] (cdotslb) [left=of 2mb]{$\cdots$};
        \node[gauge, label=below:$2$] (2lb) [left=of cdotslb]{};
        \node[gauge, label=right:$1$] (1tb) [above=of 2rb]{};

        \draw[-] (2lb)--(cdotslb)--(2mb)--(cdotsrb)--(2rb)--(1rb) (2rb)--(1tb);

         \draw (2lb) to [out=135, in=225,looseness=8] (2lb);

        \draw [decorate, 
    decoration = {brace,
        raise=15pt,
        amplitude=5pt}] (1rb) --  (2lb) node[pos=0.5,below=20pt,black]{$k+p-1$};
        
        \node[] (topghost) [below=of 1lt]{};
        \node[] (bottomghost) [left=of 1lb]{};
        \draw [->] (topghost) to [out=-150,in=150,looseness=1] (bottomghost);
        
    \end{tikzpicture}
    \caption{$\surm(2)$ \hyperref[fig:Chain]{chain polymerisation} of \Quiver{fig:H2km2Z2} with the finite $D_{p+2}$ quiver to give the additive $\frac{1}{2}D^{(1)}_{2k+2p-1}$ quiver, \Quiver{fig:nminBkpm3Two}.}
    \label{fig:nminBkpm3Two}
\end{figure}
The Coulomb branch of \Quiver{fig:H2km2Z2} is $\mathbb C^2/\mathbb Z_2\times \mathbb H^{2k-1}$, which can be seen from explicit computation of the Hilbert series or argued from its Hasse diagram. This quiver is of the form of a Dynkin diagram for a generalised Cartan matrix which admits \textit{subadditive} but not \textit{additive functions}\footnote{If the Cartan matrix admits subadditive functions that are not additive, one cannot define Coxeter labels by definition. For this reason the labelling of the gauge nodes of \Quiver{fig:H2km2Z2} is judicious to find additive quivers whose gauge nodes are labelled by dual Coxeter labels.}. 

The product may be taken with the $D_{p+2}$ quiver, whose Coulomb branch is $\mathbb H^{2p}$. When followed by an $\surm(2)$ \hyperref[fig:Chain]{chain polymerisation}, as shown in \Figref{fig:nminBkpm3Two}, this produces another magnetic quiver \Quiver{fig:nminBkpm3Two} for $\overline{n.min. B_{k+p-1}}$. This quiver takes the form of the additive quiver $\frac{1}{2}D^{(1)}_{2k+2p-1}$.
The conclusion on the Coulomb branch is that:
\begin{equation}
   \left(\mathbb C^2/\mathbb Z_2\times \mathbb H^{2k-1}\right)\times \mathbb H^{2p}///\surm(2)=\overline{n.min. B_{k+p-1}}.
\end{equation}
The Higgs branch Hilbert series for quiver \Quiver{fig:H2km2Z2} is easily computed as:
\begin{equation}
    \hs\left[\mathcal H(\text{\Quiver{fig:H2km2Z2}})\right]=\pe\left[[1]_{\surm(2)}t+[2]_{\surm(2)}t^2-t^4\right].
\end{equation}
This identifies the Higgs branch as $\mathcal H(\text{\Quiver{fig:H2km2Z2}})=\mathbb C^2/\mathbb Z_2\times \mathbb H$, which is of dimension 2 and is independent of the number of nodes.

The Higgs branch Hilbert series for the quiver \Quiver{fig:nminBkpm3Two} is also computed as:
\begin{equation}
    \hs\left[\mathcal H(\text{\Quiver{fig:nminBkpm3Two}})\right]=\pe\left[[1]_{\surm(2)}t+[2]_{\surm(2)}t^2+[1]_{\surm(2)}t^{2(k+p-1)-1}-t^{4(k+p-1)}\right].
\end{equation}
This indicates the Higgs branch is a complete intersection and of dimension 3 \cite{Hanany:2010qu,2019arXiv190305822F}. In fact this Higgs branch can be identified as the product of a Slodowy slice of the GNO dual group, with a free field, $\mathcal H(\text{\Quiver{fig:nminBkpm3Two}})=\mathcal S^{C_{k+p-1}}_{\mathcal N,(2(k+p-2),1^2)}\times \mathbb H$. The $\surm(2)$ chain polymerisation gives the expected increase of Higgs branch dimension by $\textrm{rank}\;\surm(2)=1$.


\subsection{Additive $\frac{1}{2}A^{(1)}_{2k+2p-3}$}

\begin{figure}[h!]
    \centering
    \begin{tikzpicture}
        \node[gauge, label=below:$2$](2lt) []{};
        \node[] (cdotst) [right=of 2lt]{$\cdots$};
        \node[gauge, label=below: $2$] (2rt) [right=of cdotst] {};
        \node[gauge, label=below:$1$] (1t) [right=of 2rt]{};

        \draw[-] (2lt)--(cdotst)--(2rt)--(1t);
        \draw (2lt) to [out=135, in=225,looseness=8] (2lt);

        \draw [decorate, 
    decoration = {brace,
        raise=15pt,
        amplitude=5pt}] (1t) --  (2lt) node[pos=0.5,below=20pt,black]{$k+1$};

        \node[] (times) [below=of 2rt]{$\times$};

        \node[gauge, label=below:$2$] (2l) [below=of times]{};
        \node[gauge, label=below:$1$] (1l) [left=of 2l]{};
        \node[] (cdots) [right=of 2l]{$\cdots$};
        \node[gauge, label=below:$2$] (2r) [right=of cdots]{};

        \draw[-] (1l)--(2l)--(cdots)--(2r);

        \draw (2r) to [out=-45, in=45,looseness=8] (2r);
         
        \draw [decorate, 
    decoration = {brace,
        raise=15pt,
        amplitude=5pt}] (2r) --  (1l) node[pos=0.5,below=20pt,black]{$p+1$};

        \node[] (ghost) [below=of 2l]{};
        
        \node[gauge, label=below:$2$] (2mb) [below=of ghost]{};
        \node[] (cdotsrb) [right=of 2mb]{$\cdots$};
        \node[gauge, label=below:$2$] (2rb) [right=of cdotsrb]{};
        \node[] (cdotslb) [left=of 2mb]{$\cdots$};
        \node[gauge, label=below:$2$] (2lb) [left=of cdotslb]{};

        \draw[-] (2lb)--(cdotslb)--(2mb)--(cdotsrb)--(2rb);

         \draw (2lb) to [out=135, in=225,looseness=8] (2lb);

         \draw (2rb) to [out=-45, in=45,looseness=8] (2rb);

        \draw [decorate, 
    decoration = {brace,
        raise=15pt,
        amplitude=5pt}] (2rb) --  (2lb) node[pos=0.5,below=20pt,black]{$k+p-1$};

        \node[] (topghost) [below=of 1lt]{};
        \node[] (bottomghost) [left=of 1lb]{};
        \draw [->] (topghost) to [out=-150,in=150,looseness=1] (bottomghost);
        
    \end{tikzpicture}
    \caption{$\surm(2)$ \hyperref[fig:Chain]{chain polymerisation} of \Quiver{fig:H2km2Z2} with itself to give the additive $\frac{1}{2}A^{(1)}_{2k+2p-3}$ quiver, \Quiver{fig:Cslice}.}
    \label{fig:Cslice}
\end{figure}
The quiver type \Quiver{fig:H2km2Z2}, whose Coulomb branch is $\mathbb C^2/\mathbb Z_2\times \mathbb H^{2k-1}$, may be $\surm(2)$ \hyperref[fig:Chain]{chain polymerised} with itself as shown in \Figref{fig:Cslice}. This produces an additive quiver $\frac{1}{2}A^{(1)}_{2k+2p-3}$, \Quiver{fig:Cslice}. In this case the gauge nodes are specified by twice the dual Coxeter labels. This quiver is a magnetic quiver for $\mathcal S^{C_{2k+2p-2}}_{(4,2^{2k+2p-4}),(2^{2k+2p-4})}$. There are insufficient fugacities to give refined expressions for the Coulomb branch Hilbert series. However, there is agreement at the level of the unrefined Hilbert series.
The conclusion on the Coulomb branch is that:
\begin{equation}
   \left(\mathbb C^2/\mathbb Z_2\times \mathbb H^{2k-1}\right)\times  \left(\mathbb C^2/\mathbb Z_2\times \mathbb H^{2p-1}\right)///\surm(2)=\mathcal S^{C_{2k+2p-2}}_{(4,2^{2k+2p-4}),(2^{2k+2p-2})}.
\end{equation}

Note also that $\mathcal C(\text{\Quiver{fig:Cslice}})=\overline{min. D_{k+p}}/(\mathbb Z_2\times \mathbb Z_2)$, which can be seen from discrete gauging \cite{Hanany:2018vph,Hanany:2018dvd,Bourget:2020bxh,Hanany:2023uzn}.
The Higgs branch Hilbert series for \Quiver{fig:Cslice} is also easily computed. These do not have a simple expression as a PE and so the final result is quoted as \begin{equation}
    \mathcal H(\text{\Quiver{fig:Cslice}})=\mathbb H^2\times \mathcal S^{C_{2k+2p-2}}_{((2k+2p-2)^2),((2k+2p-3)^2,1^2)}.
\end{equation} This Higgs branch is of dimension 5 which is consistent with the increase in the Higgs branch dimension by $\textrm{rank}\;\surm(2)=1$.

A particularly interesting point is that the Coulomb branch and the (non-trivial part of) the Higgs branch of \Quiver{fig:Cslice} are not dual under the Barbash-Vogan map. In fact, the Higgs and Coulomb branch are slices in the same nilcone of a $C$-type algebra\footnote{The partitions in this case are dual under the metaplectic Lustig-Spaltenstein map, which is the usual Barbasch-Vogan map with an additional $-$ collapse.}. This is an example of the \textit{metaplectic} Lustig-Spaltenstein duality \cite{Moeglin2002,2014arXiv1412.8742J,2020arXiv201016089B}.

\subsection{Additive $\frac{1}{2}D^{(2)}_{2k+2p-1}$}

\begin{figure}[h!]
    \centering
    \begin{tikzpicture}
        \node[gauge, label=below:$2$](2lt) []{};
        \node[] (cdotst) [right=of 2lt]{$\cdots$};
        \node[gauge, label=below: $2$] (2rt) [right=of cdotst] {};
        \node[gauge, label=below:$1$] (1t) [right=of 2rt]{};

        \draw[-] (2lt)--(cdotst)--(2rt)--(1t);
        \draw (2lt) to [out=135, in=225,looseness=8] (2lt);

        \draw [decorate, 
    decoration = {brace,
        raise=15pt,
        amplitude=5pt}] (1t) --  (2lt) node[pos=0.5,below=20pt,black]{$k+1$};

        \node[] (times) [below=of 2rt]{$\times$};

        \node[gauge, label=below:$2$] (2l) [below=of times]{};
        \node[gauge, label=below:$1$] (1l) [left=of 2l]{};
        \node[] (cdots) [right=of 2l]{$\cdots$};
        \node[gauge, label=below:$2$] (2r) [right=of cdots]{};
        \node[gauge, label=below:$1$] (1r) [right=of 2r]{};

        \draw[-] (1l)--(2l)--(cdots)--(2r);

         \draw [line width=1pt, double distance=3pt,
             arrows = {-Latex[length=0pt 2 0]}] (2r) -- (1r);

        \draw [decorate, 
    decoration = {brace,
        raise=15pt,
        amplitude=5pt}] (1r) --  (1l) node[pos=0.5,below=20pt,black]{$p+1$};

        \node[] (ghost) [below=of 2l]{};
        
        \node[gauge, label=below:$2$] (2mb) [below=of ghost]{};
        \node[] (cdotsrb) [right=of 2mb]{$\cdots$};
        \node[gauge, label=below:$2$] (2rb) [right=of cdotsrb]{};
        \node[gauge, label=below:$1$] (1rb) [right=of 2rb]{};
        \node[] (cdotslb) [left=of 2mb]{$\cdots$};
        \node[gauge, label=below:$2$] (2lb) [left=of cdotslb]{};

        \draw[-] (2lb)--(cdotslb)--(2mb)--(cdotsrb)--(2rb);

         \draw [line width=1pt, double distance=3pt,
             arrows = {-Latex[length=0pt 2 0]}] (2rb) -- (1rb);

         \draw (2lb) to [out=135, in=225,looseness=8] (2lb);

        \draw [decorate, 
    decoration = {brace,
        raise=15pt,
        amplitude=5pt}] (1rb) --  (2lb) node[pos=0.5,below=20pt,black]{$k+p-1$};

        \node[] (topghost) [below=of 1lt]{};
        \node[] (bottomghost) [left=of 1lb]{};
        \draw [->] (topghost) to [out=-150,in=150,looseness=1] (bottomghost);
        
    \end{tikzpicture}
    \caption{$\surm(2)$ \hyperref[fig:Chain]{chain polymerisation} of \Quiver{fig:H2km2Z2} with the finite $B_{p+1}$ quiver to give the additive $\frac{1}{2}D^{(2)}_{2k+2p-1}$ quiver, \Quiver{fig:D31orb}.}
    \label{fig:D31orb}
\end{figure}
The quiver \Quiver{fig:H2km2Z2}, whose Coulomb branch is $\mathbb C^2/\mathbb Z_2\times \mathbb H^{2k-1}$, may be $\surm(2)$ \hyperref[fig:Chain]{chain polymerised} with the finite $B_{p+1}$ quiver as shown in \Figref{fig:D31orb}; this produces another quiver \Quiver{fig:D31orb}, which is the additive quiver $\frac{1}{2}D^{(2)}_{2k+2p-1}$. This quiver is a magnetic quiver for $\overline{\mathcal O}^{D_{k+p-1}}_{(3,1^{2k+2p-5})}$. There are insufficient fugacities to give refined expressions for the Coulomb branch Hilbert series. However, there is agreement at the level of the unrefined Hilbert series.
The conclusion on the Coulomb branch is that:
\begin{equation}
   \left(\mathbb C^2/\mathbb Z_2\times \mathbb H^{2k-1}\right)\times \mathbb H^{2p-1}///\surm(2)=\overline{\mathcal O}^{D_{k+p-1}}_{(3,1^{2k+2p-5})}.
\end{equation}
The Higgs branch Hilbert series for \Quiver{fig:D31orb} cannot be computed due to the non-simply laced edge so no comment can be made about the Higgs branch.

\subsection{Two $\frac{1}{2}$ M5 branes probing $E_6$ Klein Singularity}
Consider an M-theory background with two $\frac{1}{2}$ M5 branes, spanning the directions $x^{0,1,2,3,4,5}$ and separated along the $x^6$ direction on an $E_6$ Klein singularity which spans the directions $x^{0,1,2,3,4,5,6}$. Each $\frac{1}{2}$ M5 brane comes from two $\frac{1}{4}$ M5 branes with zero separation along the $x^6$ direction. The F-theory dual description of four separated $\frac{1}{4}$ M5 branes consists of three curves with self-intersection numbers $-1,\;-3,\;-1$ where the $(-3)$-curve supports an $\surm(3)$ gauge group \cite{Aspinwall:1998xj,DelZotto:2014hpa,Ohmori:2015pua}. The F-theory description of the two separated $\frac{1}{2}$ M5 branes is derived from blowing down both the left and right $(-1)$-curves. In doing so the central $(-3)$-curve becomes a $(-1)$-curve, with a tensor multiplet, which still supports $\surm(3)$ gauge symmetry. The gauge coupling for the two $\sprm(0)$ becomes infinite. The gauge coupling for the $\surm(3)$ is finite as there is still non-trivial separation between the two $\frac{1}{2}$ M5 branes. However now there is coupling of two copies of the rank-1 E-string theory to $\surm(3)$. The brane system and electric quiver are summarised in \Figref{fig:M5}.

\begin{figure}[h!]
    \centering
    \begin{subfigure}{0.45\textwidth}
    \centering
        \begin{tikzpicture}
        \node[label=above:$\frac{1}{2}$ M5] (2) at (2,0){$\times$};
        \node[] (3) at (3.5,0){$\times$};

        \node[] (E6) at (0,0){$E_6$};

        \draw[-] (0.5,0)--(5,0);

        \draw[-] (5.5,0)--(6.5,0) (6.4,0.1)--(6.5,0)--(6.4,-0.1);
        \draw[-] (5.5,0)--(5.5,1) (5.4,0.9)--(5.5,1)--(5.6,0.9);

        \node[] (x6) at (7,0){$x^6$};
        \node[] (x78910) at (5.5,1.5){$x^{7,8,9,10}$};
        
        \end{tikzpicture}
        \caption{}
        \label{fig:M5Branes}
    \end{subfigure}
    \centering
    \begin{subfigure}{0.45\textwidth}
    \centering
    \begin{tikzpicture}
        \node[gaugeb] (spol) {};
        \node[align=center,anchor=north] (lab) at (spol.south) {$\sprm(0)$};
        \node[gauge] (surm3) [right=of spol]{};
        \node[align=center,anchor=north] (lab2) at (surm3.south) {$\surm(3)$};
        \node[gaugeb] (spor) [right=of surm3]{};
        \node[align=center,anchor=north] (lab3) at (spor.south) {$\sprm(0)$};
        \node[flavour, label=left:$E_6$] (l) [above=of spol]{};
        \node[flavour, label=right:$E_6$] (r) [above=of spor]{};

        \draw[-] (l)--(spol)--(surm3)--(spor)--(r);
        
    \end{tikzpicture}
        \caption{}
        \label{fig:M5Electric}
    \end{subfigure}
    \caption{Two $\frac{1}{2}$ M5 branes probing an $E_6$ Klein singularity. The directions $x^{0,1,2,3,4,5}$ are suppressed. The corresponding electric quiver \Quiver{fig:M5Electric} comes from the F-theory dual description.}
    \label{fig:M5}
\end{figure}

The Higgs branch of this theory can also be realised through magnetic quivers. The magnetic quiver for the rank-1 E-string theory is the $E_8^{(1)}$ quiver. The collapse of a $(-3)$-curve implies the gauging of a continuous subgroup of the Higgs branch global symmetry \cite{Hanany:2022itc}. For this example, the operation on magnetic quivers is the $\surm(3)$ \hyperref[fig:Chain]{chain polymerisation} of two $E_8^{(1)}$ quivers, as shown in \Figref{fig:E6E6Quiver}. This results in the quiver \Quiver{fig:E6E6Quiver} which is a magnetic quiver for $6d\;\mathcal{N}=(1,0)$ theory coming from two separated $\frac{1}{2}$ M5 branes probing $E_6$ Klein singularity. This magnetic quiver has been known in the physics community for some time but has not been documented before.

\begin{figure}[h!]
    \centering
    \begin{tikzpicture}
         \node[gauge, label=below:$1$] (1l) []{};
            \node[gauge, label=below:$2$] (2l)[right=of 1l]{};
            \node[gauge, label=below:$3$] (3l) [right=of 2l]{};
            \node[gauge, label=below:$4$] (4l) [right=of 3l]{};
            \node[gauge, label=below:$5$] (5l) [right=of 4l]{};
            \node[gauge, label=below:$6$] (6l) [right=of 5l]{};
            \node[gauge, label=below:$4$] (4r) [right=of 6l]{};
            \node[gauge, label=below:$2$] (2r) [right=of 4r]{};
            \node[gauge, label=right:$3$] (3) [above=of 6l]{};

            \draw[-] (1l)--(2l)--(3l)--(4l)--(5l)--(6l)--(4r)--(2r) (6l)--(3);

            \node[] (times) [above=of 3l]{$\times$};

            \node[gauge, label=below:$3$] (3t) [above=of times]{};
            \node[gauge, label=below:$2$] (2t)[right=of 3t]{};
            \node[gauge, label=below:$1$] (1t) [right=of 2t]{};
            \node[gauge, label=below:$4$] (4t) [left=of 3t]{};
            \node[gauge, label=below:$5$] (5t) [left=of 4t]{};
            \node[gauge, label=below:$6$] (6t) [left=of 5t]{};
            \node[gauge, label=below:$4$] (4tl) [left=of 6t]{};
            \node[gauge, label=below:$2$] (2tl) [left=of 4tl]{};
            \node[gauge, label=right:$3$] (3tt) [above=of 6t]{};

            \draw[-] (1t)--(2t)--(3t)--(4t)--(5t)--(6t)--(4tl)--(2tl) (6t)--(3tt);

            \node[gauge,label=right:$3$] (3btr) [below=of 6l]{};
            \node[gauge, label=below:$6$] (6br) [below=of 3btr]{};
            \node[gauge, label=below:$4$] (4br) [right=of 6br]{};
            \node[gauge, label=below:$2$] (2br) [right=of 4br]{};
            \node[gauge, label=below:$5$] (5br) [left=of 6br]{};
            \node[gauge, label=below:$4$] (4bmr) [left=of 5br]{};
            \node[gauge, label=below:$3$] (3b) [left=of 4bmr]{};
            \node[gauge, label=below:$4$] (4bml) [left=of 3b]{};
            \node[gauge, label=below:$5$] (5bl) [left=of 4bml]{};
            \node[gauge, label=below:$6$] (6bl) [left=of 5bl]{};
            \node[gauge, label=below:$4$] (4bl) [left=of 6bl]{};
            \node[gauge, label=below:$2$] (2bl) [left=of 4bl]{};
            \node[gauge, label=left:$3$] (3btl) [above=of 6bl]{};

            \draw[-] (2bl)--(4bl)--(6bl)--(5bl)--(4bml)--(3b)--(4bmr)--(5br)--(6br)--(4br)--(2br) (3btr)--(6br) (3btl)--(6bl);

             \node[] (topghost) [below=of 6t]{};
            \node[] (bottomghost) [left=of 3btl]{};
            \draw [->] (topghost) to [out=-150,in=150,looseness=1] (bottomghost);

    \end{tikzpicture}
    \caption{$\surm(3)$ \hyperref[fig:Chain]{chain polymerisation} of two $E_8^{(1)}$ quivers to produce \Quiver{fig:E6E6Quiver}. This quiver is a magnetic quiver for $6d\;\mathcal{N}=(1,0)$ theory for two separated $\frac{1}{2}$ M5 branes probing Klein $E_6$ singularity.}
    \label{fig:E6E6Quiver}
\end{figure}

The dimension of the moduli space is fifty which is too large to compute exact Hilbert series with the capabilities of a typical computer. For this reason the unrefined perturbative Hilbert series and the PL of the HWG are given up to order $t^{10}$, \begin{align}
    \hs_{\text{unref}}\left[\mathcal C(\text{\Quiver{fig:E6E6Quiver}})\right]&=1+156 t^2+13859 t^4+893669 t^6+45609733 t^8+1923636761 t^{10}+O(t^{11})\\
    \pl\left[\hs_{\text{unref}}\left[\mathcal C(\text{\Quiver{fig:E6E6Quiver}})\right]\right]&=156 t^2+1613 t^4-2915 t^6-627017 t^8-1911458 t^{10}+O(t^{11})\\
    \pl\left[\hwg\left[\mathcal C(\text{\Quiver{fig:E6E6Quiver}})\right]\right]&=(\mu_6+\mu_6')t^2+(1 + \mu_1 \mu_5 + \mu_6 + \mu_1 \mu_1' + \mu_5 \mu_5' + \mu_1' \mu_5' + \mu_6')t^4+O(t^6),
\end{align}

where the $\mu_i$ and $\mu_i'$ are highest weight fugacities for the two different factors of $E_6$.
\FloatBarrier


\section{Cyclic Polymerisation for $\surm(N)$ instanton Moduli Spaces on $\mathbb C^2/\mathbb Z_l$}
\label{sec:instantonconstruc}
The CFHM and Kronheimer-Nakajima quivers, reviewed in Section \ref{sec:intantonreview}, are quivers which take the form of a "necklace" of unitary gauge nodes with some flavours. Here, these quivers are constructed  (in their unframed form) from  \hyperref[fig:Cyclic]{cyclic polymerisations} of various starting quivers. These starting quivers are drawn from magnetic quivers for the moduli spaces of free fields, nilpotent orbit closures and slices in the affine Grassmanian of type A.

In turn, this provides geometric constructions for the moduli spaces of $\surm$ instantons on A-type singularities as hyper-Kähler quotients of Coulomb branches.
\subsection{1 $\surm(N)$ instanton on $\mathbb C^2$}
\label{sec:FiniteAU12}
\begin{figure}[h!]
    \centering
    \begin{tikzpicture}
        \node[gauge, label=below:$1$] (1l) []{};
        \node[gauge, label=below:$1$] (1ml) [right=of 1l]{};
        \node[] (cdotsl) [right=of 1ml]{$\cdots$};
        \node[gauge,label=below:$1$] (1mr) [right=of cdotsl]{};
        \node[gauge, label=below:$1$] (1r) [right=of 1mr]{};

        \draw[-] (1l)--(1ml)--(cdotsl)--(1mr)--(1r);

         \draw [decorate, decoration = {brace, raise=10pt, amplitude=5pt}] (1l) --  (1r) node[pos=0.5,above=15pt,black]{$N+2$};

        \node[] (ghost) [below=of 1ml]{};
        \node[gauge, label=above:$1$] (1t) [below =of cdotsl]{};
       \node[gauge, label=below:$1$] (1lr) [below =of ghost]{};
        \node[] (cdotslr) [right=of 1lr]{$\cdots$};
        \node[gauge, label=below:$1$] (1mlr) [right=of cdotslr]{};
        \node[gauge,label=below:$1$] (1mrr) [right=of 1mlr]{};

        \draw[-] (1mrr)--(1mlr)--(cdotslr)--(1lr)-- (1t)--(1mlr);

        \draw [decorate, decoration = {brace, raise=10pt, amplitude=5pt}] (1mlr) --  (1lr) node[pos=0.5,below=15pt,black]{$N-1$};
    
        \node[] (topghost) [left=of 1l]{};
        \node[] (bottomghost) [left=of 1lr]{};
        \draw [->] (topghost) to [out=-150,in=150,looseness=1] (bottomghost);
    
    \end{tikzpicture}
    \caption{$\urm(1)$ \hyperref[fig:Cyclic]{cyclic polymerisation} of the finite $A_{N+2}$ quiver to produce the CFHM quiver \Quiver{fig:FiniteAU12}.}
    \label{fig:FiniteAU12}
\end{figure}

The starting point is the finite $A_{N+2}$ quiver, whose Coulomb branch is $\mathbb H^{N+1}$. A $\urm(1)$ \hyperref[fig:Cyclic]{cyclic polymerisation} is performed by superimposing a $\urm(1)$ at one end with the penultimate $\urm(1)$ at the other end. This is shown in \Figref{fig:FiniteAU12} and produces quiver \Quiver{fig:FiniteAU12}, which is a CFHM quiver (being a special case of the KN quiver). 

In terms of the Coulomb branch one finds the following result:
\begin{equation}
    \mathbb H\times \mathbb H^{N}///\urm(1)=\mathcal M_{1,\surm(N)}^{\mathbb C^2}\label{eq:FiniteA1U1ADHM},
\end{equation}where the left-hand-side is written in a way which makes the $\surm(N)$ symmetry apparent.

In addition, this result \eqref{eq:FiniteA1U1ADHM}, can be viewed as the ADHM construction \cite{Atiyah:1978ri} for the moduli space of 1 $\surm(N)$ instanton on $\mathbb C^2$.

The Higgs branch of the finite $A_{N+2}$ quiver is trivial and is of dimension 0. The Higgs branch of \Quiver{fig:FiniteAU12} is $\mathbb C^2/\mathbb Z_N$ and is of dimension 1. The conclusion on the Higgs branch is: \begin{equation}
    \mathbb C^2/\mathbb Z_N///\urm(1)=\{0\},
\end{equation}
which is confirmed from Weyl integration.

\subsection{2 $\surm(N)$ instantons on $\mathbb C^2$}
\label{sec:FiniteDU2}
\begin{figure}[h!]
    \centering
    \begin{tikzpicture}
    \node[gauge, label=below:$1$] (1l) []{};
    \node[gauge, label=below:$2$] (2l) [right=of 1l]{};
    \node[] (cdots) [right=of 2l] {$\cdots$};
    \node[gauge, label=below:$2$] (2r) [right=of cdots]{};
    \node[gauge, label=right:$1$] (1t) [above right=of 2r]{};
    \node[gauge, label=right:$1$] (1b) [below right=of 2r]{};

    \draw[-] (1l)--(2l)--(cdots)--(2r)--(1t);
    \draw[-] (1b)--(2r);

    \draw [decorate, decoration = {brace, raise=10pt, amplitude=5pt}] (2r) --  (2l) node[pos=0.5,below=15pt,black]{$N+1$};

    \node[gauge,label=above:$1$] (1r) [below left=of 1b]{};
    \node[gauge, label=below:$2$] (2tr) [below =of 1r]{};
     \node[] (cdotsres) [below=of 2tr]{$\cdots$};
    \node[gauge, label=below:$2$] (2rl) [left=of cdotsres]{};
    \node[gauge, label=below:$2$] (2rr) [right=of cdotsres]{};

    \draw[-] (1r)--(2tr)--(2rl)--(cdotsres)--(2rr)--(2tr);
    \draw [decorate, decoration = {brace, raise=10pt, amplitude=5pt}] (2rr) --  (2rl) node[pos=0.5,below=15pt,black]{$N-1$};

    \node[] (topghost) [left=of 1l]{};
    \node[] (bottomghost) [left=of 2tr]{};
    \draw [->] (topghost) to [out=-150,in=150,looseness=1] (bottomghost);
    \end{tikzpicture}
    \caption{$\urm(2)$ \hyperref[fig:Cyclic]{cyclic polymerisation} of the finite $D_{N+4}$ quiver to produce \Quiver{fig:FiniteDkU2} for $N\geq 1$.}
    \label{fig:FiniteDkU2}
\end{figure}
The starting point is the finite $D_{N+4}$ quiver, whose Coulomb branch is $\mathbb H^{2N+4}$ for $N\geq 0$. The cases for $N\geq 1$ are amenable to a cyclic polymerisation for a $\urm(2)$ hyper-Kähler quotient. Doing the $\urm(2)$ \hyperref[fig:Cyclic]{cyclic polymerisation}, as shown in \Figref{fig:FiniteDkU2}, produces quiver \Quiver{fig:FiniteDkU2}, which is also well known from the CFHM construction \cite{Cremonesi:2014xha}. The conclusion on the Coulomb branch is that:
\begin{equation}
    \mathbb H^{2N+4}///\urm(2)=\begin{cases}\mathcal M_{2,\surm(N)}^{\mathbb C^2},&\quad N>1\\\mathrm{Sym}^2\left(\mathbb C^2\right),&\quad N=1. \end{cases}
\label{eq:FiniteDkU2}
\end{equation}
The result \eqref{eq:FiniteDkU2} is confirmed by Weyl integration. It can be viewed as reproducing the ADHM construction \cite{Atiyah:1978ri} for the moduli space of $2\;\surm(N)$ instantons on $\mathbb C^2$ from a Coulomb branch.

The Higgs branch of the finite $D_{N+4}$ quiver is trivial and of dimension 0. The Higgs branch of the CFHM quiver \Quiver{fig:FiniteDkU2} is $\text{Sym}^2\left(\mathbb C^2/\mathbb Z_N\right)$, which is of dimension 2. There is the expected increase of the Higgs branch dimension after polymerisation by $\textrm{rank}\;\urm(2)=2$.

\subsection{$k\;\surm(N)$ instantons on $\mathbb C^2$}
\label{sec:FreeFieldUk}
\begin{figure}[h!]
    \centering
    \begin{tikzpicture}
        \node[gauge, label=below:$1$] (1l) []{};
        \node[gauge, label=below:$2$] (2l) [right=of 1l]{};
        \node[] (cdotsl) [right=of 2l] {$\cdots$};
        \node[gauge, label=below:$k$] (kl) [right=of cdotsl]{};
        \node[] (cdotsm) [right=of kl]{$\cdots$};
        \node[gauge, label=below:$k$,fill=black] (kr) [right=of cdotsm]{};
        \node[] (cdotsr) [right=of kr]{$\cdots$};
        \node[gauge, label=below:$2$] (2r) [right=of cdotsr]{};
        \node[gauge, label=below:$1$] (1r) [right=of 2r]{};
        \node[gauge, label=left:$1$] (1t) [above=of kl]{};

        \draw[-] (1l)--(2l)--(cdotsl)--(kl)--(cdotsm)--(kr)--(cdotsr)--(2r)--(1r) (kl)--(1t);

            \draw [decorate, decoration = {brace, raise=10pt, amplitude=5pt}] (kr) --  (kl) node[pos=0.5,below=15pt,black]{$N+1$};
    \end{tikzpicture}
    \caption{Magnetic quiver \Quiver{fig:FreeField} for $\mathbb H^{k^2+Nk}$. The node of rank $k$ filled in black is "ugly", with balance $-1$, all other nodes are "good", with non-negative balance.}
    \label{fig:FreeField}
\end{figure}
\begin{figure}[h!]
    \centering
    \begin{tikzpicture}
        \node[gauge, label=below:$1$] (1l) []{};
        \node[gauge, label=below:$2$] (2l) [right=of 1l]{};
        \node[] (cdotsl) [right=of 2l] {$\cdots$};
        \node[gauge, label=below:$k$] (kl) [right=of cdotsl]{};
        \node[] (cdotsm) [right=of kl]{$\cdots$};
        \node[gauge, label=below:$k$] (kr) [right=of cdotsm]{};
        \node[] (cdotsr) [right=of kr]{$\cdots$};
        \node[gauge, label=below:$2$] (2r) [right=of cdotsr]{};
        \node[gauge, label=below:$1$] (1r) [right=of 2r]{};
        \node[gauge, label=left:$1$] (1t) [above=of kl]{};

        \draw[-] (1l)--(2l)--(cdotsl)--(kl)--(cdotsm)--(kr)--(cdotsr)--(2r)--(1r) (kl)--(1t);

            \draw [decorate, decoration = {brace, raise=10pt, amplitude=5pt}] (kr) --  (kl) node[pos=0.5,below=15pt,black]{$N+1$};

        \node[gauge, label=right:$k$] (kt) [below=of kr]{};
        \node[] (vdots) [below =of kt]{$\rvdots$};
        \node[gauge, label=right:$k$] (kb) [below =of vdots]{};
        \node[gauge, label=below:$k$] (kr) [left=of vdots]{};
        \node[gauge, label=below:$1$] (1r) [left=of kr]{};

        \draw[-] (1r)--(kr)--(kb)--(vdots)--(kt)--(kr);

         \draw [decorate, decoration = {brace, raise=10pt, amplitude=5pt}] (kt) --  (kb) node[pos=0.5,right=15pt,black]{$N-1$};

         \node[] (topghost) [left=of 1l]{};
    \node[] (bottomghost) [left=of 1r]{};
    \draw [->] (topghost) to [out=-150,in=-150,looseness=1] (bottomghost);
    \end{tikzpicture}
    \caption{$\urm(k)$ \hyperref[fig:Cyclic]{cyclic polymerisation} of \Quiver{fig:FreeField} to produce quiver \Quiver{fig:FreeFieldUk}.}
    \label{fig:FreeFieldUk}
\end{figure}
The construction of the moduli spaces of 1 and 2 $\surm(N)$ instantons on $\mathbb C^2$, presented in the previous two Subsections \ref{sec:FiniteAU12} and \ref{sec:FiniteDU2} respectively, can be generalised as follows.

The starting quiver \Quiver{fig:FreeField}, shown in \Figref{fig:FreeField}, consists of "good" nodes and a single "ugly" $\urm(k)$ node (filled in black). Application of the dualisation algorithm \cite{Gaiotto:2008ak} to \Quiver{fig:FreeField} terminates the quiver, and so the Coulomb branch is the free field $\mathbb H^{k^2+Nk}$, as is simple to check by explicit computation of its Hilbert series for specific values of $k$ and $N$. Carrying out a $\urm(k)$ \hyperref[fig:Cyclic]{cyclic polymerisation} on \Quiver{fig:FreeField}, as shown in \Figref{fig:FreeFieldUk}, produces the CFHM quiver \Quiver{fig:FreeFieldUk}. The Coulomb branch of this is well known as the moduli space of $k\;\surm(N)$ instantons $\mathcal M_{k,\surm(N)}^{\mathbb C^2}$. The conclusion on the Coulomb branch is that:
\begin{equation}
    \mathbb H^{k^2+Nk}///\urm(k)=\mathcal M_{k,\surm(N)}^{\mathbb C^2}.\label{eq:FreeFieldUk}
\end{equation}
The result \eqref{eq:FreeFieldUk} has been verified for $k=3$ for $N=1,2,3,4$. This is in addition to the cases for $k=1,2$ verified in Sections \ref{sec:FiniteAU12} and \ref{sec:FiniteDU2} respectively.
This construction of $\mathcal M_{k,\surm(N)}^{\mathbb C^2}$ is significant since it realises the ADHM construction of $k\;\surm(N)$ instantons on $\mathbb C^2$, but as a hyper-Kähler quotient on a Coulomb branch.

The Higgs branch of \Quiver{fig:FreeField} is trivial as it is terminated by the dualisation algorithm, which indicates that there are no Higgs branch moduli. The Higgs branch of \Quiver{fig:FreeFieldUk} is $\textrm{Sym}^k\left(\mathbb C^2/\mathbb Z_N\right)$, which is of dimension $k$. There is again the expected increase of the Higgs branch dimension after polymerisation by $\textrm{rank}\;\urm(k)=k$.

\subsection{$\text{Sym}^k\left(\mathbb C^2/\mathbb Z_2\right)$}
\label{sec:A2k1Uk}

\begin{figure}[h!]
    \centering
    \begin{tikzpicture}

    \node[gauge, label=below:$1$] (1l) []{};
    \node[gauge, label=below:$2$] (2l) [right=of 1l]{};
    \node[] (cdotsl) [right=of 2l]{$\cdots$};
    \node[gauge, label=below:$k$] (kl) [right=of cdotsl]{};
    \node[gauge, label=below:$k$] (kr) [right=of kl]{};
    \node[] (cdotsr) [right=of kr]{$\cdots$};
    \node[gauge, label=below:$2$] (2r) [right=of cdotsr]{};
    \node[gauge, label=below:$1$] (1r) [right=of 2r]{};
    \node[gauge, label=above:$1$] (1t) [above right= 1cm and 0.5cm of kl]{};

    \draw[-] (1l)--(2l)--(cdotsl)--(kl)--(kr)--(cdotsr)--(2r)--(1r);
    \draw[-] (kl)--(1t)--(kr);

    \node[] (space) [below=of kr]{};

    \node[gauge, label=below:$k$] (kres)[below=of space] {};
    \node[gauge, label=below:$1$] (1res) [left=of kres]{};

    \draw[double,double distance=3pt,line width=0.4pt] (1res)--(kres);
    \draw (kres) to [out=45, in=-45,looseness=8] (kres);

    \node[](topghost) [left=of 1l]{};
        \node[] (bottomghost) [left=of 1res]{};
        \draw [->] (topghost) to [out=-135,in=-180,looseness=1] (bottomghost);
        
    \end{tikzpicture}
    \caption{$\urm(k)$ \hyperref[fig:Cyclic]{cyclic polymerisation} of the unframed magnetic quiver $MQ\left(\overline{\mathcal O}^{A_{2k}}_{(2^k,1)}\right)$ for $k\geq 2$ to produce \Quiver{fig:ADHMUk}.}
    \label{fig:ADHMUk}
\end{figure}

The $\urm(k)$ \hyperref[fig:Cyclic]{cyclic polymerisation} on the unitary magnetic quiver for the nilpotent orbit closure $\overline{\mathcal O}^{A_{2k}}_{(2^k,1)}$, $MQ\left(\overline{\mathcal O}^{A_{2k}}_{(2^k,1)}\right)$, for $k\geq 2$ is shown in \Figref{fig:ADHMUk}. This produces the quiver \Quiver{fig:ADHMUk}, which becomes an ADHM quiver when framed. The Coulomb branch is $\text{Sym}^k\left(\mathbb C^2/\mathbb Z_2\right)$. The conclusion is,
\begin{equation}
    \overline{\mathcal O}^{A_{2k}}_{(2^k,1)}///\urm(k)=\text{Sym}^k\left(\mathbb C^2/\mathbb Z_2\right).
    \label{eq:ADHMUk}
\end{equation}
The result \eqref{eq:ADHMUk} has been verified up to $k=4$ through Weyl integration. 

From Lustig-Spaltenstein duality the Higgs branch of the quiver at the top of \Figref{fig:ADHMUk} is $\mathcal S^{A_{2k}}_{\mathcal N,(k+1,k)}$, which is of dimension $k$. The Higgs branch of the ADHM quiver \Quiver{fig:ADHMUk} is $\mathcal M_{k,\surm(2)}^{\mathbb C^2}$ which is of dimension $2k$. There is the expected increase of the Higgs branch dimension after polymerisation by $\textrm{rank}\;\urm(k)=k$.
\subsection{2 $\surm(N)$ instantons on $\mathbb C^2/\mathbb Z_2$}
\label{sec:nminAU2}
\begin{figure}[h!]
    \centering
     \begin{tikzpicture}[main/.style = {draw, circle}]
        \node[gauge, label=below:$1$] (1) []{};
        \node[gauge, label=below:$2$] (2L) [right=of 1]{};
        \node[draw=none,fill=none] (2M) [right=of 2L]{$\cdots$};
        \node[gauge, label=below:$2$] (2R) [right=of 2M]{};
        \node[gauge, label=below:$1$] (1R) [right=of 2R]{};
        \node[gauge, label=above:$1$] (1T) [above= of 2M]{};
        \draw[-] (1)--(2L)--(2M)--(2R)--(1R);
        \draw[-] (2L)--(1T)--(2R);
        \draw [decorate, decoration = {brace, raise=10pt, amplitude=5pt}] (2R) --  (2L) node[pos=0.5,below=15pt,black]{$N+1$};

        \node[gauge,label=right:$1$] (1rest) [below=of 2M]{};
        \node[gauge, label=right:$2$] (2rest) [below=of 1rest]{};
        \node[] (cdotsres) [below=of 2rest]{$\cdots$};
        \node[gauge, label=below:$2$] (2resl) [left=of cdotsres]{};
        \node[gauge, label=below:$2$] (2resr) [right=of cdotsres]{};
        
        \draw[-] (2rest)--(2resl)--(cdotsres)--(2resr)--(2rest);

        \draw[double,double distance=3pt,line width=0.4pt] (1rest)--(2rest);

        \node[](topghost) [left=of 1]{};
        \node[] (bottomghost) [left=of 2resl]{};
        \draw [->] (topghost) to [out=-150,in=150,looseness=1] (bottomghost);

        \draw [decorate, decoration = {brace, raise=10pt, amplitude=5pt}] (2resr) --  (2resl) node[pos=0.5,below=15pt,black]{$N-1$};
    \end{tikzpicture}
    \caption{$\urm(2)$ \hyperref[fig:Cyclic]{cyclic polymerisation} of the unframed magnetic quiver $MQ\left(\overline{n. min. A_{N+3}}\right)$ for $N\geq1$ to produce \Quiver{fig:nminAkU2}.}
    \label{fig:nminAkU2}
\end{figure}

The starting point is the magnetic quiver, $\mathcal Q_{M}\left(\overline{n. min. A_{N+3}}\right)$, for $\overline{n.min. A_{N+3}}$ for $N\geq 1$. The $\urm(2)$ \hyperref[fig:Cyclic]{cyclic polymerisation} of $\overline{n.min. A_{N+3}}$ for $N\geq 1$ is shown in \Figref{fig:nminAkU2} to produce quiver \Quiver{fig:nminAkU2}. This is a Kronheimer-Nakajima quiver whose Coulomb branch is $\mathcal M_{2,\surm(N)}^{\mathbb C^2/\mathbb Z_2}$. The conclusion is that,
\begin{equation}
    \overline{n.min. A_N}///\urm(2)=\mathcal M_{2,\surm(N)}^{\mathbb C^2/\mathbb Z_2},\quad N\geq 1.
    \label{eq:nminAkU2}
\end{equation}
The result \eqref{eq:nminAkU2} has been verified with Weyl integration up to $N=4$.

From Lustig-Spaltenstein duality the Higgs branch of the quiver at the top of \Figref{fig:nminAkU2} is $\mathcal S^{A_{N+3}}_{\mathcal N,(N+2,2)}$, which is of dimension 2. The Higgs branch of the quiver \Quiver{fig:nminAkU2} is $\mathcal M_{2,\surm(2)}^{\mathbb C^2/\mathbb Z_N}$ which is of dimension 4. There is the expected increase of the Higgs branch dimension after polymerisation by $\textrm{rank}\;\urm(2)=2$.

\subsection{$k\;\surm(N)$ instantons on $\mathbb C^2/\mathbb Z_l$}
\label{sec:FreeAGUK}
\begin{figure}[h!]
    \centering
    \begin{tikzpicture}
         \node[gauge, label=below:$1$] (1l) []{};
        \node[gauge, label=below:$2$] (2l) [right=of 1l]{};
        \node[] (cdotsl) [right=of 2l] {$\cdots$};
        \node[gauge, label=below:$k$] (kl) [right=of cdotsl]{};
        \node[gauge, label=below:$k$] (kll) [right=of kl]{};
        \node[] (cdotsm) [right=of kll]{$\cdots$};
        \node[gauge, label=below:$k$] (krr) [right=of cdotsm]{};
        \node[gauge, label=below:$k$] (kr) [right=of krr]{};
        \node[] (cdotsr) [right=of kr]{$\cdots$};
        \node[gauge, label=below:$2$] (2r) [right=of cdotsr]{};
        \node[gauge, label=below:$1$] (1r) [right=of 2r]{};
        \node[gauge, label=above:$1$] (1t) [above=of cdotsm]{};

        \draw[-] (1l)--(2l)--(cdotsl)--(kl)--(kll)--(cdotsm)--(krr)--(kr)--(cdotsr)--(2r)--(1r);

        \draw[double,double distance=3pt,line width=0.4pt] (kl)--(1t) node[pos=0.5,above=5pt]{$l-l'$};

        \draw[double,double distance=3pt,line width=0.4pt] (kr)--(1t) node[pos=0.5,above=5pt]{$l'$};

            \draw [decorate, decoration = {brace, raise=10pt, amplitude=5pt}] (krr) --  (kll) node[pos=0.5,below=15pt,black]{$N-1$};

    \end{tikzpicture}
    \caption{Magnetic quiver \Quiver{fig:A2KNm1AG} for $\overline{\left[\mathcal W_{A_{2k+N-1}}\right]}^{[0^{k-1},l-l',0^{N-1},l',0^{k-1}]}_{[0^{k-1},l-l'-1,0^{N-1},l'-1,0^{k-1}]}$ for $N\geq 1$ and $l>l'\geq 1$.}
    \label{fig:A2KNm1AG}
\end{figure}
\begin{figure}[h!]
    \centering
    \begin{subfigure}{\textwidth}
    \centering
         \begin{tikzpicture}
        \node[gauge, label=below:$1$] (1l) []{};
        \node[gauge, label=below:$2$] (2l) [right=of 1l]{};
        \node[] (cdotsl) [right=of 2l] {$\cdots$};
        \node[gauge, label=below:$k$] (kl) [right=of cdotsl]{};
        \node[] (cdotsm) [right=of kl]{$\cdots$};
        \node[gauge, label=below:$k$,fill=black] (kr) [right=of cdotsm]{};
        \node[] (cdotsr) [right=of kr]{$\cdots$};
        \node[gauge, label=below:$2$] (2r) [right=of cdotsr]{};
        \node[gauge, label=below:$1$] (1r) [right=of 2r]{};
        \node[gauge, label=left:$1$] (1t) [above=of kl]{};

        \draw[-] (1l)--(2l)--(cdotsl)--(kl)--(cdotsm)--(kr)--(cdotsr)--(2r)--(1r);

        \draw[double,double distance=3pt,line width=0.4pt] (kl)--(1t) node[pos=0.5,right=5pt]{$l$};

            \draw [decorate, decoration = {brace, raise=10pt, amplitude=5pt}] (kr) --  (kl) node[pos=0.5,below=15pt,black]{$N+1$};
    \end{tikzpicture}
    \caption{}
     \label{fig:FreeFieldL}
    \end{subfigure}
    \centering
    \begin{subfigure}{\textwidth}
    \centering
        \begin{tikzpicture}
        \node[gauge, label=below:$1$] (1l) {};
        \node[gauge, label=below:$2$] (2l) [right=of 1l]{};
        \node[] (cdotsl) [right=of 2l]{$\cdots$};
        \node[gauge, label=below:$k$] (k) [right=of cdotsl]{};
        \node[] (cdotsr) [right=of k]{$\cdots$};
        \node[gauge, label=below:$2$] (2r) [right=of cdotsr]{};
        \node[gauge, label=below:$1$] (1r) [right=of 2r]{};
        \node[gauge, label=left:$1$] (1t) [above=of k]{};

         \draw[double,double distance=3pt,line width=0.4pt] (k)--(1t) node[pos=0.5,right=5pt]{$l$};

         \draw[-] (1l)--(2l)--(cdotsl)--(k)--(cdotsr)--(2r)--(1r);
        \end{tikzpicture}
    \caption{}
    \label{fig:A2k-1AG}
    \end{subfigure}
    \caption{Magnetic quiver \Quiver{fig:FreeFieldL} for $\overline{\left[\mathcal W_{A_{2k-1}}\right]}^{[0^{k-1},l,0^{k-1}]}_{[0^{k-1},l-2,0^{k-1}]}\times \mathbb H^{Nk}$ for $N\geq 1$. This quiver specialises \Quiver{fig:A2KNm1AG} to the case $l'=0$. The node of rank $k$ filled in black is "ugly", with balance $-1$, all other nodes are "good", with non-negative balance. Application of the dualisation algorithm to \Quiver{fig:FreeFieldL} yields \Quiver{fig:A2k-1AG}, which is a magnetic quiver for $\overline{\left[\mathcal W_{A_{2k-1}}\right]}^{[0^{k-1},l,0^{k-1}]}_{[0^{k-1},l-2,0^{k-1}]}$.}
   
\end{figure}
\begin{figure}[h!]
    \centering
    \begin{tikzpicture}
     \node[gauge, label=below:$1$] (1l) []{};
        \node[gauge, label=below:$2$] (2l) [right=of 1l]{};
        \node[] (cdotsl) [right=of 2l] {$\cdots$};
        \node[gauge, label=below:$k$] (kl) [right=of cdotsl]{};
        \node[gauge, label=below:$k$] (kll) [right=of kl]{};
        \node[] (cdotsm) [right=of kll]{$\cdots$};
        \node[gauge, label=below:$k$] (krr) [right=of cdotsm]{};
        \node[gauge, label=below:$k$] (kr) [right=of krr]{};
        \node[] (cdotsr) [right=of kr]{$\cdots$};
        \node[gauge, label=below:$2$] (2r) [right=of cdotsr]{};
        \node[gauge, label=below:$1$] (1r) [right=of 2r]{};
        \node[gauge, label=above:$1$] (1t) [above=of cdotsm]{};

        \draw[-] (1l)--(2l)--(cdotsl)--(kl)--(kll)--(cdotsm)--(krr)--(kr)--(cdotsr)--(2r)--(1r);

        \draw[double,double distance=3pt,line width=0.4pt] (kl)--(1t) node[pos=0.5,above=5pt]{$l-l'$};

        \draw[double,double distance=3pt,line width=0.4pt] (kr)--(1t) node[pos=0.5,above=5pt]{$l'$};

            \draw [decorate, decoration = {brace, raise=10pt, amplitude=5pt}] (krr) --  (kll) node[pos=0.5,below=15pt,black]{$N-1$};

        \node[gauge, label=right:$k$] (kt) [below=of krr]{};
        \node[] (vdots) [below =of kt]{$\rvdots$};
        \node[gauge, label=right:$k$] (kb) [below =of vdots]{};
        \node[gauge, label=below:$k$] (kr) [left=of vdots]{};
        \node[gauge, label=below:$1$] (1r) [left=of kr]{};

        \draw[-] (kr)--(kb)--(vdots)--(kt)--(kr);

        \draw[double,double distance=3pt,line width=0.4pt] (kr)--(1r) node[pos=0.5,above=5pt]{$l$};

         \draw [decorate, decoration = {brace, raise=10pt, amplitude=5pt}] (kt) --  (kb) node[pos=0.5,right=15pt,black]{$N-1$};

         \node[] (topghost) [left=of 1l]{};
    \node[] (bottomghost) [left=of 1r]{};
    \end{tikzpicture}
    \caption{$\urm(k)$ \hyperref[fig:Cyclic]{cyclic polymerisation} of \Quiver{fig:A2KNm1AG} to produce quiver \Quiver{fig:FreeFieldUkL}.}
    \label{fig:FreeFieldUkL}
\end{figure}

Following the construction of the moduli space of 2 $\surm(N)$ instanton on $\mathbb C^2/\mathbb Z_2$ in Section \ref{sec:nminAU2}, this construction is generalised for the moduli space of $k\;\surm(N)$ instantons on $\mathbb C^2/\mathbb Z_l$. In this case, the embedding of $\mathbb Z_l$ into the gauge group $\surm(N)$ preserves the gauge group.

The starting quiver is \Quiver{fig:A2KNm1AG}. The Coulomb branch of \Quiver{fig:A2KNm1AG} depends on various choices of $l$, $l'$, and $N$. Firstly, $N\geq 1$ otherwise one cannot perform $\urm(k)$ cyclic polymerisation. If $l=1$ and $l'=0$ one recovers \Quiver{fig:FreeField}, the Coulomb branch is $\mathbb H^{k^2+Nk}$. If $l>1$ and $l'=0$ then the Coulomb branch is $\overline{\left[\mathcal W_{A_{2k-1}}\right]}^{[0^{k-1},l,0^{k-1}]}_{[0^{k-1},l-2,0^{k-1}]}\times \mathbb H^{Nk}$. The appearance of the free part can be seen from application of the dualisation algorithm on \Quiver{fig:FreeFieldL}; with $l\geq 2$ there is a non-trivial part to the moduli space which is given by the Coulomb branch of \Quiver{fig:A2k-1AG}, $\overline{\left[\mathcal W_{A_{2k-1}}\right]}^{[0^{k-1},l,0^{k-1}]}_{[0^{k-1},l-2,0^{k-1}]}$\cite{Bourget:2021siw}\footnote{To see this, frame \Quiver{fig:A2k-1AG} on the top node of $\urm(1)$. Then remember that the gauge nodes form the finite Dynkin diagram of $A_{2k-1}$ and that the top coweight is given by the flavours on each gauge node and the bottom by the balance of each gauge node.}. Finally for $l>l'\geq 1$ the Coulomb branch is $\overline{\left[\mathcal W_{A_{2k+N-1}}\right]}^{[0^{k-1},l-l',0^{N-1},l',0^{k-1}]}_{[0^{k-1},l-l'-1,0^{N-1},l'-1,0^{k-1}]}$.

The $\urm(k)$ \hyperref[fig:Cyclic]{cyclic polymerisation} on \Quiver{fig:A2KNm1AG} produces \Quiver{fig:FreeFieldUkL}, as shown in \Figref{fig:FreeFieldUkL}. This quiver \Quiver{fig:FreeFieldUkL} is a Kronheimer-Nakajima quiver \cite{Kronheimer1990}. The Coulomb branch is the moduli space of $k\;\surm(N)$ instantons on $\mathbb C^2/\mathbb Z_l$, $\mathcal M_{k,\surm(N)}^{\mathbb C^2/\mathbb Z_l}$. From the different choices of $l$ and $l'$ the following conclusions on the Coulomb branch are\footnote{The case for $l=1$ and $l'=0$ was studied in Section \ref{sec:FreeFieldUk} and will not be restated here.}: \begin{align}
    \overline{\left[\mathcal W_{A_{2k+N-1}}\right]}^{[0^{k-1},l-l',0^{N-1},l',0^{k-1}]}_{[0^{k-1},l-l'-1,0^{N-1},l'-1,0^{k-1}]}///\urm(k)&=\mathcal M_{k,\surm(N)}^{\mathbb C^2/\mathbb Z_l},\quad l>l'\geq 1\\
    \overline{\left[\mathcal W_{A_{2k-1}}\right]}^{[0^{k-1},l,0^{k-1}]}_{[0^{k-1},l-2,0^{k-1}]}\times \mathbb H^{Nk}///\urm(k)&=\mathcal M_{k,\surm(N)}^{\mathbb C^2/\mathbb Z_l},\quad l>1,l'=0.
\end{align}

Conclusions about the Higgs branch of \Quiver{fig:FreeFieldL} and \Quiver{fig:A2KNm1AG} are also made. Firstly, since \Quiver{fig:FreeFieldL} can be dualised to \Quiver{fig:A2k-1AG} the Higgs branch of \Quiver{fig:A2k-1AG} is also the Higgs branch of \Quiver{fig:FreeFieldL}. This complication does not arise for \Quiver{fig:A2KNm1AG} for $l>l'\geq 1$. As \Quiver{fig:A2k-1AG} and \Quiver{fig:A2KNm1AG} are magnetic quivers for affine Grassmannian slices, the Higgs branch cannot be found from an order inverting duality as was possible for magnetic quivers for slices in nilcones. Instead, the Coulomb branch of the $3d$ mirror quiver can be found, which is constructed from the type IIB brane system with a series of brane moves and S-duality \cite{Hanany:1996ie}. The $3d$ mirrors of 
\Quiver{fig:A2k-1AG} and \Quiver{fig:A2KNm1AG} are \Quiver{fig:A2k-1AGMirror} and \Quiver{fig:A2KNm1AG} respectively, show in \Figref{fig:Mirrors}. The Coulomb branch of \Quiver{fig:A2k-1AGMirror} is $\overline{\left[\mathcal W_{A_{l-1}}\right]}^{[k,0^{l-3},k]}_{[0^{l-1}]}$ and the Coulomb branch of \Quiver{fig:A2KNm1AGMirror} is $\overline{\left[\mathcal W_{A_{l-1}}\right]}^{[k,0^{l-l'-2},N,0^{l'-2},k]}_{[0^{l-l'-1},N,0^{l'-1}]}$. In summary, \begin{align}
    \mathcal H(\text{\Quiver{fig:FreeFieldL}})=\mathcal H(\text{\Quiver{fig:A2k-1AG}})=\mathcal C(\text{\Quiver{fig:A2k-1AGMirror}})&=\overline{\left[\mathcal W_{A_{l-1}}\right]}^{[k,0^{l-3},k]}_{[0^{l-1}]},\\\mathcal H(\text{\Quiver{fig:A2KNm1AG}})=\mathcal C(\text{\Quiver{fig:A2KNm1AGMirror}})&=\overline{\left[\mathcal W_{A_{l-1}}\right]}^{[k,0^{l-l'-2},N,0^{l'-2},k]}_{[0^{l-l'-1},N,0^{l'-1}]}.
\end{align} These moduli spaces are both of dimension $k(l-1)$ and the Higgs branch of \Quiver{fig:FreeFieldUkL} is of dimension $kl$ and so the Higgs branch dimension has increased by $\textrm{rank}\;\urm(k)=k$ after polymerisation as expected.

\begin{figure}[h!]
    \centering
    \begin{subfigure}{\textwidth}
    \centering
         \begin{tikzpicture}
       \node[gauge, label=below:$k$] (kl) [right=of k]{};
         \node[] (cdots) [right=of kl]{$\cdots$};
         \node[gauge, label=below:$k$] (kr) [right=of cdots]{};
         \node[draw, label=left:$k$] (kfl) [above=of kl]{};
         \node[draw, label=right:$k$] (kfr) [above=of kr]{};

         \draw[-] (kfl)--(kl)--(cdots)--(kr)--(kfr);

          \draw [decorate, decoration = {brace, raise=20pt, amplitude=5pt}] (kr) --  (kl) node[pos=0.5,below=25pt,black]{$l-1$};

    \end{tikzpicture}
    \caption{}
     \label{fig:A2k-1AGMirror}
    \end{subfigure}
    \centering
    \begin{subfigure}{\textwidth}
    \centering
        \begin{tikzpicture}
        \node[gauge, label=below:$k$] (kl) []{};
        \node[gauge, label=below:$k$] (kll)[right=of kl]{};
        \node[] (cdotsl) [right=of kll]{$\cdots$};
        \node[gauge, label=below:$k$] (klr) [right=of cdotsl]{};
        \node[gauge, label=below:$k$] (km) [right=of klr]{};
        \node[gauge, label=below:$k$] (krl) [right=of km]{};
        \node[] (cdotsr) [right=of krl]{$\cdots$};
        \node[gauge, label=below:$k$] (krr) [right=of cdotsr]{};
        \node[gauge, label=below:$k$] (kr) [right=of krr]{};
        \node[draw, label=left:$k$] (kfl) [above=of kl]{};
        \node[draw, label=right:$k$] (kfr) [above=of kr]{};
        \node[draw, label=left:$N$] (kfN) [above=of km]{};

        \draw[-] (kfl)--(kl)--(kll)--(cdotsl)--(klr)--(km)--(kfN) (km)--(krl)--(cdotsr)--(krr)--(kr)--(kfr);

         \draw [decorate, decoration = {brace, raise=20pt, amplitude=5pt}] (klr) --  (kl) node[pos=0.5,below=25pt,black]{$l-l'-1$};
          \draw [decorate, decoration = {brace, raise=20pt, amplitude=5pt}] (kr) --  (krl) node[pos=0.5,below=25pt,black]{$l'-1$};
        
        \end{tikzpicture}
    \caption{}
    \label{fig:A2KNm1AGMirror}
    \end{subfigure}
    \caption{(a) Magnetic quiver \Quiver{fig:A2k-1AGMirror} for $\overline{\left[\mathcal W_{A_{l-1}}\right]}^{[k,0^{l-3},k]}_{[0^{l-1}]}$. This quiver is $3d$ mirror to \Quiver{fig:A2k-1AG}. (b) Magnetic quiver \Quiver{fig:A2KNm1AGMirror} for $\overline{\left[\mathcal W_{A_{l-1}}\right]}^{[k,0^{l-l'-2},N,0^{l'-2},k]}_{[0^{l-l'-1},N,0^{l'-1}]}$. This quiver is $3d$ mirror to \Quiver{fig:A2KNm1AG}.}
    \label{fig:Mirrors}
    \end{figure}
\subsection{$\mathcal M_{(2,1),(N,0)}^{\mathbb C^2/\mathbb Z_2;\surm(N)}$}
\label{sec:nminAU1}
\begin{figure}[h!]
    \centering
     \begin{tikzpicture}[main/.style = {draw, circle}]
        \node[gauge, label=below:$1$] (1) []{};
        \node[gauge, label=below:$2$] (2L) [right=of 1]{};
        \node[draw=none,fill=none] (2M) [right=of 2L]{$\cdots$};
        \node[gauge, label=below:$2$] (2R) [right=of 2M]{};
        \node[gauge, label=below:$1$] (1R) [right=of 2R]{};
        \node[gauge, label=above:$1$] (1T) [above= of 2M]{};
        \draw[-] (1)--(2L)--(2M)--(2R)--(1R);
        \draw[-] (2L)--(1T)--(2R);
        \draw [decorate, decoration = {brace, raise=10pt, amplitude=5pt}] (2R) --  (2L) node[pos=0.5,below=15pt,black]{$N-1$};

        \node[gauge,label=right:$1$] (1rest) [below=of 2M]{};
        \node[] (cdotsres) [below=of 1rest]{$\cdots$};
        \node[gauge, label=below:$2$] (2resl) [left=of cdotsres]{};
        \node[gauge, label=below:$2$] (2resr) [right=of cdotsres]{};
        \node[gauge, label=below:$1$] (1resb) [below=of cdotsres]{};

        \draw[-] (2resl)--(cdotsres)--(2resr)--(1rest)--(2resl)--(1resb)--(2resr);

        \node[](topghost) [left=of 1]{};
        \node[] (bottomghost) [left=of 2resl]{};
        \draw [->] (topghost) to [out=-150,in=150,looseness=1] (bottomghost);
    \end{tikzpicture}
    \caption{$\urm(1)$ \hyperref[fig:Cyclic]{cyclic polymerisation} of the unframed magnetic quiver for $\overline{n. min. A_{N+1}}$, where $N\geq2$, to produce \Quiver{fig:nminAkU1}.}
    \label{fig:nminAkU1}
\end{figure}

The $\urm(1)$ \hyperref[fig:Cyclic]{cyclic polymerisation} of the unframed unitary magnetic quiver for $\overline{n.min. A_{N+1}}$ for $N\geq 2$ is shown in \Figref{fig:nminAkU1}. This produces the diamond shaped quiver \Quiver{fig:nminAkU1} \cite{Dey:2013fea}, whose Coulomb branch has $\surm(N)\times \surm(2)$ global symmetry. When framed on a $\urm(1)$ gauge node this becomes a Kronheimer-Nakajima quiver, whose Coulomb branch is $\mathcal M_{(2,1),(N,0)}^{\mathbb C^2/\mathbb Z_2;\surm(N)}$ with $\surm(N)$ gauge group.\footnote{Since the gauge node ranks are not the same in this example, the quiver data is used to specify the moduli space of instantons. There is a dictionary between the quiver data and geometric data of the instanton solution in \cite{Witten:2009xu,Cherkis:2009hpw,Dey:2013fea}.}

The conclusion is that:
\begin{equation}
    \overline{n. min. A_{N+1}}///\urm(1)=\mathcal M_{(2,1),(N,0)}^{\mathbb C^2/\mathbb Z_2;\surm(N)},\quad N\geq2 \label{eq:nminAkU1}.
\end{equation}

\begin{table}[h!]
    \centering
    \begin{tabular}{cc}
    \toprule
        $N$ & $\hs[\mathcal C(\mathcal Q_{\ref{fig:nminAkU1}})]$\\\midrule
        
         2 & $\frac{1 + 3 t^2 + 11 t^4 + 10 t^6 + 11 t^8 + 
 3 t^{10} + t^{12}}{(1 - t^2)^3 (1 - t^4)^3}$\\ 
         3 & $\frac{1 + 6 t^2 + 39 t^4 + 104 t^6 + 218 t^8 + 248 t^{10} + 218 t^{12} + 
         104 t^{14} + 39 t^{16} + 6 t^{18} + t^{20}}{(1 - t^2)^5 (1 - t^4)^5}$\\
         4 & $\frac{1 + 11 t^2 + 103 t^4 + 509 t^6 + 1741 t^8 + 3911 t^{10} + 6427 t^{12} + 
 7442 t^{14} +\pal + t^{28}}{(1 - t^2)^7 (1 - t^4)^7}$\\
        5 & $\frac{1 + 18 t^2 + 233 t^4 + 1760 t^6 + 8930 t^8 + 31412 t^{10} + 
 81122 t^{12} + 156464 t^{14} + 231184 t^{16} + 262512 t^{18} +\pal + t^{36}}{(1 - t^2)^9 (1 - t^2)^9}$\\\bottomrule
    \end{tabular}
    \caption{Unrefined Hilbert series for the Coulomb branch of \Quiver{fig:nminAkU1} for $N=1,2,3,4$.}
    \label{tab:nminAkU1HS}
\end{table}

\begin{table}[h!]
    \centering
    \begin{tabular}{cc}
    \toprule
        $N$ & $\hs[\mathcal H(\mathcal Q_{\ref{fig:nminAkU1}})]$\\\midrule
        
        2 &$ \frac{1 + 3 t^2 + 11 t^4 + 10 t^6 + 11 t^8 + 3 t^{10} + t^{12}}{(1 - t^2)^3 (1 - t^4)^3}$\\
        3&$\frac{1 - 2 t + 3 t^2 - 2 t^3 + 3 t^4 + 3 t^8 - 2 t^9 + 3 t^{10} - 
 2 t^{11} + t^{12}}{(1 - t)^2 (1 - t^4) (1 - t^3)^2 (1 - t^6)}$\\ 
 4& $\frac{1 + 4 t^4 + 4 t^6 + 5 t^8 + 5 t^{10} + 4 t^{12} + 
 4 t^{14} + t^{18}}{(1 - t^2)^2 (1 - t^4)^3 (1 - t^8)}$\\
 5 & $\frac{1 - 2 t + 3 t^2 - 4 t^3 + 7 t^4 - 8 t^5 + 9 t^6 - 6 t^7 + 5 t^8 - 
 4 t^9 + 6 t^{10} + \pal + t^{20}}
 {(1 - t)^2 (1 - t^4)  (1 - t^5)^2 (1 - t^{10})}$\\\bottomrule
    \end{tabular}
    \caption{Unrefined Hilbert series for the Higgs branch of \Quiver{fig:nminAkU1} for $N=2,3,4,5$.}
    \label{tab:nminAkU1HSHiggs}
\end{table}
The result \eqref{eq:nminAkU1} has been checked for $2\leq N\leq 5$. Unrefined Hilbert series for the Coulomb branch of \Quiver{fig:nminAkU1} are presented in Table \ref{tab:nminAkU1HS} for $N=2,3,4,5$.

The Higgs branch of \Quiver{fig:nminAkU1} is also easily computed, and is presented in Table \ref{tab:nminAkU1HSHiggs} for $N=2,3,4,5$. The Higgs branch global symmetry is $\urm(1)\times \urm(1)$ and the dimension of the moduli space is 3. The Higgs branch of \Quiver{fig:nminAkU1} may be identified more easily through the Coulomb branch of its $3d$ mirror \Quiver{fig:nminAkU1mirror} shown in \Figref{fig:nminAkU1mirror}. Note that for the case of $N=2$ \Quiver{fig:nminAkU1} is self-mirror. When framed on a $\urm(1)$ gauge node, of which there are two choices, then \Quiver{fig:nminAkU1mirror} is of Kronheimer-Nakajima type. To preserve the duality between number of flavours and number of gauge nodes under $3d$ mirror symmetry the $\urm(1)$ with edge multiplicity $N$ is framed. This identifies the moduli space as $\mathcal M^{\mathbb C^2/\mathbb Z_{N};\surm(2)}_{(2^{N-1},1),(1,0^{N-3},1,0)}$. The conclusion on the Higgs branch is that \begin{equation}
   \mathcal M^{\mathbb C^2/\mathbb Z_{N};\surm(2)}_{(2^{N-1},1),(1,0^{N-3},1,0)}///\urm(1)=\mathcal S^{A_{N+1}}_{\mathcal N,(N,2)},\quad N\geq1,
\end{equation}
which is verified with Weyl integration.

\begin{figure}
    \centering
    \begin{tikzpicture}
    \node[gauge,label=below:$1$] (1) []{};
    \node[gauge, label=below:$2$] (2) [right=of 1]{};
    \node[gauge, label=below:$1$] (F) [right=of 2]{};

     \draw[double,double distance=3pt,line width=0.4pt] (1)--(2);
     \draw[double,double distance=3pt,line width=0.4pt] (2)--(F)node[pos=0.5,above=5pt]{$N$};
    \end{tikzpicture}
    \caption{$3d$ mirror quiver to \Quiver{fig:nminAkU1}.}
    \label{fig:nminAkU1mirror}
\end{figure}

\subsection{Generic Kronheimer-Nakajima Quivers}
\label{sec:genericKN}
\begin{figure}[h!]
    \centering
    \begin{tikzpicture}
         \node[gauge, label=below:$1$] (1l) []{};
        \node[] (cdotsl) [right=of 1l]{$\cdots$};
        \node[gauge, label=below:$k_1$] (k1) [right=of cdotsl]{};
        \node[gauge, label=below:$k_2$] (k2) [right=of k1]{};
        \node[] (cdotsm) [right=of k2]{$\cdots$};
        \node[gauge, label=below:$k_N$] (kN) [right=of cdotsm]{};
        \node[gauge, label=below:$k_1$] (k1r) [right=of kN]{};
        \node[] (cdotsr) [right=of k1r]{$\cdots$};
        \node[gauge, label=below:$1$] (1r) [right=of cdotsr]{};
        \node[gauge, label=above:$1$] (1t) [above=of cdotsm]{};

        \draw[-] (1l)--(cdotsl)--(k1)--(k2)--(cdotsm)--(kN)--(k1r)--(cdotsr)--(1r);

        \draw[double,double distance=3pt,line width=0.4pt] (k1)--(1t) node[pos=0.5,above=5pt]{$l_1-l'$};
        \draw[double,double distance=3pt,line width=0.4pt] (k2)--(1t) node[pos=0.5,right=5pt]{$l_2$};
        \draw[double,double distance=3pt,line width=0.4pt] (kN)--(1t) node[pos=0.5,left=5pt]{$l_N$};
        \draw[double,double distance=3pt,line width=0.4pt] (k1r)--(1t) node[pos=0.5,above=5pt]{$l'$};
    \end{tikzpicture}
    \caption{Unframed linear quiver \Quiver{fig:AAGGen}.}
    \label{fig:AAGGen}
\end{figure}
\begin{figure}[h!]
    \centering
    \begin{tikzpicture}
        \node[gauge, label=below:$1$] (1l) []{};
        \node[] (cdotsl) [right=of 1l]{$\cdots$};
        \node[gauge, label=below:$k_1$] (k1) [right=of cdotsl]{};
        \node[gauge, label=below:$k_2$] (k2) [right=of k1]{};
        \node[] (cdotsm) [right=of k2]{$\cdots$};
        \node[gauge, label=below:$k_N$] (kN) [right=of cdotsm]{};
        \node[gauge, label=below:$k_1$] (k1r) [right=of kN]{};
        \node[] (cdotsr) [right=of k1r]{$\cdots$};
        \node[gauge, label=below:$1$] (1r) [right=of cdotsr]{};
        \node[gauge, label=above:$1$] (1t) [above=of cdotsm]{};

        \draw[-] (1l)--(cdotsl)--(k1)--(k2)--(cdotsm)--(kN)--(k1r)--(cdotsr)--(1r);

        \draw[double,double distance=3pt,line width=0.4pt] (k1)--(1t) node[pos=0.5,above=5pt]{$l_1-l'$};
        \draw[double,double distance=3pt,line width=0.4pt] (k2)--(1t) node[pos=0.5,right=5pt]{$l_2$};
        \draw[double,double distance=3pt,line width=0.4pt] (kN)--(1t) node[pos=0.5,left=5pt]{$l_N$};
        \draw[double,double distance=3pt,line width=0.4pt] (k1r)--(1t) node[pos=0.5,above=5pt]{$l'$};

        \node[gauge, label=above:$k_1$] (K1) [below =of cdotsm]{};
        \node[] (ghostl) [below left=of K1]{};
        \node[] (ghostr) [below right=of K1]{};
        \node[gauge, label=below:$1$] (1m) [below= of K1]{};
        \node[gauge, label=below left:$k_2$] (K2)[below left=of ghostl]{};
        \node[gauge, label=below right:$k_N$] (KN)[below right=of ghostr]{};
        \node[] (cdotsres) at ($(K2)!0.5!(KN)$) {$\cdots$};

        \draw[-] (K1)--(K2)--(cdotsres)--(KN)--(K1);

        \draw[double,double distance=3pt,line width=0.4pt] (K1)--(1m) node[pos=0.5,left=0pt]{$l_1$};
        \draw[double,double distance=3pt,line width=0.4pt] (K2)--(1m) node[pos=0.5,right=10pt]{$l_2$};
        \draw[double,double distance=3pt,line width=0.4pt] (KN)--(1m) node[pos=0.5,left=10pt]{$l_N$};

        \node[] (topghost) [left=of 1]{};
    \node[] (bottomghost) [left=of K2]{};
    \draw [->] (topghost) to [out=-150,in=150,looseness=1] (bottomghost);
    \end{tikzpicture}
    \caption{Construction of a generic Kronheimer-Nakajima quiver from $\urm(k_1)$ \hyperref[fig:Cyclic]{cyclic polymerisation}.}
    \label{fig:GenericKN}
\end{figure}

Having constructed specific cases of the Kronheimer-Nakajima quivers, a generic construction from cyclic polymerisation for any "good" or "ugly" Kronheimer-Nakajima quiver is presented.

The starting quiver \Quiver{fig:AAGGen}, drawn in \Figref{fig:AAGGen}, consists of two legs of gauge nodes up to $\urm(k_1)$ with the middle of the quiver containing other unitary gauge nodes. All of the gauge nodes in the middle are connected to the $\urm(1)$ gauge node at the top with some multiplicity of edges. When framed at the $\urm(1)$ at the top this quiver becomes linear with flavours on top. The Coulomb branch of this quiver depends on the particular values of $\urm(k_i)$ and $l_i$ for $i=1,\cdots,N$ so it is difficult to give a generic name to this moduli space.

However, suppose that the $k_i$ and $l_i$ are chosen so that \Quiver{fig:AAGGen} is "good", then the Coulomb branch is an affine Grassmannian slice of $A_{2k_1+N-1}$. The coweights corresponding to the flavours and balance are easily computed but not presented here for brevity. If \Quiver{fig:AAGGen} is "ugly" then most likely, the moduli space consists of a slice in the affine Grassmannian of $A_{r}$ for $r<2k_1+N-1$ with a free part. 

In either case of \Quiver{fig:AAGGen} being "good" or "ugly", $\urm(k_1)$ \hyperref[fig:Cyclic]{cyclic polymerisation} may be performed as shown in \Figref{fig:GenericKN} to give the most generic unitary Kronheimer-Nakajima quiver \Quiver{fig:KNFrameSame},(when framed on the central $\urm(1)$ gauge node). The conclusion is that: \begin{equation}
    \mathcal C\left(\text{\Quiver{fig:AAGGen}}\right)///\urm(k_1)=\mathcal M_{(k'_1,\cdots,k'_l),(l'_1,\cdots,l'_l)}^{\mathbb C^2/\mathbb Z_l},\quad l=\sum_{i=1}^Nl_i.
\end{equation}

This schematic construction generalises to all unitary Kronheimer-Nakajima quivers and hence gives the construction of all $\surm$ instantons on $\mathbb C^2/\mathbb Z_l$ (where $l=\sum_i l_i$), for any data specifying the instanton bundle as a hyper-Kähler quotient on the Coulomb branch.

\section{$3d$ Mirror Symmetry and Hyper-Kähler Quotients}
\label{sec:3dmirror}
$3d$ mirror symmetry can give rise to interesting relationships between hyper-Kähler quotients on the Higgs/Coulomb branch of one theory and hyper-Kähler quotients on the Coulomb/Higgs branch of the dual theory. Recall that taking the hyper-Kähler quotient of the Higgs branch of a quiver involves gauging a subgroup of the flavour symmetry, which is trivial to carry out diagramatically. The results in Section \ref{sec:FreeFieldUk} and Section \ref{sec:FreeAGUK} illustrate such relationships involving instanton moduli spaces, and these are summarised graphically in \Figref{fig:ADHMCommute} and \Figref{fig:ADHMCommuteL}, respectively.

\Figref{fig:ADHMCommute} shows the relationships involving \Quiver{fig:FreeField} (top left), whose Coulomb branch is the moduli space of free fields $\mathbb H^{k^2+Nk}$, and whose Higgs branch is trivial. The Dynkin diagram symmetry of \Quiver{fig:FreeField} is $\surm(k)\times \surm(N+k)\times \urm(1)$. This is broken to $\urm(k)\times \surm(N)$ by the choice of embedding used for the polymerisation, which identifies the two subsets of $\surm(k)$ simple roots, portrayed by matching colours.

The $3d$ mirror is in the top right of \Figref{fig:ADHMCommute}, whose Coulomb branch is trivial and whose Higgs branch is the free theory $\mathbb H^{k^2+Nk}$, with a manifest $\urm(k)\times \surm(N)$ flavour symmetry.

$\urm(k)$ cyclic polymerisation of quiver \Quiver{fig:FreeField} (top left) produces the CFHM quiver \Quiver{fig:FreeFieldUk} (bottom left). The action on the Coulomb branch of \Quiver{fig:FreeField} is a $\urm(k)$ hyper-Kähler quotient. When a corresponding $\urm(k)$ hyper-Kähler quotient is applied to the flavour symmetry of the top right theory in \Figref{fig:ADHMCommute}, this produces the ADHM quiver (bottom right), whose Higgs branch is $\mathcal M_{k,\surm(N)}^{\mathbb C^2}$.

Notably, the quiver \Quiver{fig:FreeFieldUk} and the ADHM quiver are also $3d$ mirror dual. Hence, the operations of a $\urm(k)$ hyper-Kähler quotient of the Coulomb branch of \Quiver{fig:FreeField}, or the Higgs branch of its dual, commute with $3d$ mirror symmetry.

\Figref{fig:ADHMCommuteL} summarises the relationships involving \Quiver{fig:FreeFieldL} (top left). As discussed in Section \ref{sec:FreeAGUK}, the Coulomb branch is a product space between an affine Grassmannian slice described by the magnetic quiver \Quiver{fig:A2k-1AG} and an $\mathbb H^{Nk}$ free field. Its Higgs branch is determined by \Quiver{fig:A2k-1AG}.

The $3d$ mirror for \Quiver{fig:FreeFieldL} is the product quiver shown in the top right of \Figref{fig:ADHMCommuteL}. This $3d$ mirror is obtained by applying S-duality transformations to the magnetic quivers for the constituents of the product space. As before, the identification of gauge nodes entails that the global symmetry is $\surm(N)\times \urm(k)$ in both cases.

The $\urm(k)$ cyclic polymerisation of \Quiver{fig:FreeFieldL} to yield \Quiver{fig:FreeFieldUkL} is shown in the left-hand column of \Figref{fig:ADHMCommuteL}. The action on the Coulomb branch is a $\urm(k)$ hyper-Kähler quotient. When a corresponding $\urm(k)$ hyper-Kähler quotient is applied to the Higgs branch of the product quiver in the top right of \Figref{fig:ADHMCommuteL}, by gauging the common $\urm(k)$ flavour symmetry, this produces the Kronheimer-Nakajima quiver (bottom right), whose Higgs branch is $\mathcal M_{k,\surm(N)}^{\mathbb C^2/\mathbb Z_l}$.

Notably, the $3d$ mirror of \Quiver{fig:FreeFieldUkL} (when framed on the $\urm(1)$ gauge node) also gives the Kronheimer-Nakajima quiver (bottom right). Hence, $3d$ mirror symmetry commutes with the $\urm(k)$ hyper-Kähler quotient on the Coulomb branches and the $\urm(k)$ hyper-Kähler quotient on the Higgs branches of these theories.

In both \Figref{fig:ADHMCommute} and \Figref{fig:ADHMCommuteL}, the action on the Higgs branch in the left-hand column is an increase in dimension by $k$, whereas in the right-hand column there is an increase in dimension of the Coulomb branch by $k$.

\begin{landscape}
    
    \begin{figure}
        \centering
         \scalebox{0.9}{\begin{tikzpicture}
        \node (a) at (0,0) {
    $\begin{tikzpicture}
        \node[gauge, label=below:$1$,fill=green] (1l) []{};
        \node[gauge, label=below:$2$,fill=orange] (2l) [right=of 1l]{};
        \node[] (cdotsl) [right=of 2l] {$\cdots$};
        \node[gauge, label=below:$k-1$,fill=purple] (km1l) [right=of cdotsl]{};
        \node[gauge, label=below:$k$] (kl) [right=of km1l]{};
        \node[] (cdotsm) [right=of kl]{$\cdots$};
        \node[gauge, label=below:$k$] (kr) [right=of cdotsm]{};
        \node[gauge, label=below:$k-1$,fill=green] (km1r) [right=of kr]{};
        \node[] (cdotsr) [right=of km1r]{$\cdots$};
        \node[gauge, label=below:$2$,fill=orange] (2r) [right=of cdotsr]{};
        \node[gauge, label=below:$1$,fill=purple] (1r) [right=of 2r]{};
        \node[gauge, label=left:$1$] (1t) [above=of kl]{};

        \draw[-] (1l)--(2l)--(cdotsl)--(km1l)--(kl)--(cdotsm)--(kr)--(km1r)--(cdotsr)--(2r)--(1r) (kl)--(1t);

            \draw [decorate, decoration = {brace, raise=10pt, amplitude=5pt}] (kr) --  (kl) node[pos=0.5,below=15pt,black]{$N+1$};
    \end{tikzpicture}$};
  
    \node (b) at (14,0) {
    $\begin{tikzpicture}
         \node[draw, label=below:$\surm(N)$] (n) []{};
        \node[draw, label=below:$\urm(k)$] (k) [right=of n]{};
        \draw[-] (n)--(k);

         \draw[-] (k) to [out=-45, in=45,looseness=8] (k);
    \end{tikzpicture}$
    };

    \node (c) at (0,-10) {$\begin{tikzpicture}
        \node[gauge, label=left:$1$] (1) []{};
        \node[gauge, label=below:$k$] (kt) [below=of 1]{};
        \node[] (cdots) [below=of kt]{$\cdots$};
        \node[gauge, label=below:$k$] (kl) [left=of cdots]{};
        \node[gauge, label=below:$k$] (kr) [right=of cdots]{};

        \draw[-] (1)--(kt)--(kl)--(cdots)--(kr)--(kt);
        
         \draw [decorate, decoration = {brace, raise=10pt, amplitude=5pt}] (kr) --  (kl) node[pos=0.5,below=15pt,black]{$N-1$};
    \end{tikzpicture}$};

    \node (d) at (14,-10) {$\begin{tikzpicture}
         \node[draw, label=below:$\surm(N)$] (n) []{};
        \node[gauge, label=below:$\urm(k)$] (k) [right=of n]{};
        \draw[-] (n)--(k);

         \draw[-] (k) to [out=-45, in=45,looseness=8] (k);
    \end{tikzpicture}$};
    
    \draw[<->] (a)-- (b) node[pos=0.5, above]{$3d$ mirror (framed)};
    \draw[->] (a)--(c) node[pos=0.5, left]
     {$\begin{array}{c}
    \urm(k) \text{ hyper-Kähler}\\
    \text{quotient on } \mathcal{C}
    \end{array}$};
    \draw[->] (b)--(d)node[pos=0.5, left]
     {$\begin{array}{c}
    \urm(k) \text{ hyper-Kähler}\\
    \text{quotient on } \mathcal{H}
    \end{array}$};
    \draw[<->] (c)--(d) node[pos=0.5, below]{ $3d$ mirror (framed)};
    
      \end{tikzpicture}}
        \caption{A commutative diagram of \Quiver{fig:FreeField}, an electric quiver for the $\mathbb H^{k^2+Nk}$ free theory, the CFHM quiver \Quiver{fig:FreeFieldUk}, and the ADHM quiver for the moduli space of $k\;\surm(N)$ instantons on $\mathbb C^2$, under $3d$ mirror symmetry and $\urm(k)$ hyper-Kähler quotient on either the Higgs and Coulomb branches. The non-white gauge nodes in \Quiver{fig:FreeField} indicate simple root identification of nodes of the same colour.}
    \label{fig:ADHMCommute}
    \end{figure}
  
\end{landscape}

\begin{landscape}
    \begin{figure}
        \centering
        \scalebox{0.9}{
         \begin{tikzpicture}
        \node (a) at (0,0) {
    $\begin{tikzpicture}
        \node[gauge, label=below:$1$,fill=green] (1l) []{};
        \node[gauge, label=below:$2$,fill=orange] (2l) [right=of 1l]{};
        \node[] (cdotsl) [right=of 2l] {$\cdots$};
        \node[gauge, label=below:$k-1$,fill=purple] (km1l) [right=of cdotsl]{};
        \node[gauge, label=below:$k$] (kl) [right=of km1l]{};
        \node[] (cdotsm) [right=of kl]{$\cdots$};
        \node[gauge, label=below:$k$] (kr) [right=of cdotsm]{};
        \node[gauge, label=below:$k-1$,fill=green] (km1r) [right=of kr]{};
        \node[] (cdotsr) [right=of km1r]{$\cdots$};
        \node[gauge, label=below:$2$,fill=orange] (2r) [right=of cdotsr]{};
        \node[gauge, label=below:$1$,fill=purple] (1r) [right=of 2r]{};
        \node[gauge, label=left:$1$] (1t) [above=of kl]{};

        \draw[-] (1l)--(2l)--(cdotsl)--(km1l)--(kl)--(cdotsm)--(kr)--(km1r)--(cdotsr)--(2r)--(1r);

        \draw[double,double distance=3pt,line width=0.4pt] (kl)--(1t) node[pos=0.5,right=5pt]{$l$};

            \draw [decorate, decoration = {brace, raise=10pt, amplitude=5pt}] (kr) --  (kl) node[pos=0.5,below=15pt,black]{$N+1$};
    \end{tikzpicture}$};
  
    \node (b) at (14,0) {
    $\begin{tikzpicture}

        \node[gauge, label=below:$k$] (kl) []{};
        \node[] (cdots) [right=of kl]{$\cdots$};
        \node[gauge, label=below:$k$] (kr) [right=of cdots]{};
        \node[draw, label=left:$k$] (kfl) [above right= ({sqrt(2)/2} and {sqrt(2)/2)})of kl]{};
         \node[draw, label=right:$k$] (kfr) [above left=({sqrt(2)/2} and {sqrt(2)/2)})of kr]{};
         
        \node[draw, label=above right:$\urm(k)$] (k) [above=of cdots]{};

        \node[draw, label=left:$\surm(N)$] (n) [above=of k]{};
        \draw[-] (n)--(k);

         \draw[-] (kfl)--(kl)--(cdots)--(kr)--(kfr);

         \draw [decorate, decoration = {brace, raise=10pt, amplitude=5pt}] (kr) --  (kl) node[pos=0.5,below=15pt,black]{$l-1$};

         \draw[dashed,blue] \convexpath{kfl,k,kfr}{0.3cm};
    \end{tikzpicture}$
    };

    \node (c) at (0,-10) {$\begin{tikzpicture}
         \node[gauge, label=left:$1$] (1) []{};
        \node[gauge, label=below:$k$] (kt) [below=of 1]{};
        \node[] (cdots) [below=of kt]{$\cdots$};
        \node[gauge, label=below:$k$] (kl) [left=of cdots]{};
        \node[gauge, label=below:$k$] (kr) [right=of cdots]{};

        \draw[-] (1)--(kt)--(kl)--(cdots)--(kr)--(kt);
        
         \draw [decorate, decoration = {brace, raise=10pt, amplitude=5pt}] (kr) --  (kl) node[pos=0.5,below=15pt,black]{$N-1$};

        \draw[double,double distance=3pt,line width=0.4pt] (kt)--(1) node[pos=0.5,left=5pt]{$l$};
    \end{tikzpicture}$};

    \node (d) at (14,-10) {$\begin{tikzpicture}

          \node[draw, label=left:$\surm(N)$] (1) []{};
        \node[gauge, label=below:$k$] (kt) [below=of 1]{};
        \node[] (cdots) [below=of kt]{$\cdots$};
        \node[gauge, label=below:$k$] (kl) [left=of cdots]{};
        \node[gauge, label=below:$k$] (kr) [right=of cdots]{};

        \draw[-] (1)--(kt)--(kl)--(cdots)--(kr)--(kt);
        
         \draw [decorate, decoration = {brace, raise=10pt, amplitude=5pt}] (kr) --  (kl) node[pos=0.5,below=15pt,black]{$l-1$};
    \end{tikzpicture}$};

    \draw[<->] (a)-- (b) node[pos=0.5, above]{$3d$ mirror (framed)};
    \draw[->] (a)--(c) node[pos=0.5, left]
    {$\begin{array}{c}
    \urm(k) \text{ hyper-Kähler}\\
    \text{quotient on } \mathcal{C}
    \end{array}$};
    \draw[->] (b)--(d)node[pos=0.5, left] {$\begin{array}{c}
    \urm(k) \text{ hyper-Kähler}\\
    \text{quotient on } \mathcal{H}
    \end{array}$};
    \draw[<->] (c)--(d) node[pos=0.5, below]{$3d$ mirror (framed)};
    
      \end{tikzpicture}}
        \caption{A commutative diagram of \Quiver{fig:FreeFieldL}, \Quiver{fig:ADHMCommuteL}, and two Kronheimer-Nakajima quivers, under $3d$ mirror symmetry and $\urm(k)$ hyper-Kähler quotient on either the Higgs and Coulomb branches. The non-white gauge nodes in \Quiver{fig:FreeFieldL} indicate simple root identification of nodes of the same colour.}
    \label{fig:ADHMCommuteL}
    \end{figure}
\end{landscape}

\section{Cyclic polymerisation of $ADE$-shaped Quivers}
\label{sec:ADEOrb}
\subsection{$\urm(1)$ cyclic polymerisation of $A_N$ for $N\geq 2$}
\begin{figure}[h!]
    \centering
    \begin{tikzpicture}
        \node[gauge, label=below:$1$] (1ll) []{};
        \node[] (cdotsl) [right=of 1ll]{$\cdots$};
        \node[gauge, label=below:$1$] (1lll) [right=of cdotsl]{};
        \node[gauge, label=below:$1$] (1nlrl) [right=of 1lll]{};
        \node[] (cdotsm) [right=of 1nlrl]{$\cdots$};
        \node[gauge, label=below:$1$] (1nlrr) [right=of cdotsm]{};
        \node[gauge, label=below:$1$] (1rl) [right=of 1nlrr]{};
        \node[] (cdotsr) [right=of 1rl]{$\cdots$};
        \node[gauge, label=below:$1$] (1rr) [right=of cdotsr]{};

        \draw[-] (1ll)--(cdotsl)--(1lll)--(1nlrl)--(cdotsm)--(1nlrr)--(1rl)--(cdotsr)--(1rr);

         \draw [decorate, decoration = {brace, raise=10pt, 
         amplitude=5pt}] (1ll) --  (1lll) node[pos=0.5,above=15pt,black]{$L$};
         \draw [decorate, decoration = {brace, raise=10pt, amplitude=5pt}] (1rl) --  (1rr) node[pos=0.5,above=15pt,black]{$R$};
         \draw [decorate, decoration = {brace, raise=10pt, amplitude=5pt}] (1nlrl) --  (1nlrr) node[pos=0.5,above=15pt,black]{$N-L-R$};

         \node[] (cdotsrest) [below=of cdotsm]{$\cdots$};
         \node[gauge, label=left:$1$] (1tl) [left=of cdotsrest]{};
         \node[gauge, label=right:$1$] (1tr) [right=of cdotsrest]{};
         \node[gauge, label=below:$1$] (1m) [below =of cdotsrest]{};
         \node[gauge, label=below:$1$] (1ml) [left=of 1m]{};
         \node[] (cdotslres) [left=of 1ml]{$\cdots$};
         \node[gauge, label=below:$1$] (1lres) [left=of cdotslres]{};
         \node[gauge, label=below:$1$] (1mr) [right=of 1m]{};
         \node[] (cdotsrres) [right=of 1mr]{$\cdots$};
         \node[gauge, label=below:$1$] (1rres) [right=of cdotsrres]{};

         \draw[-] (1rres)--(cdotsrres)--(1mr)--(1m)--(1ml)--(cdotslres)--(1lres) (1m)--(1tl)--(cdotsrest)--(1tr)--(1m);

          \draw [decorate, decoration = {brace, raise=10pt, 
         amplitude=5pt}] (1tl) --  (1tr) node[pos=0.5,above=15pt,black]{$N-L-R$};
         \draw [decorate, decoration = {brace, raise=10pt, 
         amplitude=5pt}] (1ml) --  (1lres) node[pos=0.5,below=15pt,black]{$L-1$};
         \draw [decorate, decoration = {brace, raise=10pt, 
         amplitude=5pt}] (1rres) --  (1mr) node[pos=0.5,below=15pt,black]{$R-1$};
        
    \end{tikzpicture}
    \caption{$\urm(1)$ \hyperref[fig:Cyclic]{cyclic polymerisation} of the finite $A_{N+2}$ quiver to produce \Quiver{fig:FiniteAU1}.}
    \label{fig:FiniteAU1}
\end{figure}

The $\urm(1)$ \hyperref[fig:Cyclic]{cyclic polymerisation} of the finite A-type quiver, discussed in Section \ref{sec:FiniteAU12}, is not the only way to perform $\urm(1)$ \hyperref[fig:Cyclic]{cyclic polymerisation}. The most general $\urm(1)$ \hyperref[fig:Cyclic]{cyclic polymerisation} of the finite $A_{N}$ quiver involves taking the $L^{\textrm{th}}\;\urm(1)$ gauge node from the left and the $R^{\textrm{th}}\;\urm(1)$ gauge node from the right and superimposing them. This shown in \Figref{fig:FiniteAU1}, and produces \Quiver{fig:FiniteAU1}, which depends on the choice of $N$, $L$, and $R$. The quiver \Quiver{fig:FiniteAU1} is unframed, and if the $\urm(1)$ is ungauged at the superimposed nodes, the Coulomb branch is easily read as $\overline{min. A_{N-L-R}}\times \mathbb H^{L+R-2}$. The conclusion is that:
\begin{equation}
    \mathbb H^{N-1}///\urm(1)=\overline{min. A_{N-L-R}}\times \mathbb H^{L+R-2},\quad 1\leq L,R\leq \lceil N/2\rceil.\label{eq:FiniteA1U1}
\end{equation}

For certain choices of $L$ and $R$ the framings that can be applied to \Quiver{fig:FiniteAU1} give equivalent but different interpretations of the final moduli space.

For $L=R=1$, \Quiver{fig:FiniteAU1} takes the form of the affine $A^{(1)}_{N-2}$ Dynkin diagram and the Coulomb branch is $\overline{min. A_{N-2}}$. This case is encompassed by \eqref{eq:FiniteA1U1}. For the case, $L=1$ and $R=2$ (or vice versa, but taking the case as given), \Quiver{fig:FiniteAU1} takes the form of an affine $A^{(1)}_{N-2}$ Dynkin diagram with an extension of a $\urm(1)$ gauge node. This is a known CFHM quiver \cite{Cremonesi:2014xha} for the moduli space of 1 $\surm(N-1)$ instanton on $\mathbb C^2$ and was studied in Section \ref{sec:FiniteAU12}. For $L=1$ and $R\geq 3$ (or vice versa), framing \Quiver{fig:FiniteAU1} at the superimposed $\urm(1)$, or adjacent to the superimposed $\urm(1)$, gives two interpretations of the same moduli space. These are:
\begin{equation}
    \mathbb H^{N-1}///\urm(1)=\overline{min. A_{N-1-R}}\times \mathbb H^{R-1}=\mathcal M_{1,\surm(N-R)}^{\mathbb C^2}\times \mathbb H^{R-2},\quad 2\leq R\leq N-2.
\end{equation}

The Higgs branch of the finite $A_N$ quiver is trivial and of dimension 0. 
In all cases, the Higgs branch of the polymerised quiver is the $A_{N-L-R}$ singularity, which is of dimension 1. The relationship on the Higgs branch is:
\begin{equation}
    A_{N-L-R}///\urm(1)=\{0\},
\end{equation}
which can be verified with Weyl integration for all valid choices of $L$ and $R$.

\subsection{$\urm(2)$ cyclic polymerisation of $D_{N+4}^{(1)}$ for $N\geq1$}
\label{subsec:DOrb}
\begin{figure}[h!]
    \centering
    \begin{tikzpicture}
    \node[gauge, label=below:$1$] (1l) []{};
    \node[gauge, label=below:$2$] (2l) [right=of 1l]{};
    \node[] (cdots) [right=of 2l]{$\cdots$};
    \node[gauge, label=below:$2$] (2r) [right=of cdots]{};
    \node[gauge, label=below:$1$] (1r) [right=of 2r]{};
    \node[gauge,label=left:$1$] (1lt) [above=of 2l]{};
    \node[gauge, label=right:$1$] (1rt) [above=of 2r]{};

    \draw[-] (1l)--(2l)--(cdots)--(2r)--(1r);
    \draw[-] (1lt)--(2l);
    \draw[-] (1rt)--(2r);

    \draw [decorate, 
    decoration = {brace,
        raise=15pt,
        amplitude=5pt}] (2r) --  (2l) node[pos=0.5,below=20pt,black]{$N+1$};

    \node[] (space) [below=of cdots]{};

    \node[gauge, label=below:$2$] (2tres) [below=of space]{};
    \node[gauge, label=left:$1$] (1ltres) [above left=of 2tres]{};
    \node[gauge, label=right:$1$] (1rtres) [above right=of 2tres]{};
    \node[gauge, label=below:$2$] (2rtres) [below right=of 2tres]{};
    \node[gauge, label=below:$2$] (2ltres) [below left=of 2tres]{};
    \node[] (cdotsres) at ($(2rtres)!0.5!(2ltres)$) {$\cdots$};

    \draw[-] (1ltres)--(2tres)--(2rtres)--(cdotsres)--(2ltres)--(2tres)--(1rtres);

    \draw [decorate, 
    decoration = {brace,
        raise=15pt,
        amplitude=5pt}] (2rtres) --  (2ltres) node[pos=0.5,below=20pt,black]{$N-1$};

    \node[] (topghost) [left=of 1l]{};
    \node[] (bottomghost) [left=of 2tres]{};
    \draw[->] (topghost) to [out=-135,in=-135,looseness=1] (bottomghost);
     
    \end{tikzpicture}
    \caption{$\urm(2)$ \hyperref[fig:Cyclic]{cyclic polymerisation} of the unframed magnetic quiver for $\overline{min. D_{N+4}}$ for $N\geq1$ to produce quiver \Quiver{fig:minDkU2}.}
    \label{fig:minDkU2}
\end{figure}

The $\urm(2)$ \hyperref[fig:Cyclic]{cyclic polymerisation} of the unframed magnetic quiver for $\overline{min. D_{N+4}}$ for $N\geq 1$ is shown in \Figref{fig:minDkU2}. This produces quiver \Quiver{fig:minDkU2}, which is a loop of $N$ gauge nodes of $\urm(2)$, with one of these having a bouquet of two $\urm(1)$ gauge nodes attached. From consideration of the balance of the gauge nodes, the Coulomb branch of \Quiver{fig:minDkU2} appears to have an $\surm(N)\times \sorm(4)$ global symmetry, with the $\surm(N)$ contribution coming from the $N-1$ gauge nodes of $\urm(2)$ in the loop and the $\sorm(4)$ contribution coming from the bouquet of two $\urm(1)$. In fact, as can be confirmed from the Coulomb branch Hilbert series, tabulated for $N=1,2,3$ in Table \ref{tab:minDkU2HS}, there is an enhancement of the global symmetry to $\surm(N)\times \sorm(5)$, with the $\sorm(4)$ expected from the bouquet being enhanced to $\sorm(5)$.
The conclusion is that,
\begin{equation}
    \overline{min. D_{N+4}}///\urm(2)=\mathcal C\left(\text{\Quiver{fig:minDkU2}}\right),
    \quad N\geq1.
\label{eq:minDkU2}
\end{equation}
The result \eqref{eq:minDkU2} is confirmed by Weyl integration on a case-by-case basis up to $N=3$.

Although the quiver \Quiver{fig:minDkU2} is not a Kronheimer-Nakajima quiver, it is related. Applying a trivial $\urm(1)$ \hyperref[fig:Cyclic]{cyclic polymerisation} to the bouquet of $\urm(1)$ gauge nodes in \Quiver{fig:minDkU2} produces \Quiver{fig:nminAkU2}.
This $\urm(1)$ cyclic polymerisation 
commutes diagrammatically with the $\urm(2)$ cyclic polyermerisation in \Figref{fig:minDkU2}. Consequently, the Coulomb branch relationship in \eqref{eq:nminAkU2} follows from \eqref{eq:minDkU2} by application of a $\urm(1)$ hyper-Kähler quotient to both sides.

The $\urm(2)$ cyclic polymerisation in \Figref{fig:minDkU2} also gives the expected increase in Higgs branch dimension. The Higgs branch of the affine $D^{(1)}_{N+4}$ quiver is the Kleinian $D_{N+4}$ singularity of dimension 1. The Higgs branch of \Quiver{fig:minDkU2} has dimension 3. The Higgs branch Hilbert series of \Quiver{fig:minDkU2} for $N=1,2,3$ are tabulated in Table \ref{tab:minDkU2HSHiggs}.

In general, the Coulomb and Higgs branches of \Quiver{fig:minDkU2} do not have any particular names as moduli spaces. The exception is for the case of $N=1$, where the Coulomb branch is $\overline{n.min. B_2}$ and the Higgs branch is $\mathbb{H}\times \mathcal S^{C_2}_{\mathcal N,(2^2,1)}$.

\begin{table}[h!]
    \centering
    \begin{tabular}{cc}
    \toprule
        $N$ & $\hs[\mathcal C(\mathcal Q_{\ref{fig:minDkU2}})]$\\\midrule
        
         1 & $\frac{1 + 4 t^2 + 4 t^4 + t^6}{(1 - t^2)^6}$ \\ 
         2 & $\frac{1 + 8 t^2 + 40 t^4 + 107 t^6 + 199 t^8 + 234 t^{10} +
 \pal
 + t^{20}}
 {(1 - t^2)^5 (1 - t^4)^5}$\\
         3 & $\frac{1 + 11 t^2 + 93 t^4 + 422 t^6 + 
   1393 t^8 + 3049 t^{10} + 4981 t^{12} + 5732 t^{14} + \pal + t^{28}}{(1 - t^2)^7 (1 - t^4)^7}$  \\\bottomrule
    \end{tabular}
    \caption{Unrefined Hilbert series for the Coulomb branch of \Quiver{fig:minDkU2} for $N=1,2,3$.}
    \label{tab:minDkU2HS}
\end{table}

\begin{table}[h!]
    \centering
    \begin{tabular}{cc}
    \toprule
        $N$ & $\hs[\mathcal H(\mathcal Q_{\ref{fig:minDkU2}})]$\\\midrule
        
         1 & $\frac{1 - t^8}{(1 - t)^2 (1 - t^2)^3 (1 - t^3)^2}$\\ 
         2 & $\frac{1 + t^2 + 6 t^4 + 5 t^6 + 5 t^8 + 
 6 t^{10} + t^{12} + t^{14}}
 {(1 - t^2)^2 (1 - t^4)^4}$\\
         3 & $\frac{1 + t^3 + 2 t^4 + 2 t^5 + 3 t^6 + 4 t^7 + 3 t^8 + 3 t^9 +
    \pal
         + t^{19}}
 {(1 - t^2)  (1 - t^4) (1 - t^3) (1 - 
   t^5)^2 (1 - t^6)}$  \\\bottomrule
    \end{tabular}
    \caption{Unrefined Hilbert series for the Higgs branch of \Quiver{fig:minDkU2} for $N=1,2,3$.}
    \label{tab:minDkU2HSHiggs}
\end{table}




        

\subsection{$\urm(2)$ cyclic polymerisation of $E_6$}
\label{sec:FiniteE6U2}
\begin{figure}[h!]
    \centering
    \begin{tikzpicture}
    \node[gauge, label=below:$1$] (1l) []{};
    \node[gauge, label=below:$2$] (2l) [right=of 1l]{};
    \node[gauge, label=below:$3$] (3) [right=of 2l]{};
    \node[gauge, label=below:$2$] (2r) [right=of 3]{};
    \node[gauge, label=below:$1$] (1r) [right=of 2r]{};
    \node[gauge, label=right:$2$] (2t) [above=of 3]{};

    \draw[-] (1l)--(2l)--(3)--(2r)--(1r);
    \draw[-] (2t)--(3);

    \node[gauge, label=right:$2$] (2tr) [below=of 3]{};
    \node[gauge, label=below:$3$] (3r) [below=of 2tr]{};
    \node[gauge, label=below:$2$] (2rr) [right=of 3r]{};

    \draw[-] (2tr)--(3r);
    \draw[double,double distance=3pt,line width=0.4pt] (3r)--(2rr);

     \node[] (topghost) [left=of 1l]{};
    \node[] (bottomghost) [left=of 3r]{};
    \draw [->] (topghost) to [out=-150,in=150,looseness=1] (bottomghost);
    
    \end{tikzpicture}
    \caption{$\urm(2)$ \hyperref[fig:Cyclic]{cyclic polymerisation} of the finite $E_6$ quiver to produce \Quiver{fig:FiniteE6U2}.}
    \label{fig:FiniteE6U2}
\end{figure}
When treated as a magnetic quiver, the finite $E_6$ quiver has a Coulomb branch which is $\mathbb H^{10}$. Performing $\urm(2)$ \hyperref[fig:Cyclic]{cyclic polymerisation} on it forms a loop as shown in \Figref{fig:FiniteE6U2}. The global symmetry of the Coulomb branch of the finite $E_6$ quiver is $\surm(6)$, whereas the global symmetry of the Coulomb branch of the unframed \Quiver{fig:FiniteE6U2} is $\surm(2)\times \urm(1)$, as confirmed by the refined Hilbert series. For brevity only the unrefined Hilbert series is presented below:
\begin{equation}
    \hs\left[\mathcal C\left(\text{\Quiver{fig:FiniteE6U2}}\right)\right]=\frac{1 + t + 3 t^2 + 6 t^3 + 8 t^4 + 6 t^5 + 8 t^6 + 6 t^7 + 
 3 t^8 + t^9 + t^{10}}{(1 - t)^5 (1 - t^2)^4 (1 - t^3)^3}.
\end{equation}
From the dualisation algorithm \cite{Gaiotto:2008ak}, the Coulomb branch of \Quiver{fig:FiniteE6U2} is expected to be $\mathcal M^{\mathbb C^2}_{2,\surm(2)}\times \mathbb H^2$, consistent with the Hilbert series above.
The conclusion on the Coulomb branch is that:
\begin{equation}
    \mathbb H^{10}///\urm(2)=\mathcal M_{2,\surm(2)}^{\mathbb C^2}\times \mathbb H^2.\label{eq:FiniteE6U2}
\end{equation}
The result \eqref{eq:FiniteE6U2} is related to the result \eqref{eq:FiniteDkU2} for the case of $N=2$, but with a factor of $\mathbb H^2$ "multiplying" both sides. Note that in the left hand side of \eqref{eq:FiniteE6U2} the multiplication of the $\mathbb H^2$ free field commutes with the $\urm(2)$ hyper-Kähler quotient.

The Higgs branch relationship in this example is the same as in Section \ref{sec:FiniteDU2} since the Higgs branches of these quivers are the same.

\FloatBarrier
\subsection{$\urm(1)$ cyclic polymerisation of $E_6^{(1)}$}
\label{subsec:EOrbU1}
\begin{figure}[h!]
    \centering
    \begin{tikzpicture}
    \node[gauge, label=below:$1$] (1l) []{};
    \node[gauge, label=below:$2$] (2l) [right=of 1l]{};
    \node[gauge, label=below:$3$] (3) [right=of 2l]{};
    \node[gauge, label=below:$2$] (2r) [right=of 3]{};
    \node[gauge, label=below:$1$] (1r) [right=of 2r]{};
    \node[gauge, label=right:$2$] (2t) [above=of 3]{};
    \node[gauge, label=right:$1$] (1t) [above=of 2t]{};

    \draw[-] (1l)--(2l)--(3)--(2r)--(1r);
    \draw[-] (3)--(2t)--(1t);

    \node[] (space) [below =of 1r]{};

    \node[gauge, label=below:$1$] (1res) [below =of space]{};
    \node[gauge, label=below:$2$] (2res) [left=of 1res]{};
    \node[gauge, label=below:$3$] (3res) [left=of 2res]{};
    \node[gauge, label=below:$2$] (2lres) [left=of 3res]{};
    \node[gauge, label=right:$2$] (2tres) [above=of 3res]{};
    \node[gauge, label=left:$1$] (1tres) [left=of 2tres]{};

    \draw[-] (3res)--(2tres)--(1tres)--(2lres)--(3res)--(2res)--(1res);

    \node[] (topghost) [left=of 1l]{};
    \node[] (bottomghost) [left=of 2lres]{};

     \draw [->] (topghost) to [out=-150,in=150,looseness=1] (bottomghost);
    
    \end{tikzpicture}
    \caption{$\urm(1)$ \hyperref[fig:Cyclic]{cyclic polymerisation} of the affine $E^{(1)}_6$ quiver to produce \Quiver{fig:minE6U1}.}
    \label{fig:minE6U1}
\end{figure}

The $\urm(1)$ \hyperref[fig:Cyclic]{cyclic polymerisation} of the affine $E^{(1)}_6$ quiver, whose Coulomb branch is $\overline{min. E_6}$, produces quiver \Quiver{fig:minE6U1} as shown in \Figref{fig:minE6U1}. This is an unframed magnetic quiver for the height two nilpotent orbit closure $\overline{\mathcal O}^{D_5}_{(2^4,1^2)}$ \cite{Hanany:2016gbz}. The conclusion on the Coulomb branch is that:
\begin{equation}
    \overline{min. E_6}///\urm(1)=\overline{\mathcal O}^{D_5}_{(2^4,1^2)}. 
\label{eq:minE6U1}
\end{equation}
This is confirmed by Weyl integration using the fugacity map specified in \Figref{fig:CyclicEmbed}.

The Higgs branch of the affine $E^{(1)}_6$ quiver is the Kleinian singularity $E_6$, which is of dimension 1. The Higgs branch Hilbert series of \Quiver{fig:minE6U1} is:
\begin{equation}
    \hs[\mathcal H(\text{\Quiver{fig:minE6U1}})]=\pe\left[t^2+(q+q^{-1})t^4+t^6+(q+q^{-1})t^8-t^{12}-t^{16}\right],
    \label{eq:E6u1}
\end{equation}
where $q$ is a fugacity for $\urm(1)$. The Hilbert series above identifies the moduli space as the Slodowy slice $\mathcal S_{\mathcal N,(5^2)}^{D_5}$, which is a complete intersection of dimension 2. The relationship between orbits and slices of $D_5$ is that of Lustig-Spaltenstein duality, since the $D$ partitions $(5^2)$ and $(2^4,1^2)$ are dual to each other under the Barbasch-Vogan map.

The relationship on the Higgs branch of \Quiver{fig:minE6U1} is that:
\begin{equation}
    \mathcal S_{\mathcal N,(5^2)}^{D_5}///\urm(1)=E_6.
\end{equation}
This is verified from Weyl integration and is also identifiable from the Hilbert series \ref{eq:E6u1}.
\FloatBarrier

\subsection{$\urm(2)$ cyclic polymerisation of $E_6^{(1)}$}
\label{subsec:EOrb1}
\begin{figure}[h!]
    \centering
    \begin{tikzpicture}
    \node[gauge, label=below:$1$] (1l) []{};
    \node[gauge, label=below:$2$] (2l) [right=of 1l]{};
    \node[gauge, label=below:$3$] (3) [right=of 2l]{};
    \node[gauge, label=below:$2$] (2r) [right=of 3]{};
    \node[gauge, label=below:$1$] (1r) [right=of 2r]{};
    \node[gauge, label=right:$2$] (2t) [above=of 3]{};
    \node[gauge, label=right:$1$] (1t) [above=of 2t]{};

    \draw[-] (1l)--(2l)--(3)--(2r)--(1r);
    \draw[-] (3)--(2t)--(1t);

    \node[] (space) [below =of 1r]{};

    \node[gauge, label=below:$1$] (1res) [below =of space]{};
    \node[gauge, label=below:$2$] (2res) [left=of 1res]{};
    \node[gauge, label=below:$3$] (3res) [left=of 2res]{};
    \node[gauge, label=below:$2$] (2lres) [left=of 3res]{};

    \draw[-] (1res)--(2res)--(3res);
    \draw[double,double distance=3pt,line width=0.4pt] (3res)--(2lres);

    \node[] (topghost) [left=of 1l]{};
    \node[] (bottomghost) [left=of 2lres]{};

     \draw [->] (topghost) to [out=-150,in=150,looseness=1] (bottomghost);
    
    \end{tikzpicture}
    \caption{$\urm(2)$ \hyperref[fig:Cyclic]{cyclic polymerisation} of the affine $E^{(1)}_6$ quiver to produce \Quiver{fig:minE6U2}.}
    \label{fig:minE6U2}
\end{figure}

The $\urm(2)$ \hyperref[fig:Cyclic]{cyclic polymerisation} of the affine $E^{(1)}_6$ quiver, whose Coulomb branch is $\overline{min. E_6}$, produces quiver \Quiver{fig:minE6U2}, as shown in \Figref{fig:minE6U2}. From the consideration of the balance of the gauge nodes, the Coulomb branch global symmetry appears to be $\surm(4)$. However, there is an enhancement of the global symmetry to $\sorm(7)$ \cite{Gledhill:2021cbe}.

Applying the monopole formula to \Quiver{fig:minE6U2} and projecting the resulting Coulomb branch Hilbert series onto highest weight fugacities of $\sorm(7)$, yields the following HWG:
\begin{equation}    
    \hwg\left[\mathcal C\left(\text{\Quiver{fig:minE6U2}}\right)\right]=PE\left[\mu_2 t^2 + (\mu_1^2 + \mu_3^2) t^4 + \mu_1 \mu_3^2 t^6 + \mu_2^2 t^8 + 
 \mu_1 \mu_2 \mu_3^2 t^{10} - \mu_1^2 \mu_2^2 \mu_3^4 t^{20}\right].
\label{eq:HWGminE6U2}
\end{equation}
This is the HWG for the 7-dimensional orbit of $B_3$, $\overline{\mathcal O}_{(3^2,1)}^{B_3}$ \cite{Hanany:2016gbz}.

Since the embedding given in \Figref{fig:CyclicEmbed} realises an $\surm(4)$ Coulomb branch global symmetry instead of the enhanced $\sorm(7)$ global symmetry, a more direct embedding is used to verify  \eqref{eq:HWGminE6U2} from Weyl integration. This embedding is $E_6\hookleftarrow \sorm(10)\times \urm(1)\hookleftarrow \sorm(7)\times \surm(2)\times \urm(1)$ which decomposes the adjoint as \begin{equation}
    \mu_6\rightarrow \mu_2 + \nu^2 + \mu_1 \nu^2 + \mu_3 (\nu/q + \nu q)+1,
\end{equation}where the $\mu_6$ on the left hand side is for $E_6$, the $\mu_{1,2,3}$ on the right hand side are for $\sorm(7)$, the $\nu$ is for $\surm(2)$ and the $\urm(1)$ fugacity is $q$. There is agreement with \eqref{eq:HWGminE6U2} if the Weyl integration is done w.r.t the $\surm(2)\times \urm(1)$.

The conclusion on the Coulomb branch is that,
\begin{equation}
    \overline{min. E_6}///\urm(2)=\overline{\mathcal O}_{(3^2,1)}^{B_3}.
\label{eq:minE6U2}
\end{equation}
It is worth noting that this orbit is of height 4 and that \Quiver{fig:minE6U2} is a magnetic quiver for it that was previously unknown.

This quiver \Quiver{fig:minE6U2} is another example of a magnetic quiver whose nodes do not form a balanced Dynkin diagram, but which nevertheless has a global symmetry enhanced to that of a simple group and a Coulomb branch which forms an orbit.


Hasse diagrams can provide an additional way to identify moduli spaces. The Hasse diagram for the Coulomb branch of \Quiver{fig:minE6U2} is shown in \Figref{fig:OrbBHasse}. This was computed using the method of quiver subtraction \cite{Cabrera:2018ann,Grimminger:2020dmg} and decorations \cite{Bourget:2022ehw,Bourget:2022tmw}. The final slice to the trivial leaf is $b_3$ consistent with the $\sorm(7)$ global symmetry. Indeed, the Hasse diagram shown in \Figref{fig:OrbBHasse} is the same as the Hasse diagram for $\overline{\mathcal O}_{(3^2,1)}^{B_3}$ \cite{Kraft1982OnGroups}, as expected.

The Higgs branch of the affine $E^{(1)}_6$ quiver is the Kleinian singularity $E_6$, which is of dimension 1. The Hilbert series of the Higgs branch of \Quiver{fig:minE6U2} is computed as:
\begin{equation}
    \hs[\mathcal H(\text{\Quiver{fig:minE6U2}})]=\pe\left[[2]_{\surm(2)}t^2+[4]_{\surm(2)}t^4-t^8-t^{12}\right].
\end{equation}
This is a complete intersection of dimension 3 and is the Hilbert series for the Slodowy slice $\mathcal H(\text{\Quiver{fig:minE6U2}})=\mathcal S^{C_3}_{\mathcal N,(2^3)}$, as tabulated in \cite{Cabrera:2018ldc}. The result is another example of Lustig-Spaltenstein duality, since the $C_3$ partition $(2^3)$ is dual to the $B_3$ partition $(3^2,1)$ under the Barbasch-Vogan map. There is the expected increase in Higgs branch dimension by $\textrm{rank}\;\urm(2)=2$.



\begin{figure}
    \centering
             \begin{tikzpicture}
        \node (a) at (0,0) {
    $\begin{tikzpicture}
        \node[gauge,label=below:$1$] (1) at (0,0) {};
        \node[gauge,label=below:$2$] (2) at (1,0) {};
        \node[gauge,label=below:$3$] (3) at (2,0){};
        \node[gauge,label=below:$2$] (2r) at (3,0){};

        \draw[-] (1)--(2)--(3);
        \draw[transform canvas={yshift=1.3pt}] (3)--(2r);
        \draw[transform canvas={yshift=-1.3pt}] (3)--(2r);
        \end{tikzpicture}$};
        \node (b) at (0,-3.5) {$\begin{tikzpicture}
        \node[gauge,label=below:$1$] (1) at (0,0) {};
        \node[gauge, label=below:$2$] (2) at (1,0){};
        \node[gauge, label=below:$2$] (2r) at (2,0){};
        \node[gauge, label=below:$1$] (1r) at (3,0){};
        \node[gauge, label=left:$1$] (1t) at (1,1){};

        \draw[-] (1)--(2)--(2r) (2)--(1t);
        \draw[transform canvas={yshift=1.3pt}] (1r)--(2r);
        \draw[transform canvas={yshift=-1.3pt}] (1r)--(2r);

        \draw[purple] (1t) circle (0.6cm);
        \draw[purple] \convexpath{1r,2r}{0.3cm};
    \end{tikzpicture}$};
    \node (c) at (0,-7) {
    $\begin{tikzpicture}
        \node[gauge, label=below:$1$] (1) at (0,0){};
        \node[gauge, label=below:$2$] (2) at (1,0){};
        \node[gauge, label=below:$1$] (1br) at (2,0){};
        \node[gauge, label=left:$1$] (1l) at (1-0.707,0.707){};
        \node[gauge, label=right:$1$] (1r) at (1+0.707, 0.707){};

        \draw[-] (1)--(2)--(1br) (1l)--(2)--(1r);
        \draw[purple] (1l) circle (0.6cm);
        \draw[purple] (1r) circle (0.6cm);
    \end{tikzpicture}$};
    \node (d) at (0,-10.5) {
    $\begin{tikzpicture}
        \node[gauge, label=below:$1$] (1) at (0,0){};
        \node[gauge, label=below:$2$] (2) at (1,0){};
        \node[gauge, label=below:$1$] (1r) at (2,0){};
        \node[gauge, label=above:$1$] (1t) at (1,1){};

        \draw[-] (1)--(2)--(2r);
         \draw[transform canvas={xshift=1.3pt}] (2)--(1t);
        \draw[transform canvas={xshift=-1.3pt}] (2)--(1t);
        \draw (1-0.2,0.5-0.1)--(1,0.5+0.1)--(1+0.2,0.5-0.1);
        
    \end{tikzpicture}$
    };
    \node (e) at (0,-13.5) {$\emptyset$};

    \draw[] (a)--(b) node[pos=0.5, label=right:$A_1$] {};

     \draw[transform canvas ={xshift={1.3pt}}](b)--(c) node[pos=0.5, label=right:$2A_1$]{};
     \draw[transform canvas ={xshift={-1.3pt}}](b)--(c) {};
    
    \draw[](c) --(d) node[pos=0.5, label=right:$A_1$] {}--(e) node[pos=0.5, label=right:$b_3$] {};
    \end{tikzpicture}
    \caption{Hasse diagram for the Coulomb branch of \Quiver{fig:minE6U2} which is $\overline{\mathcal O}_{(3^2,1)}^{B_3}$. Decorations are given by the \textcolor{purple}{purple} circles. The slice $2A_1$ means $A_1\cup A_1$.}
    \label{fig:OrbBHasse}
\end{figure}

\FloatBarrier
\subsection{$\urm(1)$ cyclic polymerisation of $E_7^{(1)}$}
\label{subsec:EOrbU2}
\begin{figure}[h!]
    \centering
    \begin{tikzpicture}
    \node[gauge, label=below:$1$] (1l) []{};
    \node[gauge, label=below:$2$] (2l) [right=of 1l]{};
    \node[gauge, label=below:$3$] (3l) [right=of 2l]{};
    \node[gauge, label=below:$4$] (4) [right=of 3l]{};
    \node[gauge, label=below:$3$] (3r) [right=of 4]{};
    \node[gauge, label=below:$2$] (2r) [right=of 3r]{};
    \node[gauge, label=below:$1$] (1r) [right=of 2r]{};
    \node[gauge, label=right:$2$] (2t) [above=of 4]{};

    \draw[-] (1l)--(2l)--(3l)--(4)--(3r)--(2r)--(1r);
    \draw[-] (4)--(2t);

    \node[gauge,label=right:$2$] (2rest) [below=of 4]{};
    \node[gauge, label=below:$4$] (4res) [below =of 2rest]{};
    \node[gauge, label=below:$3$] (3resl) [left=of 4res]{};
    \node[gauge, label=below:$2$] (2resl) [left=of 3resl]{};
    \node[gauge, label=below:$3$] (3resr) [right=of 4res]{};
    \node[gauge, label=below:$2$] (2resr) [right=of 3resr]{};
    \node[gauge, label=below:$1$] (1res) [below=of 4res]{};

    \draw[-] (2rest)--(4res)--(3resl)--(2resl)--(1res)--(2resr)--(3resr)--(4res);

     \node[] (topghost) [left=of 1l]{};
    \node[] (bottomghost) [left=of 2resl]{};

     \draw [->] (topghost) to [out=-150,in=150,looseness=1] (bottomghost);   
    \end{tikzpicture}
    \caption{$\urm(1)$ \hyperref[fig:Cyclic]{cyclic polymerisation} of the $E_7^{(1)}$ quiver to produce \Quiver{fig:minE7U1}.}
    \label{fig:minE7U1}
\end{figure}

The $\urm(1)$ \hyperref[fig:Cyclic]{cyclic polymerisation} of the $E^{(1)}_7$ affine quiver, whose Coulomb branch is $\overline{min. E_7}$, produces quiver \Quiver{fig:minE7U1}, as shown in \Figref{fig:minE7U1}. This is a magnetic quiver for the height two nilpotent orbit closure $\overline{n.min. E_6}$.

The conclusion on the Coulomb branch is that,
\begin{equation}
    \overline{min. E_7}///\urm(1)=\overline{n.min. E_6}.
\label{eq:minE7U1}
\end{equation}

The Higgs branch of the $E^{(1)}_7$ quiver is the $E_7$ singularity which is of dimension 1. The Higgs branch Hilbert series of \Quiver{fig:minE7U1} is:\begin{equation}
    \hs[\mathcal H(\text{\Quiver{fig:minE7U2}})]=\pe\left[t^2+(q+q^{-1}) t^6+t^8+(q+q^{-1}) t^{12}-t^{18}-t^{24}\right],
\end{equation}where $q$ is a $\urm(1)$ charge. The above Hilbert series identifies the moduli space as the Slodowy slice $\mathcal S^{E_6}_{\mathcal N,D_5}$ which is a complete intersection of dimension 2. This is as expected from Lustig-Spaltenstein duality since this $E_6$ orbit and slice are dual to each other \cite{2023arXiv231000521J}. The relationship on the Higgs branch is that:
\begin{equation}
    \mathcal S^{E_6}_{\mathcal N,D_5}///\urm(1)=E_7.
\end{equation}
This is verified from Weyl integration and is also identifiable from the above Hilbert series.
\FloatBarrier
\FloatBarrier
\subsection{$\urm(2)$ cyclic polymerisation of $E_7^{(1)}$}
\label{subsec:EOrb2}
\begin{figure}[h!]
    \centering
    \begin{tikzpicture}
    \node[gauge, label=below:$1$] (1l) []{};
    \node[gauge, label=below:$2$] (2l) [right=of 1l]{};
    \node[gauge, label=below:$3$] (3l) [right=of 2l]{};
    \node[gauge, label=below:$4$] (4) [right=of 3l]{};
    \node[gauge, label=below:$3$] (3r) [right=of 4]{};
    \node[gauge, label=below:$2$] (2r) [right=of 3r]{};
    \node[gauge, label=below:$1$] (1r) [right=of 2r]{};
    \node[gauge, label=right:$2$] (2t) [above=of 4]{};

    \draw[-] (1l)--(2l)--(3l)--(4)--(3r)--(2r)--(1r);
    \draw[-] (4)--(2t);

    \node[gauge,label=right:$2$] (2rest) [below=of 4]{};
    \node[gauge, label=below:$4$] (4res) [below =of 2rest]{};
    \node[gauge, label=left:$3$] (3resl) [left=of 4res]{};
    \node[gauge, label=below:$2$] (2resb) [below=of 4res]{};
    \node[gauge, label=right:$3$] (3resr) [right=of 4res]{};

    \draw[-] (2rest)--(4res)--(3resl)--(2resb)--(3resr)--(4res);

     \node[] (topghost) [left=of 1l]{};
    \node[] (bottomghost) [left=of 3resl]{};

     \draw [->] (topghost) to [out=-150,in=150,looseness=1] (bottomghost);

    \end{tikzpicture}
    \caption{$\urm(2)$ \hyperref[fig:Cyclic]{cyclic polymerisation} of the $E_7^{(1)}$ quiver to produce \Quiver{fig:minE7U2}.}
    \label{fig:minE7U2}
\end{figure}
The action of $\urm(2)$ \hyperref[fig:Cyclic]{cyclic polymerisation} on the $E^{(1)}_7$ affine quiver, whose Coulomb branch is $\overline{min. E_7}$, produces quiver \Quiver{fig:minE7U2} as shown in \Figref{fig:minE7U2}. From consideration of the balance of gauge nodes, it would appear that the Coulomb branch global symmetry of \Quiver{fig:minE7U2} is $\sorm(8)$. However there is an enhancement to $\sorm(9)$ \cite{Gledhill:2021cbe}, which can be seen in the Hilbert series. For brevity the unrefined HS is:
\begin{equation}
    \hs\left[\mathcal C\left(\text{\Quiver{fig:minE7U2}}\right)\right]=\frac{\left(\begin{aligned}1 &+ 23 t^2 + 347 t^4 + 3531 t^6 + 
   26458 t^8 + 149457 t^{10} + 655957 t^{12} \\&+ 2278292 t^{14} + 
   6376015 t^{16} + 14556113 t^{18}+ 27400071 t^{20} + 42825385 t^{22}\\& + 
   55882015 t^{24} + 61039230 t^{26} + \pal+ t^{52}\end{aligned}\right)}{(1 - t^2)^{13} (1 - t^4)^{13}},
\end{equation}
and the coefficient of the $t^2$ term is consistent with the dimension of the $B_4$ adjoint representation.

The conclusion on the Coulomb branch is that:
\begin{equation}
    \overline{min. E_7}///\urm(2)=\mathcal C\left(\text{\Quiver{fig:minE7U2}}\right).
\label{eq:minE7U2}
\end{equation}
Once again, the embedding in \Figref{fig:CyclicEmbed} does not capture the enhancement of the Coulomb branch global symmetry. The above result is confirmed with Weyl integration using the following and more direct embedding of $E_7\hookleftarrow \sorm(12) \times \surm(2)\hookleftarrow \sorm(9) \times \surm(2) \times \surm(2) \hookleftarrow \sorm(9) \times \surm(2) \times \urm(1)$. This decomposes the fundamental of $E_7$ as,
\begin{equation}
    \mu_6\rightarrow \mu_4 \nu + \nu^2/q + \nu^2 q + \mu_1 (1/q + q),
\end{equation}
where on the right hand side the $\mu_i$ and $\nu$ are highest weight fugacities for $\sorm(9)$ and $\surm(2)$, respectively, and $q$ is a $\urm(1)$ charge. The Weyl integration is performed w.r.t $\surm(2)\times \urm(1)$.

The Higgs branch of the $E^{(1)}_7$ quiver is the $E_7$ singularity which is of dimension 1. The Higgs branch Hilbert series of \Quiver{fig:minE7U2} is computed as \begin{equation}
    \hs[\mathcal H(\text{\Quiver{fig:minE7U2}})]=\frac{1 + t^6 + 3 t^8 + t^{10} + t^{12} + t^{14} + t^{16} + 
 3 t^{18} + t^{20} + t^{26}}{(1 - t^2) (1 - t^4)^3  (1 - t^6) (1 - t^{12})}.
\end{equation}This Higgs branch is not a complete intersection but is of dimension 3, consistent with the expected increase in Higgs branch dimension by $\textrm{rank}\;\urm(2)=2$.

The Hasse diagram of the Coulomb branch of \Quiver{fig:minE7U2} is shown in \Figref{fig:E7U2Hasse}. The appearance of the slice $2a_3$ indicates that the Coulomb branch is not a slice in the nilcone of any semi-simple Lie algebra.

\begin{figure}
    \centering
             \begin{tikzpicture}
        \node (a) at (0,0) {
    $\begin{tikzpicture}
       \node[gauge,label=right:$2$] (2t) at (0,0) {};
        \node[gauge, label=below:$4$] (4) at (0,-1){};
        \node[gauge, label=left:$3$] (3l) at (-1,-1){};
        \node[gauge, label=below:$2$] (2b) at (0,-2){};
        \node[gauge, label=right:$3$] (3r) at (1,-1){};

    \draw[-] (2t)--(4)--(3l)--(2b)--(3r)--(4);
        \end{tikzpicture}$};
        \node (b) at (0,-5) {$\begin{tikzpicture}
        \node[gauge, label=right:$1$] (1)  at (0,0) {};
        \node[gauge, label=right:$2$] (2) at (0,-1){};
        \node[gauge, label=below:$3$] (3) at (0,-2){};
        \node[gauge, label=left:$2$] (2l) at (-1,-2){};
        \node[gauge, label=right:$2$] (2r) at (1,-2){};
        \node[gauge, label=below:$1$] (1b) at (0,-3){};

        \draw[-] (1)--(2)--(3)--(2l)--(1b)--(2r)--(3);

        \draw[purple] (1) circle (0.6cm);

        \draw[purple] \convexpath{2l,2r,1b}{0.3cm};
    \end{tikzpicture}$};
    \node (c) at (0,-10) {
    $\begin{tikzpicture}
        \node[] (ghost) at (0,0){};
        \node[gauge, label=right:$2$] (2) at (0,-1){};
        \node[gauge, label=left:$1$] (1tl) at (-0.707,-1+0.707){};
        \node[gauge, label=right:$1$] (1tr) at (0.707,-1+0.707){};
        \node[gauge, label=below:$2$] (2b) at (0,-2){};
        \node[gauge, label=left:$1$] (1l) at (-1,-2){};
        \node[gauge, label=right:$1$] (1r) at (1,-2){};

        \draw[-] (1tl)--(2)--(1tr) (2)--(2b)--(1l) (1r)--(2b);
        \draw[purple] (1tl) circle (0.6cm);
        \draw[purple] (1tr) circle (0.6cm);
    \end{tikzpicture}$};
    \node (d) at (0,-15) {
    $\begin{tikzpicture}
       \node[gauge, label=above:$1$] (1t) at (0,0) {};
       \node[gauge, label=right:$2$] (2) at (0,-1){};
       \node[gauge, label=below:$2$] (2b) at (0,-2){};
       \node[gauge, label=left:$1$] (1l) at (-1,-2){};
        \node[gauge, label=right:$1$] (1r) at (1,-2){};

        \draw[-] (1l)--(2b)--(1r) (2)--(2b);
        \draw[transform canvas={xshift=1.3pt}] (2)--(1t);
        \draw[transform canvas={xshift=-1.3pt}] (2)--(1t);
        \draw (-0.2,0.5-0.1-1)--(0,0.5+0.1-1)--(0.2,0.5-0.1-1);

    \end{tikzpicture}$
    };
    \node (e) at (0,-20) {$\emptyset$};

    \draw[] (a)--(b) node[pos=0.5, label=right:$a_3$] {};
    \draw[transform canvas ={xshift={1.3pt}}](b)--(c) node[pos=0.5, label=right:$2a_3$]{};
    \draw[transform canvas ={xshift={-1.3pt}}](b)--(c);
    \draw[] (c)--(d) node[pos=0.5, label=right:$A_1$] {}--(e) node[pos=0.5, label=right:$b_4$] {};
    \end{tikzpicture}
    \caption{Hasse diagram for the Coulomb branch of \Quiver{fig:minE7U2} which has ${B_4}$ global symmetry. Decorations are given by the \textcolor{purple}{purple} circles.}
    \label{fig:E7U2Hasse}
\end{figure}
\FloatBarrier
\subsection{$\urm(3)$ cyclic polymerisation of $E_7^{(1)}$}
\begin{figure}[h!]
    \centering
    \begin{tikzpicture}
    \node[gauge, label=below:$1$] (1l) []{};
    \node[gauge, label=below:$2$] (2l) [right=of 1l]{};
    \node[gauge, label=below:$3$] (3l) [right=of 2l]{};
    \node[gauge, label=below:$4$] (4) [right=of 3l]{};
    \node[gauge, label=below:$3$] (3r) [right=of 4]{};
    \node[gauge, label=below:$2$] (2r) [right=of 3r]{};
    \node[gauge, label=below:$1$] (1r) [right=of 2r]{};
    \node[gauge, label=right:$2$] (2t) [above=of 4]{};

    \draw[-] (1l)--(2l)--(3l)--(4)--(3r)--(2r)--(1r);
    \draw[-] (4)--(2t);

    \node[] (space) [below=of 3l]{};

    \node[gauge, label=below:$2$] (2res) [below =of space]{};
    \node[gauge, label=below:$4$] (4res) [right=of 2res]{};
    \node[gauge, label=below:$3$] (3res) [right=of 4res]{};

    \draw[-] (2res)--(4res);
    \draw[double,double distance=3pt,line width=0.4pt] (3res)--(4res);

    \node[] (topghost) [left=of 1l]{};
    \node[] (bottomghost) [left=of 2res]{};

     \draw [->] (topghost) to [out=-150,in=150,looseness=1] (bottomghost);    
    \end{tikzpicture}
    
    \caption{$\urm(3)$ \hyperref[fig:Cyclic]{cyclic polymerisation} of the $E^{(1)}_7$ quiver to produce \Quiver{fig:minE7U3}.}
    \label{fig:minE7U3}
\end{figure}

The $\urm(3)$ \hyperref[fig:Cyclic]{cyclic polymerisation} of the affine $E^{(1)}_7$ is shown in \Figref{fig:minE7U3}. This produces quiver \Quiver{fig:minE7U3}. From consideration of the balance of gauge nodes, it would appear that the Coulomb branch global symmetry is $\surm(3)$, however yet again there is an enhancement of the global symmetry, this time to $G_2$ \cite{Gledhill:2021cbe}. This can be verified from the Hilbert series. However only the unrefined Hilbert series is presented for brevity:
\begin{equation}
    \hs\left[\mathcal C\left(\text{\Quiver{fig:minE7U3}}\right)\right]=
    \frac{\left(\begin{aligned}1 &+ 6 t^2 + 40 t^4 + 149 t^6 + 470 t^8 + 1048 t^{10} + 1927 t^{12} + 
 2668 t^{14} + 3048 t^{16}\\& +\pal+ t^{32}\end{aligned}\right)}{(1 - t^2)^{8} (1 - t^4)^8},\label{eq:minE7U3HS}
\end{equation}
and the coefficient of the $t^2$ term matches the dimension of the $G_2$ adjoint representation.

This Coulomb branch can be identified as the Slodowy intersection $\mathcal S^{E_7}_{A_3+A_2+A_1,A_2+3A_1}$. To do this, either the Hilbert series of the Slodowy intersection can be computed using the formula in \cite[Appendix B]{Hanany:2019tji} to obtain agreement with \eqref{eq:minE7U3HS}. Alternatively, the Hasse diagram of the Coulomb branch of \Quiver{fig:minE7U3} is shown in \Figref{fig:E7U3Hasse} and this again matches the Hasse diagram of $\mathcal S^{E_7}_{A_3+A_2+A_1,A_2+3A_1}$ \cite{2015arXiv150205770F}.

The conclusion on the Coulomb branch is,
\begin{equation}
    \overline{min. E_7}///\urm(3)=\mathcal S^{E_7}_{A_3+A_2+A_1,A_2+3A_1}.
    \label{eq:minE7U3}
\end{equation}
Again, the embedding in \Figref{fig:CyclicEmbed} does not realise the $G_2$ symmetry explicitly so agreement with the above results using Weyl integration is done employing the direct embedding of $E_7\hookleftarrow E_6\times \urm(1)\hookleftarrow G_2\times \surm(3)\times \urm(1)$. This decomposes the fundamental of $E_7$ as,
\begin{equation}
    \mu_6\rightarrow \mu_2\nu_1q^{-1}+\mu_2\nu_2q+\nu_1^2q+\nu_2^2q^{-1}+q^3+q^{-3},
\end{equation}
where on the right hand side the $\mu_i$ and $\nu_i$ are highest weight fugacities for $G_2$ and $\surm(3)$ respectively, and $q$ is a $\urm(1)$ charge. The Weyl integration is performed w.r.t $\surm(3)\times \urm(1)$.

The Higgs branch of the $E^{(1)}_7$ quiver is the $E_7$ singularity, which is of dimension 1. The Higgs branch Hilbert series of \Quiver{fig:minE7U3} has been computed, but for brevity only the unrefined Hilbert series is presented:\begin{equation}
    \hs[\mathcal H(\text{\Quiver{fig:minE7U3}})]=\frac{(1-t^{12}) \left(\begin{aligned}1 &+ 3 t^2 + 6 t^4 + 13 t^6 + 24 t^8 + 29 t^{10} + 
     29 t^{12}\\
     & + \pal + t^{24}\end{aligned}\right)}
     {(1 - t^4)^5 (1 - t^6)^4}.
\end{equation}
The Higgs branch is of dimension 4 but is not a complete intersection. In fact, using Lustig-Spaltenstein duality the Higgs branch can be identified as the Slodowy intersection within the $E_7$ nilcone, $\mathcal H(\text{\Quiver{fig:minE7U3}})=\mathcal S^{E_7}_{A_6,A_4+A_2}$, which is dual to the intersection on the Coulomb branch \cite{2023arXiv231000521J}. This can be confirmed from the refined Hilbert series. The Higgs branch dimension increases by $\mathrm{rank}\;\urm(3)=3$ as expected.

\begin{figure}
    \centering
             \begin{tikzpicture}
        \node (a) at (0,0) {
    $\begin{tikzpicture}
       \node[gauge, label=below:$2$] (2) at (0,0){};
       \node[gauge, label=below:$4$] (4) at (1,0){};
       \node[gauge, label=below:$3$] (3) at (2,0){};

        \draw[-] (2)--(4);
        \draw[transform canvas={yshift=1.3pt}] (4)--(3);
        \draw[transform canvas={yshift=-1.3pt}] (4)--(3);
        
        \end{tikzpicture}$};
        \node (b) at (0,-3.5) {$\begin{tikzpicture}
        \node[gauge, label=below:$1$] (1)  at (0,0) {};
        \node[gauge, label=below:$2$] (2) at (1,0){};
        \node[gauge, label=below:$3$] (3) at (2,0){};
        \node[gauge, label=below:$2$] (2r) at (3,0){};
        
        \draw[-] (1)--(2)--(3);
        \draw[transform canvas={yshift=1.3pt}] (2r)--(3);
        \draw[transform canvas={yshift=-1.3pt}] (2r)--(3);
        \draw[purple] (1) circle (0.6cm);
        \draw[purple] \convexpath{3,2r}{0.3cm};
    \end{tikzpicture}$};
    \node (c) at (0,-7) {
    $\begin{tikzpicture}
        \node[gauge, label=below:$2$] (2) at (0,0){};
        \node[gauge, label=left:$1$] (1t) at (-0.707,0+0.707){};
        \node[gauge, label=left:$1$] (1b) at (-0.707,-0.707){};
        \node[gauge, label=below:$2$] (2r) at (1,0){};
        \node[gauge, label=below:$1$] (1r) at (2,0){};

        \draw[-] (1t)--(2)--(1b) (2)--(2r);
        \draw[transform canvas={yshift=1.3pt}] (2r)--(1r);
        \draw[transform canvas={yshift=-1.3pt}] (2r)--(1r);
        \draw[purple] (1t) circle (0.6cm);
        \draw[purple] (1b) circle (0.6cm);
        \draw[purple] \convexpath{2r,1r}{0.3cm};
    \end{tikzpicture}$};
    \node (d) at (-7,-10.5) {
    $\begin{tikzpicture}
       \node[gauge, label=below:$1$] (1l) at (0,0){};
       \node[gauge, label=below:$2$] (2l) at (1,0){};
       \node[gauge, label=below:$2$] (2r) at (2,0){};
       \node[gauge, label=below:$1$] (1r) at (3,0){};
       
        \draw[-] (2l)--(2r);
        \draw[transform canvas={yshift=1.3pt}] (2r)--(1r);
        \draw[transform canvas={yshift=-1.3pt}] (2r)--(1r);
         \draw[transform canvas={yshift=1.3pt}] (2l)--(1l);
        \draw[transform canvas={yshift=-1.3pt}] (2l)--(1l);
        \draw (0.5+0.1,-0.2)--(0.5-0.1,0)--(0.5+0.1,0.2);

        \draw[purple] (1l) circle (0.6cm);
        \draw[purple] \convexpath{2r,1r}{0.3cm};

    \end{tikzpicture}$
    };
    \node (e) at (0,-14) {
    $\begin{tikzpicture}
        \node[gauge, label=below:$2$] (2) at (0,0){};
        \node[gauge, label=below:$1$] (1r) at (1,0){};
        \node[gauge, label=left:$1$] (1b) at (-0.707, -0.707){};
        
        \draw[-](1r)--(2)--(1b);

         \begin{scope}[rotate around={+45:(0,0)}]
        \node[gauge,label=left:{$1$}] (1t) at (0,1) {};
        \draw[transform canvas={xshift=1.3pt,yshift=1.3pt}] (2)--(1t);
        \draw[transform canvas={xshift=-1.3pt,yshift=-1.3pt}] (2)--(1t);
        \draw (0-0.2,0.5-0.1)--(0,0.5+0.1)--(0+0.2,0.5-0.1);
        \end{scope}

        \draw[purple] (1t) circle (0.6cm);
        \draw[purple] (1b) circle (0.6cm);
    \end{tikzpicture}$
    };
    \node (f) at (0,-17.5){
    $\begin{tikzpicture}
        \node[gauge, label=below:$1$] (1l) at (0,0){};
        \node[gauge, label=below:$2$] (2) at (1,0){};
        \node[gauge, label=below:$1$] (1r) at (2,0){};

        \draw[-] (1r)--(2);
        \draw[transform canvas={yshift=1.5pt}] (1l)--(2);
        \draw (1l)--(2);
        \draw[transform canvas={yshift=-1.5pt}] (1l)--(2);
        \draw (0.5+0.1,-0.2)--(+0.5-0.1,0)--(0.5+0.1,0+0.2);
        \end{tikzpicture}$
    };

    \node (g) at (0,-21){$\emptyset$};

    \node (h) at (0,-10.5){
    $\begin{tikzpicture}
        \node[gauge, label=below:$2$] (2) at (0,0){};
        \node[gauge, label=below:$1$] (1r) at (1,0){};
        \node[gauge, label=left:$1$] (1t) at (-0.707, 0.707){};
        \node[gauge, label=left:$1$] (1l) at (-1,0){};
        \node[gauge, label=left:$1$] (1b) at (-0.707,-0.707){};

        \draw[-] (1l)--(2)--(1b) (1t)--(2)--(1r);

        \draw[purple] (1l) circle (0.6cm){};
        \draw[purple] (1t) circle (0.6cm){};
        \draw[purple] (1b) circle (0.6cm){};
    \end{tikzpicture}$
    };
    \draw[] (a)--(b) node[pos=0.5, label=right:$A_1$] {};
    
   \draw[transform canvas ={xshift={1.3pt}}](b)--(c) node[pos=0.5, label=right:$2A_1$]{};
    \draw[transform canvas ={xshift={-1.3pt}}](b)--(c);
    
    \draw[transform canvas ={xshift={1.3pt}}](c)--(h) node[pos=0.5, label=right:$3A_1$]{};
    \draw[transform canvas ={xshift={-1.3pt}}](c)--(h);
    \draw[](c)--(h)--(e) node[pos=0.5, label=right:$A_1$]{};

    \draw[](c)--(d) node[pos=0.5, label=above:$A_1$]{}--(e) node[pos=0.5, label=right:$A_1$] {}--(f) node[pos=0.5,label=right:$m$] {}--(g) node[pos=0.5, label=right:$g_2$]{};
    
    \end{tikzpicture}
    \caption{Hasse diagram for the Coulomb branch of \Quiver{fig:minE7U3} which has $G_2$ global symmetry. Decorations are given by the \textcolor{purple}{purple} circles. The slices $2A_1$ and $3A_1$ refer to $A_1\cup A_1$ and $A_1\cup A_1\cup A_1$ respectively.}
    \label{fig:E7U3Hasse}

\end{figure}

\FloatBarrier


\section{Cyclic Polymerisation with Quiver Extensions}
\label{sec:cyclicproduct}
In this section, the \hyperref[fig:ClassSCyclicFix]{quiver extensions} described in Section \ref{subsec:ClassS} which were introduced to cure the incomplete Higgsing when gluing punctures in class $\mathcal S$ are illustrated with examples. The results of \hyperref[fig:Cyclic]{cyclic polymerisation} here include $3d$ mirrors of class $\mathcal S$ theories of higher genus but possibly also other SCFTs.

The focus lies mainly on the extension that introduces free twisted hypers, which is the most natural remedy for incomplete Higgsing, however other extensions are demonstrated for completeness.

\subsection{$A_1\;\mathcal C_{4,0}$ theory with four punctures}

\begin{figure}[h!]
    \centering
    \begin{subfigure}{0.45\textwidth}

    \begin{tikzpicture}
        \node[gauge, label=below:$1$] (1r) []{};
        \node[gauge, label=below:$2$,fill=black] (2r)[left=of 1r]{};
        \node[gauge, label=below:$2$] (2l) [left=of 2r]{};
        \node[gauge, label=left:$1$] (1l) [left=of 2l]{};
        \node[gauge, label=left:$1$] (1tl) [above left=of 2l]{};
        \node[gauge, label=left:$1$] (1bl) [below left=of 2l]{};

        \draw[-] (1r)--(2r)--(2l)--(1l) (1tl)--(2l)--(1bl);
    \end{tikzpicture}
    \caption{}
    \label{fig:minD4H2}
    \end{subfigure}
    \centering
    \begin{subfigure}{0.45\textwidth}
    \centering
    \begin{tikzpicture}
        \node[gauge, label=below:$1$] (1l) []{};
        \node[gauge, label=below:$2$] (2l) [right=of 1l]{};
        \node[gauge, label=below:$2$,fill=black] (2r) [right=of 2l]{};
        \node[gauge, label=below:$1$] (1r) [right=of 2r]{};
        \node[gauge, label=left:$1$] (1tl) [above=of 2l]{};
        \node[gauge, label=right:$1$] (1tr) [above=of 2r]{};

        \draw[-] (1l)--(2l)--(2r)--(1r) (1tl)--(2l) (2r)--(1tr);
    \end{tikzpicture}
    \caption{}
    \label{fig:minD5}
    \end{subfigure}
    \caption{Two possible extensions of the $D^{(1)}_4$ quiver.}
    \label{fig:D4Extend}
\end{figure}
\begin{figure}[h!]
    \centering
    \begin{subfigure}{0.45\textwidth}
    \centering
    \begin{tikzpicture}
    \centering
        \node[gauge, label=below:$1$] (1r) []{};
        \node[gauge, label=below:$2$,fill=black] (2r)[left=of 1r]{};
        \node[gauge, label=below:$2$] (2l) [left=of 2r]{};
        \node[gauge, label=left:$1$] (1l) [left=of 2l]{};
        \node[gauge, label=left:$1$] (1tl) [above left=of 2l]{};
        \node[gauge, label=left:$1$] (1bl) [below left=of 2l]{};

        \draw[-] (1r)--(2r)--(2l)--(1l) (1tl)--(2l)--(1bl);

        \node[gauge, label=left:$1$] (1T) [below=of 1bl]{};
        \node[gauge, label=below:$2$] (2) [below right=of 1T]{};
        \node[gauge, label=left:$1$] (1B) [below left=of 2]{};

        \draw[-] (1T) -- (2)--(1B);
        \draw[-] (2) to[out=-45,in=45,looseness=8] (2);

        \node[] (topghost) [left=of 1l]{};
        \node[] (bghost) [left=of 2]{};
        \node[] (bottomghost) [left=of bghost]{};

        \draw [->] (topghost) to [out=-150,in=150,looseness=1] (bottomghost);
    \end{tikzpicture}
        \caption{}
        \label{fig:minD4H2U2}
    \end{subfigure}
    \centering
    \begin{subfigure}{0.45\textwidth}
    \begin{tikzpicture}
    \centering
        \node[gauge, label=below:$1$] (1l) []{};
        \node[gauge, label=below:$2$] (2l) [right=of 1l]{};
        \node[gauge, label=below:$2$,fill=black] (2r) [right=of 2l]{};
        \node[gauge, label=below:$1$] (1r) [right=of 2r]{};
        \node[gauge, label=left:$1$] (1tl) [above=of 2l]{};
        \node[gauge, label=right:$1$] (1tr) [above=of 2r]{};

        \draw[-] (1l)--(2l)--(2r)--(1r) (1tl)--(2l) (2r)--(1tr);

        \node[gauge, label=left:$1$] (1T) [below=of 1l]{};
        \node[gauge, label=below:$2$] (2) [below right=of 1T]{};
        \node[gauge, label=left:$1$] (1B) [below left=of 2]{};

        \draw[-] (1T) -- (2)--(1B);
        \draw[-] (2) to[out=-45,in=45,looseness=8] (2);

        \node[] (topghost) [left=of 1l]{};
        \node[] (bghost) [left=of 2]{};
        \node[] (bottomghost) [left=of bghost]{};

        \draw [->] (topghost) to [out=-150,in=150,looseness=1] (bottomghost);
    \end{tikzpicture}
    \caption{}
    \label{fig:minD5U2}
    \end{subfigure}
    
    \caption{$\urm(2)$ \hyperref[fig:Cyclic]{cyclic polymerisation} on the quivers \Quiver{fig:minD4H2} and \Quiver{fig:minD5} to produce a magnetic quiver for $\overline{n. min. B_2}$.}
    \label{fig:D4ExtU2}
\end{figure}

The $3d$ mirror quiver for the class $\mathcal S$ theory defined on $\mathcal C_{4,0}$ with four maximal $A_1$ punctures is the $D_4^{(1)}$ quiver.

In order to glue two of these punctures with complete Higgsing, a \hyperref[fig:ClassSCyclicFix]{quiver extension} and then a $\urm(2)$ \hyperref[fig:Cyclic]{cyclic polymerisation} must be performed. There are two choices of extension of the $D_4^{(1)}$ quiver resulting in the quivers \Quiver{fig:minD4H2} and \Quiver{fig:minD5}, shown in \Figref{fig:D4Extend}, whose Coulomb branches are $\overline{min. D_4}\times \mathbb H^2$ and $\overline{min. D_5}$ respectively. The additional $\urm(2)$ gauge node from the extension is coloured in black. In \Quiver{fig:minD4H2}, the $\urm(2)$ gauge node filled in black has balance $-1$, so the Coulomb branch has an additional free part.

The $\urm(2)$ \hyperref[fig:Cyclic]{cyclic polymerisation}s on \Quiver{fig:minD4H2} and \Quiver{fig:minD5} are shown in \Figref{fig:minD4H2U2} and \Figref{fig:minD5U2}. Both result in the same magnetic quiver $\frac{1}{2}D^{(1)}_5$ for $\overline{n. min. B_2}$. This is also the $3d$ mirror quiver for the class $\mathcal S$ theory on $\mathcal C_{2,1}$ with two maximal $A_1$ punctures. The \hyperref[fig:Cyclic]{cyclic polymerisation} shown in \Figref{fig:minD5U2} was also explored in Section \ref{subsec:DOrb}.

The new result, from \Figref{fig:minD4H2U2}, is that, \begin{equation}
    \overline{min. D_4}\times \mathbb H^2///\urm(2)=\overline{n.min. B_2},
\end{equation}
which is easily checked with Weyl integration.

The Higgs branch of \Quiver{fig:minD4H2} and \Quiver{fig:minD5} are the $D_4$ and $D_5$ singularities respectively, which are both of dimension 1. The Higgs branch of the quiver after $\urm(2)$ \hyperref[fig:Cyclic]{cyclic polymerisation} is $\mathcal S^{C_{2}}_{\mathcal N,(2,1^2)}\times \mathbb H$ which is of dimension 3. This is consistent with the increase in Higgs branch dimension by $\textrm{rank}\;\urm(2)=2$.
\subsection{Theory with $\sorm(4)$ global symmetry}
\label{subsec:nminB3H23U2}

\begin{figure}[h!]
    \centering
    \begin{subfigure}{0.45\textwidth}

    \begin{tikzpicture}
        \node[gauge, label=below:$1$] (1l) []{};
        \node[gauge, label=below:$2$] (2) [right=of 1l]{};
        \node[gauge, label=below:$2$] (2r) [right=of 2]{};
        \node[gauge, label=left:$1$] (1t) [above=of 2]{};

        \draw[-] (1l)--(2)--(2r) (2)--(1t);
         \draw[-] (2r) to [out=-45, in=45,looseness=8] (2r);
        
    \end{tikzpicture}
    \caption{}
    \label{fig:nminB3}
    \end{subfigure}
    \centering
    \begin{subfigure}{0.45\textwidth}
    \centering
    \begin{tikzpicture}
        \node[gauge, label=below:$1$] (1l) []{};
        \node[gauge, label=below:$2$,fill=black] (2ext) [right=of 1l]{};
        \node[gauge, label=below:$2$] (2) [right=of 2ext]{};
        \node[gauge, label=below:$2$] (2r) [right=of 2]{};
        \node[gauge, label=left:$1$] (1t) [above=of 2]{};

        \draw[-] (1l)--(2ext)--(2)--(2r) (2)--(1t);
         \draw[-] (2r) to [out=-45, in=45,looseness=8] (2r);
    \end{tikzpicture}
    \caption{}
    \label{fig:nminB3H2}
    \end{subfigure}
    \caption{Extension of the $\frac{1}{2}D^{(1)}_7$ quiver.}
    \label{fig:B3Extend}
\end{figure}

\begin{figure}[h!]
    \centering
    \begin{tikzpicture}
       \node[gauge, label=below:$1$] (1l) []{};
        \node[gauge, label=below:$2$,fill=black] (2ext) [right=of 1l]{};
        \node[gauge, label=below:$2$] (2) [right=of 2ext]{};
        \node[gauge, label=below:$2$] (2r) [right=of 2]{};
        \node[gauge, label=left:$1$] (1t) [above=of 2]{};

        \draw[-] (1l)--(2ext)--(2)--(2r) (2)--(1t);
         \draw[-] (2r) to [out=-45, in=45,looseness=8] (2r);

        \node[] (ghost) [below =of 2r]{};

        \node[gauge, label=below:$2$] (2R) [below=of ghost]{};
        \node[gauge, label=below:$2$] (2L) [left=of 2R]{};

        \draw[-] (2L)--(2R);
        \draw[-] (2R) to [out=-45, in=45,looseness=8] (2R);
        \draw[-] (2L) to [out=135, in=225,looseness=8] (2L);

    \node[] (topghost) [left=of 1l]{};
    \node[] (bottomghost) [left=of 2L]{};
    \draw [->] (topghost) to [out=-150,in=150,looseness=1] (bottomghost);
         
    \end{tikzpicture}
    \caption{$\urm(2)$ \hyperref[fig:Cyclic]{cyclic polymerisation} on the quiver \Quiver{fig:nminB3H2} to produce quiver \Quiver{fig:nminB3H2U2}.}
    \label{fig:nminB3H2U2}
\end{figure}
The starting point this time is not a $3d$ mirror of a class $\mathcal S$ theory but instead the $\frac{1}{2}D^{(1)}_{7}$ quiver reproduced in \Figref{fig:nminB3}.

In order to gauge a diagonal $\urm(2)$ subgroup of the Coulomb branch global symmetry with complete Higgsing, a \hyperref[fig:ClassSCyclicFix]{quiver extension} and a $\urm(2)$ \hyperref[fig:Cyclic]{cyclic polymerisation} must be performed. There is only one way to extend the quiver \Quiver{fig:nminB3} which results in quiver \Quiver{fig:nminB3H2} shown in \Figref{fig:nminB3H2}. The additional $\urm(2)$ gauge node from the extension is coloured in black and has balance $-1$. This in turn gives rise to a Coulomb branch that is the product space $\overline{n. min. B_3}\times \mathbb H^2$.

The \hyperref[fig:Cyclic]{cyclic polymerisation} of \Quiver{fig:nminB3H2} by $\urm(2)$ is shown in \Figref{fig:nminB3H2U2}. The resulting quiver is the $\frac{1}{2}A^{(1)}_3$ quiver \Quiver{fig:nminB3H2U2} which is part of the family of \Quiver{fig:Cslice}. Quiver \Quiver{fig:nminB3H2U2} also appears in \cite{Giacomelli:2024sex}. The conclusion is that:
\begin{equation}
    \overline{n. min. B_3}\times \mathbb H^2///\urm(2)=\mathcal S^{C_4}_{(4,2^2),(2^4)}=\overline{min. D_3}/(\mathbb Z_2\times\mathbb Z_2)\label{eq:nminB3H2U2}.
\end{equation}
The result \eqref{eq:nminB3H2U2} is confirmed by Weyl integration.

The Higgs branch of \Quiver{fig:nminB3H2} is the same as the Higgs branch of \Quiver{fig:nminB3} which is $\mathcal S^{C_3}_{\mathcal N,(4,1^2)}$. This moduli space is of dimension 3. The Higgs branch Hilbert series of \Quiver{fig:nminB3H2U2} is computed as \begin{equation}
    \hs\left[\mathcal H(\text{\Quiver{fig:nminB3H2U2}})\right]=(1-t^4-[1;1]_{\sorm(4)}t^6+[1;1]_{\sorm(4)}t^8-t^{14})\pe\left[[1;1]_{\sorm(4)}t+[2;2]_{\sorm(4)}t^2+[1;1]_{\sorm(4)}t^4-t^8\right],
\end{equation}where $[a;b]_{\sorm(4)}=[a,0]_{\sorm(4)}+[0,b]_{\sorm(4)}$, which identifies the moduli space as $\mathcal S^{C_4}_{(4^2),(3^2,1^2)}\times \mathbb H^2$ as expected.

Note that the combination of \hyperref[fig:ClassSCyclicFix]{quiver extension} followed by \hyperref[fig:Cyclic]{cyclic polymerisation} may be applied twice on the $D^{(1)}_4$ quiver to get to \Quiver{fig:nminB3H2U2}.
\subsection{$A_2\;\mathcal C_{3,0}$ theory with three maximal punctures}
\label{subsec:E6H3U3}

\begin{figure}[h!]
    \centering
    \begin{tikzpicture}
         \node[gauge, label=below:$1$] (1l) []{};
        \node[gauge, label=below:$2$] (2l) [right=of 1l]{};
        \node[gauge, label=below:$3$] (3l) [right=of 2l]{};
        \node[gauge, label=below:$3$,fill=black] (3r) [right=of 3l]{};
        \node[gauge, label=below:$2$] (2r) [right=of 3r]{};
        \node[gauge, label=below:$1$] (1r) [right=of 2r]{};
        \node[gauge, label=left:$2$] (2t) [above=of 3l]{};
        \node[gauge, label=left:$1$] (1t) [above=of 2t]{};

        \draw[-] (1l)--(2l)--(3l)--(3r)--(2r)--(1r);
        \draw[-] (1t)--(2t)--(3l);
    \end{tikzpicture}
    \caption{The extended $E^{(1)}_6$ quiver \Quiver{fig:E6H3}.}
    \label{fig:E6H3}
\end{figure}

\begin{figure}[h!]
    \centering
    \begin{tikzpicture}
        \node[gauge, label=below:$1$] (1l) []{};
        \node[gauge, label=below:$2$] (2l) [right=of 1l]{};
        \node[gauge, label=below:$3$] (3l) [right=of 2l]{};
        \node[gauge, label=below:$3$] (3r) [right=of 3l]{};
        \node[gauge, label=below:$2$] (2r) [right=of 3r]{};
        \node[gauge, label=below:$1$] (1r) [right=of 2r]{};
        \node[gauge, label=left:$2$] (2t) [above=of 3l]{};
        \node[gauge, label=left:$1$] (1t) [above=of 2t]{};

        \draw[-] (1l)--(2l)--(3l)--(3r)--(2r)--(1r);
        \draw[-] (1t)--(2t)--(3l);

        \node[] (ghost) [below =of 1l]{};

        \node[gauge, label=below:$1$] (1L) [below=of ghost]{};
        \node[gauge, label=below:$2$] (2L) [right=of 1L]{};
        \node[gauge, label=below:$3$] (3L) [right=of 2L]{};

        \draw[-] (1L)--(2L)--(3L);

         \draw[-] (3L) to [out=-45, in=45,looseness=8] (3L);

    \node[] (topghost) [left=of 1l]{};
    \node[] (bottomghost) [left=of 1L]{};
    \draw [->] (topghost) to [out=-150,in=150,looseness=1] (bottomghost);
         
    \end{tikzpicture}
    \caption{$\urm(3)$ \hyperref[fig:Cyclic]{cyclic polymerisation} on the quiver \Quiver{fig:E6H3} to produce quiver \Quiver{fig:E6H3U3}.}
    \label{fig:E6H3U3}
\end{figure}
%

The $3d$ mirror quiver for the class $\mathcal S$ theory defined on $\mathcal C_{3,0}$ with three maximal $A_2$ punctures is the $E^{(1)}_6$ quiver. 

In order to glue two of these punctures with complete Higgsing, a \hyperref[fig:ClassSCyclicFix]{quiver extension} and a $\urm(3)$ \hyperref[fig:Cyclic]{cyclic polymerisation} must be performed. There is only one way to extend the $E^{(1)}_6$ quiver shown in \Figref{fig:E6H3} to give quiver \Quiver{fig:E6H3}. The additional $\urm(3)$ gauge node from the extension is coloured in black and has balance $-1$. This in turn gives rise to a Coulomb branch that is the product space $\overline{min. E_6}\times \mathbb H^3$.

The $\urm(3)$ \hyperref[fig:Cyclic]{cyclic polymerisation} of \Quiver{fig:E6H3} is shown in \Figref{fig:E6H3U3}. The resulting quiver \Quiver{fig:E6H3U3} is the magnetic quiver for $\overline{sub. reg. G_2}$ \cite{Cremonesi:2014vla}. This is also the $3d$ mirror for the class $\mathcal S$ theory on $\mathcal C_{1,1}$ with one maximal $A_2$ puncture. The conclusion is that:
\begin{equation}
    \overline{min. E_6}\times \mathbb H^3///\urm(3)=\overline{sub. reg. G_2}\label{eq:E6H3Uk}.
\end{equation}
The result \eqref{eq:E6H3Uk} is confirmed by Weyl integration.

The Higgs branch of \Quiver{fig:E6H3} is the $E_6$ singularity of dimension 1. The Higgs branch Hilbert series of \Quiver{fig:E6H3U3} is\begin{equation}
    \hs[\mathcal H(\text{\Quiver{fig:E6H3U3}})]=\pe\left[[1]_{\surm(2)}t+[2]_{\surm(2)}t^2+[3]_{\surm(2)}t^3-t^{12}\right],
\end{equation} from which it is clear that the moduli space is a complete intersection of dimension 4. In fact, from Lustig-Spaltenstein duality this moduli space is $\mathcal S^{G_2}_{\mathcal N,A_1}\times \mathbb H$. There is the expected increase of Higgs branch dimension by $\textrm{rank}\;\urm(3)=3$.

\subsection{$A_3\;\mathcal C_{3,0}$ theory with two maximal and one $(2^2)$ puncture}
\label{subsec:E7H4U4}

\begin{figure}[h!]
    \centering
    \begin{tikzpicture}
         \node[gauge, label=below:$1$] (1l)[]{};
        \node[gauge, label=below:$2$] (2l)[right=of 1l]{};
        \node[gauge, label=below:$3$] (3l) [right=of 2l]{};
        \node[gauge, label=below:$4$] (4l) [right=of 3l]{};
        \node[gauge, label=below:$4$,fill=black] (4r) [right=of 4l]{};
        \node[gauge, label=below:$3$] (3r) [right=of 4r]{};
        \node[gauge, label=below:$2$] (2r) [right=of 3r]{};
        \node[gauge, label=below:$1$] (1r) [right=of 2r]{};
        \node[gauge, label=left:$2$] (2t) [above=of 4l]{};

        \draw[-] (1l)--(2l)--(3l)--(4l)--(4r)--(3r)--(2r)--(1r) (4l)--(2t);
    \end{tikzpicture}
    \caption{Extended $E^{(1)}_7$ quiver \Quiver{fig:E7H4}.}
    \label{fig:E7H4}
\end{figure}

\begin{figure}[h!]
    \centering
    \begin{tikzpicture}
        \node[gauge, label=below:$1$] (1l)[]{};
        \node[gauge, label=below:$2$] (2l)[right=of 1l]{};
        \node[gauge, label=below:$3$] (3l) [right=of 2l]{};
        \node[gauge, label=below:$4$] (4l) [right=of 3l]{};
        \node[gauge, label=below:$4$] (4r) [right=of 4l]{};
        \node[gauge, label=below:$3$] (3r) [right=of 4r]{};
        \node[gauge, label=below:$2$] (2r) [right=of 3r]{};
        \node[gauge, label=below:$1$] (1r) [right=of 2r]{};
        \node[gauge, label=left:$2$] (2t) [above=of 4l]{};

        \draw[-] (1l)--(2l)--(3l)--(4l)--(4r)--(3r)--(2r)--(1r) (4l)--(2t);

        \node[] (ghost) [below=of 4l]{};
        \node[gauge, label=below:$2$] (2) [below=of ghost]{};
        \node[gauge, label=below:$4$] (4) [right=of 2]{};

        \draw[-] (2)--(4);
         \draw[-] (4) to [out=-45, in=45,looseness=8] (4);

        \node[] (topghost) [left=of 1l]{};
        \node[] (bottomghost) [left=of 2]{};
        \draw [->] (topghost) to [out=-150,in=150,looseness=1] (bottomghost);
                 
    \end{tikzpicture}
    \caption{$\urm(4)$ \hyperref[fig:Cyclic]{cyclic polymerisation} on the magnetic quiver \Quiver{fig:E7H4} to produce \Quiver{fig:E7H4U4}.}
    \label{fig:E7H4U4}
\end{figure}

The $3d$ mirror quiver for the class $\mathcal S$ theory on $\mathcal C_{3,0}$ with 2 maximal and a $(2^2)\;A_3$ puncture is the $E^{(1)}_7$ quiver. 

In order to glue the two maximal punctures with complete Higgsing a \hyperref[fig:ClassSCyclicFix]{quiver extension} and a $\urm(4)$ \hyperref[fig:Cyclic]{cyclic polymerisation} must be performed. There is only one way to extend the affine $E^{(1)}_7$ quiver which is shown in \Figref{fig:E7H4} to produce \Quiver{fig:E7H4}. The additional $\urm(4)$ gauge node from the extension is coloured in black. This gauge node of $\urm(4)$ has a balance of $-1$ and so the Coulomb branch contains a free sector, as in the previous examples. The Coulomb branch is $\overline{min. E_7}\times \mathbb H^4$.

The $\urm(4)$ \hyperref[fig:Cyclic]{cyclic polymerisation} of \Quiver{fig:E7H4} is shown in \Figref{fig:E7H4U4} and produces \Quiver{fig:E7H4U4}. The \Quiver{fig:E7H4U4} is the $3d$ mirror of the class $\mathcal S$ theory on $\mathcal C_{1,1}$ with one $(2^2)\;A_3$ puncture. There are two ways to understand this Coulomb branch geometrically; as a slice in the $\mathfrak{f}_4$ nilcone $\mathcal S^{F_4}_{F_4(a_3),A_2}$\cite{2015arXiv150205770F,2023arXiv230807398F} and as $\overline{min. D_4}/S_4$ \cite{Gledhill:2021cbe,Hanany:2023uzn}. The conclusion is that:
\begin{equation}
    \overline{min. E_7}\times \mathbb H^4///\urm(4)=\mathcal S^{F_4}_{F_4(a_3),A_2}=\overline{min. D_4}/S_4.
\end{equation}

The Higgs branch of \Quiver{fig:E7H4} is the $E_7$ singularity of dimension 1. The Higgs branch Hilbert series of \Quiver{fig:E7H4U4} is easily computed as \begin{equation}
    \hs\left[\mathcal H(\text{\Quiver{fig:E7H4U4}})\right]=(1 + t^4 + t^{10} + t^{14} - [3]_{\surm(2)}t^7 )\pe\left[[1]_{\surm(2)}t+[2]_{\surm(2)}t^2+[3]_{\surm(2)}t^3+[4]_{\surm(2)}t^4-t^4-t^{12}\right].
\end{equation}The unrefined Hilbert series is \begin{equation}
     \hs\left[\mathcal H(\text{\Quiver{fig:E7H4U4}})\right]=\frac{(1 - t^{12}) (1 + 2 t + 3 t^2 + 4 t^3 + 6 t^4 + 8 t^5 + 10 t^6 + 
   8 t^7 + 6 t^8 + 4 t^9 + 3 t^{10} + 2 t^{11} + t^{12})}{(1 - t^2)^3 (1 -
    t^4)^4 (1 - t^3)^4},
\end{equation}from which it is clear that the Higgs branch is not a complete intersection but it is of dimension 5. From Lustig-Spaltenstein duality this moduli space is $\mathcal S^{F_4}_{B_3,A_2+\tilde{A}_{1}}\times \mathbb H$. There is the expected increase in Higgs branch dimension of $\textrm{rank}\;\urm(4)=4$.

\section{Conclusions}
\label{sec:conc}
In this work the two related diagrammatic techniques of \hyperref[fig:Chain]{chain polymerisation} and \hyperref[fig:Cyclic]{cyclic polymerisation} on unitary $3d\;\mathcal N=4$ quivers are introduced. They are summarised in \Figref{fig:Chain} and \Figref{fig:Cyclic} respectively.

In both cases, there is a gauging of a subgroup of the Coulomb branch global symmetry and the action on the Coulomb branch is that of a hyper-Kähler quotient by some diagonal subgroup, $\surm(k)$ or $\urm(k)$. 

This subgroup is carried by a pair of quiver legs of $(1)-\cdots-(k-1)-(k)-$, which contribute a factor of (at least\footnote{If there has been some prior enhancement of this global symmetry, then it is explicitly broken to this $\surm(k)\times \urm(1)$ subgroup.}) $\surm(k)\times \urm(1)$ to the Coulomb branch global symmetry. The dimension of the Coulomb branch is reduced by the dimension of the quotient group, indicating complete Higgsing on the Coulomb branch.

The action on the Higgs branch is an ungauging of the flavour symmetry, which leads to an increase in Higgs branch dimension equal to the rank of the quotient group, $\textrm{rank}\;\surm(k)=k-1$, or $\textrm{rank}\;\urm(k)=k$. The reverse process of polymerisation typically involves incomplete Higgsing on the Higgs branch, since $\textrm{rank}\;\surm(k)\leq\textrm{dim}\;\surm(k)$ and $\textrm{rank}\;\urm(k)\leq\textrm{dim}\;\urm(k)$, with equality for $k=1$ where there is complete Higgsing.

Both of types of polymerisation are generalisations of examples in \cite{Benini:2010uu}, which approach these gaugings from a class $\mathcal S$ perspective and their $3d$ mirrors. Alternatively, the techniques of \hyperref[fig:Chain]{chain} and \hyperref[fig:Cyclic]{cyclic polymerisation} can be thought of as Coulomb branch global symmetry analogues of gauging diagonal flavour symmetries.

\paragraph{Chains with Free Fields}
In Section \ref{sec:Chain}, examples of \hyperref[fig:Chain]{chain polymerisation} were presented. These took magnetic quivers for moduli spaces of free fields, in the form of finite $B$ and $D$ type quivers, and $\surm(2)$ chain polymerised them with other quivers. The results are summarised in Table \ref{tab:FreeChain} (with slightly different parameterisation).

The first three examples in Table \ref{tab:FreeChain} yield constructions for the affine B and D quivers, as well as the twisted affine D type quiver, as presented in \cite{Hanany:2001iy}. These are magnetic quivers for closures of nilpotent orbits, and this construction of nilpotent orbit closures via hyper-Kähler quotients can be viewed as a Coulomb branch implementation of the Kraft-Procesi construction.

The last four examples in Table \ref{tab:FreeChain} yield more constructions for magnetic quivers for slices in nilpotent cones (and not just nilpotent orbit closures). In particular, the last three examples in Table \ref{tab:FreeChain} yield quivers which come from generalised Cartan matrices and their corresponding Dynkin diagrams.

The examples illustrate the way in which chain polymerisation acting on a (disconnected) magnetic quiver can
promote a moduli space of free fields (possibly with factors of $\mathbb C^2/\mathbb Z_2$) to the closure of a nilpotent orbit or more generally, a slice.

At the same time, the flavour symmetry of the Higgs branch can be enhanced, and the dimension increased. For example, the polymerisation action on the Higgs branch of a (simply laced) quiver may lead from a moduli space of free fields (possibly times factors of $\mathbb C^2/\mathbb Z_2$) to slices in nilcones (possibly with factors of $\mathbb H$). In some cases, there are duality relationships between (non-trivial parts of) Higgs branches and Coulomb branches.

\begin{table}[h!]
    \scalebox{0.85}{
    \centering
    \begin{tabular}{cccccc}
    \toprule
        Initial  & Chained & Initial  & Initial  & Chained & Chained  \\
          Quivers &  Quiver&  Coulomb &  Higgs &  Coulomb & Higgs \\\midrule
        
        $D_{k+2}\times D_{n-k+2}$ &$D^{(1)}_{n}$ & $\mathbb H^{2n}$&Trivial&$\overline{min. D_{n}}$&$D_{n}$ \\
        $D_{k+2}\times B_{n-k+1}$&$B^{(1)}_{n-1}$ &$\mathbb H^{2n-1}$&$-$&$\overline{min. B_{n-1}}$&$-$\\
        $B_{k+1}\times B_{n-k+1}$&$D^{(2)}_{n-1}$&$\mathbb H^{2n-2}$&$-$&$\overline{min. D_{n-1}}$&$-$\\
        $\text{\Quiver{fig:A1H2km2}}\times B_{n-k+1}$&\Quiver{fig:nminBkpm3One} & $\mathbb C^2/\mathbb Z_2\times \mathbb H^{2n-1}$&$-$&$\overline{n.min. B_{n-1}}$&$-$\\
        $\text{\Quiver{fig:H2km2Z2}}\times D_{n-k+2}$&$\frac{1}{2}D^{(1)}_{2n-1}$&$\mathbb C^2/\mathbb Z_2\times \mathbb H^{2n-1}$&$\mathbb C^2/\mathbb Z_2$&$\overline{n.min. B_{n-1}}$&$\mathcal S^{C_{n-1}}_{\mathcal N,(2(n-2),1^2)}$\\$\text{\Quiver{fig:H2km2Z2}}\times\text{\Quiver{fig:H2km2Z2}}$&$\frac{1}{2}A^{(1)}_{2n-1}$&$(\mathbb C^2/\mathbb Z_2)^2\times \mathbb H^{2n}$&$(\mathbb C^2/\mathbb Z_2)^2\times \mathbb H^2$&$\mathcal S^{C_{2n}}_{(4,2^{2n-2}),(2^{2n})}$&$\mathbb H^2\times \mathcal S^{C_{2n}}_{((2n)^2),((2n-1)^2,1^2)}$\\
        $\text{\Quiver{fig:H2km2Z2}}\times B_{n-k+1}$&$\frac{1}{2}D^{(2)}_{2n-1}$&$\mathbb C^2/\mathbb Z_2\times \mathbb H^{2n-1}$&$-$&$\overline{\mathcal O}^{D_{n-1}}_{(3,1^{2n-5})}$&$-$\\        
        \bottomrule
    \end{tabular}}
    \caption{$\surm(2)$ \hyperref[fig:Chain]{chain polymerisation} of finite $D$ and $B$ type and related quivers. Initial and chained Coulomb and Higgs branches are shown if known.}
    \label{tab:FreeChain}
\end{table}
\paragraph{Two $\frac{1}{2}$ M5 on $E_6$ Klein Singularity}
\hyperref[fig:Chain]{Chain polymerisation} was applied to find a unitary magnetic quiver for the $6d\;\mathcal N=(1,0)$ theory coming from two separated $\frac{1}{2}$ M5 branes on an $E_6$ Klein singularity. This came from the $\surm(3)$ \hyperref[fig:Chain]{chain polymerisation} of two $E_8^{(1)}$ quivers, which was originally motivated from the F-theory description.

Further coinciding these two $\frac{1}{2}$ M5 branes gives the M-theory picture for the $(E_6,E_6)$ conformal matter theory \cite{DelZotto:2014hpa}. In doing so, there is a small $E_8$ instanton transition \cite{Ganor:1996mu} which increases the Higgs branch dimension by 29. Although it is known how to realise the small $E_8$ instanton transition through magnetic quivers in certain cases \cite{Hanany:2018uhm}, this does not apply for the quiver \Quiver{fig:E6E6Quiver} and remains an open question.

\paragraph{Moduli Space of Instantons}
A significant application of \hyperref[fig:Cyclic]{cyclic polymerisation} in this work is in the construction of the moduli space of $\surm(N)$ instantons on $\mathbb C^2$ and on $\mathbb C^2/\mathbb Z_l$, as a Coulomb branch hyper-Kähler quotient. This is outlined in Section \ref{sec:instantonconstruc}. Both the ADHM and Kronheimer-Nakajima constructions of these moduli spaces can be realised from polymerised quivers.

Additionally, non-trivial relationships between free moduli spaces, nilpotent orbits, affine Grassmannian slices and moduli spaces of instantons were found in this study. Some of these are tabulated in Table \ref{tab:Instantons}.

\begin{table}[h!]
 \scalebox{0.9}{
    \centering
    \begin{tabular}{ccccccc}
    \toprule
        Initial  & Cyclic & Final  & Initial  & Initial & Final &Final  \\
          Quiver &  Poly.&  Quiver &  Coulomb& Higgs &  Coulomb & Higgs \\
         \midrule
        $A_{N+2}$&$\urm(1)$&\Quiver{fig:FiniteAU12}&$\mathbb H^{N+1}$&Trivial&$\mathcal M_{1,\surm(N)}^{\mathbb C^2}$ &$\mathbb C^2/\mathbb Z_N$\\
        $D_{N+4}$ &$\urm(2)$&\Quiver{fig:FiniteDkU2}&$\mathbb H^{2N+4}$&Trivial&$\mathcal M^{\mathbb C^2}_{2,\surm(N)}$&$\textrm{Sym}^2(\mathbb C^2/\mathbb Z_N)$\\
        \Quiver{fig:FreeField}&$\urm(k)$&\Quiver{fig:FreeFieldUk}&$\mathbb H^{k^2+Nk}$&Trivial&$\mathcal M^{\mathbb C^2}_{k,\surm(N)}$&$\textrm{Sym}^k(\mathbb C^2/\mathbb Z_N)$\\
        $\mathcal Q_M\left(\overline{\mathcal O}^{A_{2k}}_{(2^k,1)}\right)$&$\urm(k)$&\Quiver{fig:ADHMUk}&$\mathcal{O}^{A_{2k}}_{(2^k,1)}$&$\mathcal S^{A_{2k}}_{\mathcal N,(k+1,k)}$&$\textrm{Sym}^k\left(\mathbb C^2/\mathbb Z_2\right)$&$\mathcal M^{\mathbb C^2}_{k,\surm(2)}$\\
        $\mathcal Q_M\left(\overline{\mathcal O}^{A_{N+3}}_{(2^2,1^N)}\right)$&$\urm(2)$&\Quiver{fig:nminAkU2}&$\overline{\mathcal O}^{A_{N+3}}_{(2^2,1^N)}$&$\mathcal S^{A_{N+3}}_{\mathcal N,(N+2,2)}$&$\mathcal M^{\mathbb C^2/\mathbb Z_2}_{2,\surm(N)}$&$\mathcal M^{\mathbb C^2/\mathbb Z_N}_{2,\surm(2)}$\\
        $\mathcal Q_M\left(\overline{n. min. A_{N+1}}\right)$&$\urm(1)$&\Quiver{fig:nminAkU1}& $\overline{n.min. A_{N+1}}$&$\mathcal S^{A_{N+1}}_{\mathcal N,(N,2)}$ & $\mathcal M_{(2,1),(N,0)}^{\mathbb C^2/\mathbb Z_2;\surm(N)}$&$\mathcal M^{\mathbb C^2/\mathbb Z_{N};\surm(2)}_{(2^{N-1},1),(1,0^{N-3},1,0)}$\\
        \Quiver{fig:A2KNm1AG}&$\urm(k)$ &\Quiver{fig:FreeFieldUkL}&See text&See text &$\mathcal M_{k,\surm(N)}^{\mathbb C^2/\mathbb Z_l}$&$\mathcal M_{k,\surm(l)}^{\mathbb C^2/\mathbb Z_N}$\\\bottomrule
        
    \end{tabular}}
    \caption{Table of \hyperref[fig:Cyclic]{cyclic polymerisations} to produce Kronheimer-Nakajima quivers.}
    \label{tab:Instantons}
\end{table}

The polymerised quivers all have Coulomb branches that are moduli spaces of $k\;\surm(N)$ instantons on ${\mathbb C^2/\mathbb Z_l}$ for some parameters $\{ k \ge 1, l \ge 1, N \ge 1 \}$. Their respective Higgs branches are the moduli spaces of $k\;\surm(l)$ instantons on ${\mathbb C^2/\mathbb Z_N}$.\footnote{Note that $\textrm{Sym}^k\left(\mathbb C^2/\mathbb Z_N\right) \equiv \mathcal M_{k,\surm(1)}^{\mathbb C^2/\mathbb Z_N}$.}

For cases where $l=1$, the polymerisation acts on magnetic quivers for free field moduli spaces; for cases where $l=2$, the polymerisation acts on magnetic quivers for nilpotent orbits.
%
%
%
In the case of general $l$, the instanton moduli space is constructed from the $\urm(k)$ \hyperref[fig:Cyclic]{cyclic polymerisation} of quiver \Quiver{fig:A2KNm1AG}, which is a magnetic quiver for a product space between an affine Grassmannian of A-type and a free field.
In all cases, the instanton number follows from the rank of the $\urm(k)$ polymerisation.

The most general construction was discussed in Section \ref{sec:genericKN}.

The $3d$ mirrors of the unitary Kronheimer-Nakajima quivers (when framed) exchange the number of flavours and the number of gauge nodes via the $l \leftrightarrow N$ parameters. Given the Coulomb branch construction of these quivers from a hyper-Kähler quotient, it is natural to seek a $3d$ mirror dual description on the Higgs branch.

Two such examples were shown in \Figref{fig:ADHMCommute} and the more general \Figref{fig:ADHMCommuteL}. These contain commuting diagrams wherein the hyper-Kähler quotient on the Coulomb branch (in the left column of each figure) corresponds to the hyper-Kähler quotient on the Higgs branch between the $3d$ mirrors (in the right column of each figure). Notably, in these examples, $3d$ mirror symmetry is preserved by the hyper-Kähler quotients.

In these figures \Figref{fig:ADHMCommute} and \Figref{fig:ADHMCommuteL}, cyclic polymerisation (in the left column of each figure) corresponds to an increase in the Higgs branch dimension by $k$ and gauging the flavour symmetry (in the right column of each figure) corresponds to an increase in the Coulomb branch dimension by $k$.

In these examples, the geometric action of \hyperref[fig:Cyclic]{cyclic polymerisation} on the Higgs branch is seen to be dual to the geometric action of gauging a flavour symmetry on the Coulomb branch of the mirror theory.



\paragraph{Cyclic ADE Quivers}
In Section \ref{sec:ADEOrb} \hyperref[fig:Cyclic]{cyclic polymerisations} were carried out on some finite and affine ADE quivers to generate quivers with loops or multiple links. These illustrate a variety of relationships between free fields, nilpotent orbit closures and the moduli spaces of instantons. These are summarised in Table \ref{tab:ADECyclic} and Table \ref{tab:EOrbCyclic} repsectively.
\begin{table}[h!]
    \centering
    \begin{tabular}{ccccccc}
    \toprule
        Initial  & Cyclic & Final  & Initial  & Initial & Final & Final  \\
          Quiver &  Poly.&  Quiver &  Coulomb& Higgs &  Coulomb & Higgs \\\midrule
         $A_N$ & $\urm(1)$ & \Quiver{fig:FiniteAU1}& $\mathbb H^{N-1}$&Trivial& $\overline{min. A_{N-L-R}}\times \mathbb H^{L+R-2}$&$A_{N-L-R}$\\ $D_{N+4}$ &$\urm(2)$&\Quiver{fig:FiniteDkU2}&$\mathbb H^{2N+4}$&Trivial&$\mathcal M^{\mathbb C^2}_{2,\surm(N)}$&$\textrm{Sym}^2(\mathbb C^2/\mathbb Z_N)$\\
        $E_6$ &$\urm(2)$&\Quiver{fig:FiniteE6U2}&$\mathbb H^{10}$&Trivial&$\mathcal M^{\mathbb C^2}_{2,\surm(2)}\times \mathbb H^2$&$\textrm{Sym}^2(\mathbb C^2)$\\\bottomrule
    \end{tabular}
    \caption{Table of \hyperref[fig:Cyclic]{cyclic polymerisations} of finite $A_N$ and $E_6$ quivers.}
    \label{tab:ADECyclic}
\end{table}

\begin{table}[h!]
    \centering
    \begin{tabular}{ccccccc}
    \toprule
        Initial  & Cyclic & Final  & Initial  & Initial & Final &Final  \\
          Quiver &  Poly.&  Quiver &  Coulomb& Higgs &  Coulomb & Higgs \\\midrule
         $D^{(1)}_{N+4}$&$\urm(2)$&\Quiver{fig:minDkU2}&$\overline{min. D_{N+4}}$&$D_{N+4}$&$-$&$-$\\$E_6^{(1)}$&$\urm(1)$&\Quiver{fig:minE6U1}&$\overline{min. E_6}$&$E_6$& $\overline{\mathcal O}^{D_5}_{(2^4,1^2)}$&$\mathcal S^{D_5}_{\mathcal N,(5^2)}$\\$E^{(1)}_6$&$\urm(2)$&\Quiver{fig:minE6U2}&$\overline{min. E_6}$&$E_6$& $\overline{\mathcal{O}}^{B_3}_{(3^2,1)}$&$\mathcal S^{C_3}_{\mathcal N,(2^3)}$\\$E_7^{(1)}$&$\urm(1)$&\Quiver{fig:minE7U1}&$\overline{min. E_7}$&$E_7$& $\overline{n. min. E_6}$&$\mathcal S^{E_6}_{\mathcal N,D_5}$\\
        $E^{(1)}_7$ &$\urm(2)$&\Quiver{fig:minE7U2}&$\overline{min. E_7}$&$E_7$&$-$&$-$\\
        $E^{(1)}_7$ &$\urm(3)$&\Quiver{fig:minE7U3}&$\overline{min. E_7}$&$E_7$&$\mathcal S^{E_7}_{A_3+A_2+A_1,A_2+3A_1}$&$\mathcal S^{E_7}_{A_6,A_4+A_2}$\\\bottomrule
    \end{tabular}
    \caption{Table of \hyperref[fig:Cyclic]{cyclic polymerisations} of affine $D^{(1)}_{N+4}$, $E^{(1)}_6$, and $E^{(1)}_7$ quivers.}
    \label{tab:EOrbCyclic}
\end{table}

A $\urm(1)$ \hyperref[fig:Cyclic]{cyclic polymerisation} on the finite $A_N$ quiver, produces different results depending on the nodes that are identified under polymerisation. The Coulomb branch is a product space of a minimal A-type orbit with free field(s), and includes the moduli space of one SU instanton on $\mathbb C^2$ as a limiting case. The Higgs branch is somewhat simpler being an A-type Kleinian singularity in all cases.


The $D^{(1)}_{N+4}$ affine quiver is amenable to $\urm(2)$ \hyperref[fig:Cyclic]{cyclic polymerisation}. The resulting quiver can be transformed into a a Kronheimer-Nakajima quiver for $\mathcal M_{2,\surm(N)}^{\mathbb C^2/\mathbb Z_2}$ by a further $\urm(1)$ \hyperref[fig:Cyclic]{cyclic polymerisation}.


An interesting new result is the quiver \Quiver{fig:minE6U2}, obtained from the $\urm(2)$ \hyperref[fig:Cyclic]{cyclic polymerisation} of the affine $E^{(1)}_6$ quiver. The Coulomb branch of \Quiver{fig:minE6U2} is $\overline{\mathcal O}_{(3^2,1)}^{B_3}$, which is a nilpotent orbit closure of height four. While unitary magnetic quivers for nilpotent orbits closures of height two or less are known for all semi-simple Lie algebras \cite{Hanany:2017ooe}, very few magnetic quivers are given in the Literature for orbits of greater height. The Higgs branch of \Quiver{fig:minE6U2} is the Slodowy slice $\mathcal S^{C_3}_{\mathcal N,(2^3)}$, which is consistent with Lustig-Spaltenstein duality.



The quiver \Quiver{fig:minE7U2} was found from the $\urm(2)$ \hyperref[fig:Cyclic]{cyclic polymerisation} of the affine $E_7^{(1)}$ quiver. The Coulomb branch Hasse diagram of \Quiver{fig:minE7U2} contains the transition $2a_3$ (or $a_3 \cup a_3$), which is not a Kraft-Procesi transition, and signals that this moduli space does not lie in the nilcone of any semi-simple Lie algebra \cite{Kraft1980MinimalGLn} neither is it a slice in the affine Grassmannian of a simply laced algebra \cite{2003math......5095M}.

The quiver \Quiver{fig:minE7U3}, which came from the $\urm(3)$ \hyperref[fig:Cyclic]{cyclic polymerisation} of the affine $E_7^{(1)}$ quiver, is also an interesting discovery. The Coulomb branch is the Slodowy intersection in the nilcone of $\mathfrak{e_7}$, $\mathcal S^{E_7}_{A_3+A_2+A_1,A_2+3A_1}$. The Higgs branch of \Quiver{fig:minE7U3} is once again the Lustig-Spaltenstein dual intersection $\mathcal S^{E_7}_{A_6,A_4+A_2}$ \cite{2023arXiv230807398F}. 

A common feature of many of the multi-linked or looped quivers that arise from $\urm(k)$ \hyperref[fig:Cyclic]{cyclic polymerisation} is that a loop of, say $n+1$, gauge nodes may be formed with $k$ possible $a_n$ slices to subtract. The natural $S_k$ permutation of these $a_n$ slices mean that the Hasse diagram of quivers arising from cyclic polymerisation need to be analysed in terms of decorations. Additionally, it is possible that the slices in the Hasse diagram are not of Kraft-Procesi classical type.

\paragraph{Quiver Extensions}
It has been a long standing problem in the analysis of class $\mathcal S$ theories via their $3d$ mirrors, as how best to resolve the incomplete Higgsing sometimes observed in $3d$. This occurs when using a hyper-Kähler quotient to glue two maximal punctures on the same fixture to create a genus loop. This is a reason for the mismatch between the Hall-Littlewood limit of the superconformal index of class $\mathcal S$ theories on higher genus Riemann surfaces and the Hilbert series of the Coulomb branch of the $3d$ mirror quiver. As illustrated with examples in Section \ref{sec:cyclicproduct}, the technique of quiver extension, followed by \hyperref[fig:Cyclic]{cyclic polymerisation}, offers an elegant analytic solution to the incomplete Higgsing.


In order to gauge two $A_{k-1}$ maximal punctures in the $3d$ mirror of a class $\mathcal S$ theory, a quiver extension is carried out on the $3d$ mirror, by inserting an additional node of $\urm(k)$ in either of the two maximal legs of $(1)-\cdots-(k)-$. Any quiver legs corresponding to "background" punctures or other features can be attached to either $\urm(k)$ in the two legs. (In general the moduli space of the extended quiver changes non-trivially under these choices.)

A natural choice is to put all of the "background" legs and adjoint hypermultiplets onto the non-extended leg. In this case, the link to the extended leg is "ugly" and $k$ free twisted hypermultiplets are introduced, contributing a $\mathbb H^k$ free field to the Coulomb branch. These additional free twisted hypermultiplets are sufficient to ensure complete Higgsing under \hyperref[fig:Cyclic]{cyclic polymerisations}.

The different possible choices yield an equivalence class of extended quivers, all of which provide a prescription to cure this Class $\mathcal S$ incomplete Higgsing problem.


\paragraph{Open Questions}

There are a number of open questions and future directions to explore.

Firstly, by necessity a selective choice of examples has been presented here. It can be expected that the application of polymerisation techniques to quivers outside the specific classes studied herein should provide a rich source of new relationships between moduli spaces. This could be particularly fruitful when polymerisation is used in combination with established diagrammatic techniques on $3d\;\mathcal N=4$ quivers. In addition, it can be expected that polymerisation may be applicable to new physical problems through analysis with magnetic quivers.

Polymerisation requires that the subgroup of the Coulomb branch global symmetry that is being gauged is encoded in maximal legs of the quiver(s). Is it possible to identify a diagrammatic mechanism for gauging the Coulomb branch global symmetry of a quiver that is not represented by an external maximal leg?

In this work, there are examples where $3d$ mirror symmetry and Coulomb branch global (resp. flavour) symmetry gauging commute. 
Is it generally the case that when a Coulomb branch global (resp. flavour) symmetry gauging acts on a pair of $3d$ mirror symmetric quivers, the resulting quivers form a $3d$ mirror pair?



A different, and ambitious, research direction would be to interpret this gauging of the Coulomb branch global symmetry in terms of branes. Take the method of \hyperref[fig:Cyclic]{cyclic polymerisation}. The Type IIB brane system for each quiver leg is the Hanany-Witten system on flat space. The \hyperref[fig:Cyclic]{cyclic polymerisation} creates a loop in the quiver, which is interpreted as a Hanany-Witten brane system on a circle. Therefore the brane realisation of \hyperref[fig:Cyclic]{cyclic polymerisation} must involve compactifying one spacetime direction. The mechanism through which a compact direction can be created is not understood. However, the specific identification of the gauge nodes involved in polymerisation may provide a hint as to what this mechanism is. Alternatively, magnetic quivers that come from brane webs may contain loops. There may be an interpretation of polymerisation from brane webs.


Finally, a very obvious extension of this work is to orthosymplectic quivers. The main challenge is that the Coulomb branch global symmetry is not explicitly visible either in an orthosymplectic quiver or in the unrefined Coulomb branch Hilbert series constructed using the monopole formula. Nonetheless, the discussion surrounding Hall-Littlewood functions (and Slodowy slices carrying background charges) in Section \ref{subsec:analytic} is general to any semi-simple Lie group. This provides a framework for constructing and branching the refined Hilbert series corresponding to linear or star shaped orthosymplectic quivers. Thus, it is reasonable to expect that cyclic and chain polymerisation can also be implemented to gauge the Coulomb branch global symmetries of orthosymplectic quivers.

\acknowledgments

The authors would like to thank Shlomo Razamat, Gabi Zafrir, Mykola Matviichuk, and Travis Schedler for discussions. The work of AH, RJK, and GK is partially supported by STFC Consolidated Grants ST/T000791/1 and ST/X000575/1. The work of GK is supported by STFC DTP research studentship grant ST/X508433/1.

\appendix
\section{Chain Polymerisation of two $T_2$ theories}
\label{sec:triskelion}

\begin{figure}[h!]
    \centering
    \begin{tikzpicture}
        \node[gauge, label=below:$1$] (1l) []{};
        \node[gauge, label=below:$2$] (2) [right=of 1l]{};
        \node[gauge, label=below:$1$] (1r) [right=of 2]{};
        \node[gauge, label=left:$1$] (1t) [above=of 2]{};

        \draw[-] (1l)--(2)--(1r) (2)--(1t);

        \node[] (times) [below=of 2]{$\times$};

        \node[gauge, label=above:$2$] (2r) [below=of times]{};
        \node[gauge, label=below:$1$] (1lr) [left=of 2r]{};
        \node[gauge, label=below:$1$] (1rr) [right=of 2r]{};
        \node[gauge, label=below:$1$] (1b) [below=of 2r]{};

        \draw[-] (1lr)--(2r)--(1rr) (1b)--(2r);

        \node[gauge, label=left:$1$] (1TL) [below left=of 1b]{};
        \node[gauge, label=below:$2$] (2RR) [below right=of 1TL]{};
        \node[gauge, label=below:$1$] (1BL) [below left=of 2RR]{};
        \node[gauge, label=right:$1$] (1TR) [above right=of 2RR]{};
        \node[gauge, label=below:$1$] (1BR) [below right=of 2RR]{};

        \draw[-] (1TL)--(2RR)--(1TR) (1BR)--(2RR)--(1BL);

         \node[] (topghost) [left=of times]{};
        \node[] (bottomghost) [left=of 2RR]{};
        \draw [->] (topghost) to [out=-150,in=150,looseness=1] (bottomghost);
    \end{tikzpicture}
    \caption{$\surm(2)$ \hyperref[fig:Chain]{chain polymerisation} of two $D_4$ quivers to produce the $D^{(1)}_4$ quiver. This is the operation on the $3d$ mirrors of gluing two $T_2$ theories in class $\mathcal S$.}
    \label{fig:T2Gauge}
\end{figure}
The gluing of triskelion theories (restricted here to the A-type case $T_N$) has been well known in the class $\mathcal S$ literature \cite{Gaiotto:2009we} and realised from their $3d$ mirrors \cite{Benini:2010uu}.

The $3d$ mirror for the $T_N$ theory is a star-shaped quiver with three maximal legs of $\surm(N)$. The gluing of two triskelion theories is realised in the $3d$ mirror as chain polymerisation at the central node of $\urm(N)$. 

This idea has been illustrated in \Figref{fig:ClassSChain} for generic class $\mathcal S$ theories (with regular punctures). Explicitly, the gluing of two $T_2$ is shown in the $3d$ mirrors in \Figref{fig:T2Gauge} to produce the $D^{(1)}_4$ quiver.

The Coulomb branch of the $3d$ mirror of the $T_2$ theory, a.k.a the $D_4$ quiver, is $\mathbb H^4$ and the Coulomb branch of the $D^{(1)}_4$ quiver is $\overline{min. D_4}$. The action of the chain polymerisation is the hyper-Kähler quotient by $\surm(2)$. In summary,\begin{equation}
    \mathbb H^4\times \mathbb H^4///\surm(2)=\overline{min. D_4}.
\end{equation}

The Higgs branch of the $D_4$ quiver is trivial and the Higgs branch of the $D^{(1)}_4$ quiver is the $D_4$ singularity which is of dimension one. The Higgs branch dimension increases by $\textrm{rank}\;\surm(2)=1$ as expected.

\FloatBarrier

\section{Orientifolds}
\label{sec:orientifolds}
The first instance of (twisted) affine Lie algebras being constructed from brane systems was in \cite{Hanany:2001iy} and is reviewed here.

The starting point is a Type IIB background with a stack of $N$ D3 branes. The worldvolume theory is a $\urm(N)$ gauge theory. D1 branes can end between the D3 which give rise to monopole operators in the D3 worldvolume theory. This constitutes the magnetic spectrum and each "fundamental monopole" (i.e. associated to D1 ending between neighbouring D3) is associated to a co-root of the gauge algebra. The monopoles can break the $\urm(N)$ gauge symmetry to some Levi subgroup.

The electric spectrum can be seen in the S-dual picture where F1 strings can end between D3 instead. Each "fundamental F1" is associated to the root of the gauge algebra.

Adding orientifold planes to the string background reduces the number of BPS states both in the magnetic and electric spectrum. This allows for $\sorm$ and $\sprm$ gauge algebras for both the electric and magnetic spectrum.

The orientifold planes being considered here are $O3^-$, $O3^+$, $\widetilde O3^-$, and $\widetilde O3^+$. It is simplest to see the effect of orientifold planes on the electric spectrum as there is a perturbative formulation of F1 ending in the presence of orientifolds. The different orientifold planes correspond to different types of Dynkin diagram endings. The magnetic spectrum of this system may be read from the electric spectrum of the S-dual brane system. The action of S-duality on orientifold planes are given in Table \ref{tab:SDualOrientifold}. The gauge algebra for the electric and magnetic BPS states for each type of orientifold is shown in Table \ref{tab:OrientifoldEM}.
\begin{table}[h!]
    \centering
    \begin{tabular}{cc}
    \toprule
    Orientifold & S-Dual\\
    \midrule
        $O3^-$ & $O3^-$ \\
        $O3^+$ & $\widetilde O3^-$\\
        $\widetilde O3^-$ & $O3^+$\\
        $\widetilde O3^+$ & $\widetilde O3^+$\\\bottomrule
    \end{tabular}
    \caption{Orientifold planes and their S-duals.}
    \label{tab:SDualOrientifold}
\end{table}

\begin{table}[h!]
    \centering
    \begin{tabular}{ccc}
    \toprule
    Orientifold & Electric & Magnetic\\\midrule
    
        $O3^-$ & D & D\\
        $O3^+$ & C & B\\
        $\widetilde O3^-$ & B & C\\
        $\widetilde O3^+$ & C & C \\\bottomrule
    \end{tabular}
    \caption{Gauge algebra for electric and magnetic BPS states with orientifold planes.}
    \label{tab:OrientifoldEM}
\end{table}

The construction of \cite{Hanany:2001iy} proceeds with orientifold planes at either end of an interval. Dp branes are aligned with the orientifold planes and are placed in the interval. The resulting quivers from the electric and magnetic phases of the brane system take the form of affine or twisted affine Dynkin diagrams of type $BCD$. The types of orientifold plane required for electric and magnetic constructions of (twisted) affine quivers is given in Table \ref{tab:EMConstruct}.

\begin{table}[h!]
    \centering
    \begin{tabular}{ccc}
    \toprule
        Quiver & Electric & Magnetic \\\midrule
         $B^{(1)}$ & $-\widetilde -$&$-+$\\
         $C^{(1)}$ & $++$ & $\widetilde -\widetilde -$\\
         $D^{(1)}$& $--$  & $--$\\
         $A^{(2)}_{\textrm{odd}}$& $-+$ & $-\widetilde -$\\
         $A^{(2)}_{\textrm{even}}$& $+\widetilde-$& $+\widetilde-$\\
         $D^{(2)}$&$\widetilde -\widetilde -$ & $++$\\\bottomrule
    \end{tabular}
    \caption{Electric and magnetic construction for affine and twisted affine quivers. The ``O" for the orientifold has been suppressed.}
    \label{tab:EMConstruct}
\end{table}

\section{Miscellaneous examples of Cyclic Polymerisation}
\label{sec:Ornament}

\begin{figure}[h!]
    \centering
    \begin{subfigure}{0.45\textwidth}\centering\resizebox{\textwidth}{!}{
     \begin{tikzpicture}
        \node[gauge, label=below:$1$] (1l) []{};
    \node[gauge, label=below:$2$] (2l) [right=of 1l]{};
    \node[gauge, label=below:$3$] (3) [right=of 2l]{};
    \node[gauge, label=below:$2$] (2r) [right=of 3]{};
    \node[gauge, label=below:$1$] (1r) [right=of 2r]{};
    \node[gauge,label=right:$1$] (1t) [above=of 3]{};

     \draw[double,double distance=3pt,line width=0.4pt] (1t)--(3);

    \draw[-] (1l)--(2l)--(3)--(2r)--(1r);
    \draw[-] (2l) to [out=45, in=135,looseness=8] (2l);
    
     \end{tikzpicture}
     }\caption{}
    \label{fig:A5Oneloop}\end{subfigure}
    \centering
    \begin{subfigure}{0.45\textwidth}\centering\resizebox{\textwidth}{!}{
     \begin{tikzpicture}
        \node[gauge, label=below:$1$] (1l) []{};
    \node[gauge, label=below:$2$] (2l) [right=of 1l]{};
    \node[gauge, label=below:$3$] (3) [right=of 2l]{};
    \node[gauge, label=below:$2$] (2r) [right=of 3]{};
    \node[gauge, label=below:$1$] (1r) [right=of 2r]{};
    \node[gauge,label=right:$1$] (1t) [above=of 3]{};

    \draw[double,double distance=3pt,line width=0.4pt] (1t)--(3);

    \draw[-] (1l)--(2l)--(3)--(2r)--(1r);
    \draw[-] (2l) to [out=45, in=135,looseness=8] (2l);
    \draw[-] (2r) to [out=45, in=135,looseness=8] (2r);
    
     \end{tikzpicture}
     }\caption{}
    \label{fig:A5Twoloop}\end{subfigure}
    \centering
    \begin{subfigure}{0.45\textwidth}\centering\resizebox{\textwidth}{!}{
     \begin{tikzpicture}
        \node[gauge, label=below:$1$] (1l) []{};
    \node[gauge, label=below:$2$] (2l) [right=of 1l]{};
    \node[gauge, label=below:$3$] (3) [right=of 2l]{};
    \node[gauge, label=below:$2$] (2r) [right=of 3]{};
    \node[gauge, label=below:$1$] (1r) [right=of 2r]{};
    \node[gauge,label=right:$1$] (1t) [above=of 3]{};

    \draw[double,double distance=3pt,line width=0.4pt] (1t)--(3);

    \draw[-] (1l)--(2l)--(3)--(2r)--(1r);
    \draw[-] (2l) to [out=0, in=90,looseness=10] (2l) to [out=90,in=180,looseness=10] (2l);
     \end{tikzpicture}
     }\caption{}
    \label{fig:A5Twoloop2}\end{subfigure}
    \centering
    \begin{subfigure}{0.45\textwidth}\centering\resizebox{\textwidth}{!}{
     \begin{tikzpicture}
        \node[gauge, label=below:$1$] (1l) []{};
    \node[gauge, label=below:$2$] (2l) [right=of 1l]{};
    \node[gauge, label=below:$4$] (3) [right=of 2l]{};
    \node[gauge, label=below:$2$] (2r) [right=of 3]{};
    \node[gauge, label=below:$1$] (1r) [right=of 2r]{};
    \node[gauge,label=right:$1$] (1t) [above=of 3]{};

    \draw[double,double distance=4pt,line width=0.4pt] (1t)--(3);
    \draw[double,double distance=1pt,line width=0.4pt] (1t)--(3);
    \draw[-] (1l)--(2l)--(3)--(2r)--(1r);
     \end{tikzpicture}
     }\caption{}
    \label{fig:A5Four}\end{subfigure}
    \centering
    \begin{subfigure}{0.45\textwidth}
    \centering\resizebox{\textwidth}{!}{
    \begin{tikzpicture}
        \node[gauge, label=below:$1$] (1l) []{};
    \node[gauge, label=below:$2$] (2l) [right=of 1l]{};
    \node[gauge, label=below:$3$] (3) [right=of 2l]{};
    \node[gauge, label=below:$2$] (2r) [right=of 3]{};
    \node[gauge, label=below:$1$] (1r) [right=of 2r]{};
    \node[flavour,label=right:$2$] (1t) [above=of 3]{};

    \draw[-] (1t)--(3);

    \draw [line width=1pt, double distance=3pt,
             arrows = {-Latex[length=0pt 2 0]}] (2l) -- (3);
    \draw [line width=1pt, double distance=3pt,
             arrows = {-Latex[length=0pt 2 0]}] (2r) -- (3);

    \draw[-] (1l)--(2l) (2r)--(1r);
    \end{tikzpicture}
    }\caption{}
    \label{fig:A5NSL}
    \end{subfigure}
    \centering
    \begin{subfigure}{0.45\textwidth}
    \centering\resizebox{\textwidth}{!}{
    \begin{tikzpicture}
      \node[gauge, label=below:$1$] (1l) []{};
    \node[gauge, label=below:$2$] (2l) [right=of 1l]{};
    \node[gauge, label=below:$3$] (3) [right=of 2l]{};
    \node[gauge, label=below:$2$] (2r) [right=of 3]{};
    \node[gauge, label=below:$1$] (1r) [right=of 2r]{};
    \node[gauge,label=right:$1$] (1t) [above=of 3]{};

    \draw[double,double distance=3pt,line width=0.4pt] (2l)--(3);
    \draw[-] (1l)--(2l) (1t)--(3)--(2r)--(1r);

    \end{tikzpicture}
    }\caption{}
    \label{fig:A5TwoOneEdge}
    \end{subfigure}
    \centering\begin{subfigure}{0.45\textwidth}\centering\resizebox{\textwidth}{!}{
     \begin{tikzpicture}
        \node[gauge, label=below:$1$] (1l) []{};
    \node[gauge, label=below:$2$] (2l) [right=of 1l]{};
    \node[gauge, label=below:$4$] (3) [right=of 2l]{};
    \node[gauge, label=below:$2$] (2r) [right=of 3]{};
    \node[gauge, label=below:$1$] (1r) [right=of 2r]{};
    \node[gauge, label=left:$1$] (1o) [above=of 2l]{};
    \node[gauge,label=right:$1$] (1t) [above=of 3]{};

    \draw[double,double distance=3pt,line width=0.4pt] (1t)--(3);
    \draw[double,double distance=3pt,line width=0.4pt] (2l)--(3);

    \draw[-] (1l)--(2l) (3)--(2r)--(1r);
    \draw[double,double distance=3pt,line width=0.4pt] (1o)--(2l);
    \draw[-] (2r) to [out=0, in=60,looseness=14] (2r)to [out=60, in=120,looseness=14] (2r)to [out=120, in=180,looseness=14] (2r);
     \end{tikzpicture}
     }\caption{}
    \label{fig:A5threeloopnode}\end{subfigure}
    \caption{A collection of seven quivers \Quiver{fig:A5Oneloop}, \Quiver{fig:A5Twoloop}, \Quiver{fig:A5Twoloop2}, \Quiver{fig:A5Four}, \Quiver{fig:A5NSL}, \Quiver{fig:A5TwoOneEdge}, and \Quiver{fig:A5threeloopnode}.}
    \label{fig:A5Collection}
\end{figure}

The techniques of \hyperref[fig:Chain]{chain} and \hyperref[fig:Cyclic]{cyclic polymerisation} are applicable to any "good" quiver in the sense of \cite{Gaiotto:2008ak} with legs of $(1)-\cdots-(k)-$. The action on the Coulomb branch is a hyper-Kähler quotient by $\surm/\urm(k)$ respectively as proved in Section \ref{subsubsec:ChainUnitary} and Section \ref{subsubsec:CyclicUnitary}.

Many examples presented in the main body of the paper, which were chosen from a variety of physical contexts, do not explore non-trivial features of polymerisation. In this Section, a collection of miscellaneous examples are presented to illustrate these non-trivial features.

These examples are the $\urm(2)$ \hyperref[fig:Cyclic]{cyclic polymerisation} of the seven quivers \Quiver{fig:A5Oneloop}, \Quiver{fig:A5Twoloop}, \Quiver{fig:A5Twoloop2}, \Quiver{fig:A5Four}, \Quiver{fig:A5NSL}, \Quiver{fig:A5TwoOneEdge}, and \Quiver{fig:A5threeloopnode} shown in \Figref{fig:A5Collection}. Each example highlights a feature of polymerisation.

The Coulomb branches of the quivers shown in \Figref{fig:A5Collection} do not have any particular names as moduli spaces, except for \Quiver{fig:A5Four} whose Coulomb branch is the affine Grassmannian slice $\overline{[\mathcal W_{A_5}]}^{[0,0,4,0,0]}_{[0,1,0,1,0]}$\cite{Bourget:2021siw}. Again, the emphasis here is not an analysis on moduli spaces, but on testing the methods of polymerisation and the action on the Coulomb branch.
\subsection{Cyclic polymerisation with one loop on a leg}
\label{subsec:A5Oneloop}
\begin{figure}[h!]
    \centering
    \begin{tikzpicture}

    \node[gauge, label=below:$1$] (1l) []{};
    \node[gauge, label=below:$2$] (2l) [right=of 1l]{};
    \node[gauge, label=below:$3$] (3) [right=of 2l]{};
    \node[gauge, label=below:$2$] (2r) [right=of 3]{};
    \node[gauge, label=below:$1$] (1r) [right=of 2r]{};
    \node[gauge,label=right:$1$] (1t) [above=of 3]{};

    \draw[double,double distance=3pt,line width=0.4pt] (1t)--(3);

    \draw[-] (1l)--(2l)--(3)--(2r)--(1r);
    \draw[-] (2l) to [out=45, in=135,looseness=8] (2l);

    \node[gauge, label=right:$1$] (1res)[below=of 3]{};
    \node[gauge, label=below:$3$] (3res) [below=of 1res]{};
    \node[gauge, label=below:$2$] (2res) [right=of 3res]{};

    \draw[double,double distance=3pt,line width=0.4pt] (1res)--(3res);
    \draw[double,double distance=3pt,line width=0.4pt] (3res)--(2res);
    \draw[-] (2res) to [out=-45, in=45,looseness=8] (2res);

    \node[] (topghost) [left=of 1l]{};
    \node[] (bottomghost) [left=of 3res]{};
    \draw [->] (topghost) to [out=-150,in=150,looseness=1] (bottomghost);

    \end{tikzpicture}
    \caption{$\urm(2)$ \hyperref[fig:Cyclic]{cyclic polymerisation} of quiver  \Quiver{fig:A5Oneloop} to produce \Quiver{fig:A5OneloopU2}.}
    \label{fig:A5OneloopU2}
\end{figure}

The $\urm(2)$ \hyperref[fig:Cyclic]{cyclic polymerisation} of the magnetic quiver \Quiver{fig:A5Oneloop} is shown in \Figref{fig:A5OneloopU2}. This produces \Quiver{fig:A5OneloopU2}, whose Coulomb branch has an $\surm(2)\times \urm(1)$ global symmetry, as can be confirmed from the refined Hilbert series. For brevity, only the unrefined Hilbert series is presented:
\begin{equation}
    \hs\left[\mathcal C\left(\text{\Quiver{fig:A5OneloopU2}}\right)\right]=\frac{\left(\begin{aligned}1 &+ t^2 + 6 t^4 + 19 t^6 + 36 t^8 + 63 t^{10}+ 106 t^{12} \\& + 131 t^{14} + 
 138 t^{16} + \pal + t^{32}\end{aligned}\right)}{(1 - t^2)^{3} (1-t^4)^3 (1-t^6)^4}.
\label{eq:A5OneloopU2}
\end{equation}

The same Hilbert series as in \eqref{eq:A5OneloopU2} is obtained by Weyl integration of the Coulomb branch Hilbert series of \Quiver{fig:A5Oneloop} by $\urm(2)$ with the embedding specified in \Figref{fig:CyclicEmbed}.
The conclusion is that:
\begin{equation}
    \mathcal C\left(\text{\Quiver{fig:A5Oneloop}}\right)///\urm(2)=\mathcal C\left(\text{\Quiver{fig:A5OneloopU2}}\right).
\end{equation}

The adjoint hypermultiplet on the node of $\urm(2)$ in \Quiver{fig:A5Oneloop} is inert under $\urm(2)$ chain polymerisation and becomes part of \Quiver{fig:A5OneloopU2}.
\subsection{Cyclic polymerisation with one adjoint loop on each leg}
\label{subsec:A5Twoloop}
\begin{figure}[h!]
    \centering
    \begin{tikzpicture}

    \node[gauge, label=below:$1$] (1l) []{};
    \node[gauge, label=below:$2$] (2l) [right=of 1l]{};
    \node[gauge, label=below:$3$] (3) [right=of 2l]{};
    \node[gauge, label=below:$2$] (2r) [right=of 3]{};
    \node[gauge, label=below:$1$] (1r) [right=of 2r]{};
    \node[gauge,label=right:$1$] (1t) [above=of 3]{};

    \draw[double,double distance=3pt,line width=0.4pt] (1t)--(3);

    \draw[-] (1l)--(2l)--(3)--(2r)--(1r);
    \draw[-] (2l) to [out=45, in=135,looseness=8] (2l);
    \draw[-] (2r) to [out=45, in=135,looseness=8] (2r);

    \node[gauge, label=right:$1$] (1res)[below=of 3]{};
    \node[gauge, label=below:$3$] (3res) [below=of 1res]{};
    \node[gauge, label=below:$2$] (2res) [right=of 3res]{};

    \draw[double,double distance=3pt,line width=0.4pt] (1res)--(3res);
    \draw[double,double distance=3pt,line width=0.4pt] (3res)--(2res);
    \draw[-] (2res) to [out=0, in=90,looseness=8] (2res)to [out=0, in=-90,looseness=8] (2res);

    \node[] (topghost) [left=of 1l]{};
    \node[] (bottomghost) [left=of 3res]{};
    \draw [->] (topghost) to [out=-150,in=150,looseness=1] (bottomghost);

    \end{tikzpicture}
    \caption{$\urm(2)$ \hyperref[fig:Cyclic]{cyclic polymerisation} of quiver  \Quiver{fig:A5Twoloop} to produce \Quiver{fig:A5TwoloopU2}.}
    \label{fig:A5TwoloopU2}
\end{figure}
The $\urm(2)$ \hyperref[fig:Cyclic]{cyclic polymerisation} of quiver \Quiver{fig:A5Twoloop} is shown in \Figref{fig:A5TwoloopU2}. This produces quiver \Quiver{fig:A5TwoloopU2}, whose Coulomb branch has an $\surm(2)\times \urm(1)$ global symmetry, as can be confirmed from the refined Hilbert series. For brevity only the unrefined Hilbert series is presented:
\begin{equation}
    \hs\left[\mathcal C\left(\text{\Quiver{fig:A5TwoloopU2}}\right)\right]=\frac{\left(\begin{aligned}1 &+ 6 t^4 + 8 t^6 + 24 t^8 + 37 t^{10} + 68 t^{12} + 89 t^{14} \\&+ 
 134 t^{16} + 134 t^{18} + 166 t^{20} + \pal + t^{40}\end{aligned}\right)}{(1 - t^2)^{4} (1 - t^6)^3 (1 - t^8)^3}.
\label{eq:A5TwoloopU2HS}
\end{equation}

The same Hilbert series as in \eqref{eq:A5TwoloopU2HS} is obtained by Weyl integration of the Coulomb branch Hilbert series of \Quiver{fig:A5Twoloop} by $\urm(2)$ with the embedding specified in \Figref{fig:CyclicEmbed}.
The conclusion is that:
\begin{equation}
    \mathcal C\left(\text{\Quiver{fig:A5Twoloop}}\right)///\urm(2)=\mathcal C\left(\text{\Quiver{fig:A5TwoloopU2}}\right)\label{eq:A5ThreeU2}.
\end{equation}

Each adjoint hypermultiplet on the nodes of $\urm(2)$ in \Quiver{fig:A5Twoloop} are inert under $\urm(2)$ chain polymerisation and becomes part of \Quiver{fig:A5TwoloopU2}.

\subsection{Cyclic polymerisation with two adjoint loops on one leg and none on the other}
\label{subsec:A5Twoloop2}
\begin{figure}[h!]
    \centering
    \begin{tikzpicture}

    \node[gauge, label=below:$1$] (1l) []{};
    \node[gauge, label=below:$2$] (2l) [right=of 1l]{};
    \node[gauge, label=below:$3$] (3) [right=of 2l]{};
    \node[gauge, label=below:$2$] (2r) [right=of 3]{};
    \node[gauge, label=below:$1$] (1r) [right=of 2r]{};
    \node[gauge,label=right:$1$] (1t) [above=of 3]{};

    \draw[double,double distance=3pt,line width=0.4pt] (1t)--(3);

    \draw[-] (1l)--(2l)--(3)--(2r)--(1r);
   \draw[-] (2l) to [out=0, in=90,looseness=10] (2l) to [out=90,in=180,looseness=10] (2l);

    \node[gauge, label=right:$1$] (1res)[below=of 3]{};
    \node[gauge, label=below:$3$] (3res) [below=of 1res]{};
    \node[gauge, label=below:$2$] (2res) [right=of 3res]{};

    \draw[double,double distance=3pt,line width=0.4pt] (1res)--(3res);
    \draw[double,double distance=3pt,line width=0.4pt] (3res)--(2res);
    \draw[-] (2res) to [out=0, in=90,looseness=8] (2res)to [out=0, in=-90,looseness=8] (2res);

    \node[] (topghost) [left=of 1l]{};
    \node[] (bottomghost) [left=of 3res]{};
    \draw [->] (topghost) to [out=-150,in=150,looseness=1] (bottomghost);

    \end{tikzpicture}
    \caption{$\urm(2)$ \hyperref[fig:Cyclic]{cyclic polymerisation} of quiver \Quiver{fig:A5Twoloop2} to produce \Quiver{fig:A5Twoloop2U2}$=$\Quiver{fig:A5TwoloopU2}.}
    \label{fig:A5Twoloop2U2}
\end{figure}
The $\urm(2)$ \hyperref[fig:Cyclic]{cyclic polymerisation} of \Quiver{fig:A5Twoloop2} is shown in \Figref{fig:A5Twoloop2U2}. This produces a quiver that is the same as \Quiver{fig:A5TwoloopU2}, for which the Coulomb branch unrefined Hilbert series is given in \eqref{eq:A5TwoloopU2HS}.

This Hilbert series is also obtained by Weyl integration of the Coulomb branch Hilbert series of \Quiver{fig:A5Twoloop2} by $\urm(2)$ with the embedding specified in \Figref{fig:CyclicEmbed}.
The conclusion is that 
\begin{equation}
    \mathcal C\left(\text{\Quiver{fig:A5Twoloop2}}\right)///\urm(2)= \mathcal C\left(\text{\Quiver{fig:A5Twoloop}}\right)///\urm(2)=\mathcal C\left(\text{\Quiver{fig:A5TwoloopU2}}\right)\label{eq:A5FiveU2}.
\end{equation}

The results of Sections \ref{subsec:A5Oneloop}, \ref{subsec:A5Twoloop}, and \ref{subsec:A5Twoloop2} show that the number of adjoint hypermultiplets on each leg do not matter for polymerisation. The number of adjoint hypermultiplets on the node of $\urm(2)$ simply add under $\urm(2)$ \hyperref[fig:Cyclic]{cyclic polymerisation}.
\subsection{Cyclic polymerisation with two non-maximal legs}

\begin{figure}[h!]
    \centering
    \begin{tikzpicture}
        \node[gauge, label=below:$1$] (1l) []{};
    \node[gauge, label=below:$2$] (2l) [right=of 1l]{};
    \node[gauge, label=below:$4$] (3) [right=of 2l]{};
    \node[gauge, label=below:$2$] (2r) [right=of 3]{};
    \node[gauge, label=below:$1$] (1r) [right=of 2r]{};
    \node[gauge,label=right:$1$] (1t) [above=of 3]{};

    \draw[double,double distance=4pt,line width=0.4pt] (1t)--(3);
    \draw[double,double distance=1pt,line width=0.4pt] (1t)--(3);
    \draw[-] (1l)--(2l)--(3)--(2r)--(1r);

    \node[gauge, label=right:$1$] (1res)[below=of 3]{};
    \node[gauge, label=below:$4$] (3res) [below=of 1res]{};
    \node[gauge, label=below:$2$] (2res) [right=of 3res]{};

    \draw[double,double distance=4pt,line width=0.4pt] (1res)--(3res);
    \draw[double,double distance=1pt,line width=0.4pt] (1res)--(3res);
    \draw[double,double distance=3pt,line width=0.4pt] (3res)--(2res);

    \node[] (topghost) [left=of 1l]{};
    \node[] (bottomghost) [left=of 3res]{};
    \draw [->] (topghost) to [out=-150,in=150,looseness=1] (bottomghost);
    \end{tikzpicture}
    \caption{$\urm(2)$ \hyperref[fig:Cyclic]{cyclic polymerisation} of quiver \Quiver{fig:A5Four} to produce \Quiver{fig:A5FourU2}.}
    \label{fig:A5FourU2}
\end{figure}

The $\urm(2)$ \hyperref[fig:Cyclic]{cyclic polymerisation} of quiver \Quiver{fig:A5Four} is shown in \Figref{fig:A5FourU2}. This produces quiver \Quiver{fig:A5FourU2}, whose Coulomb branch has an $\surm(2)\times \urm(1)$ global symmetry, as can be confirmed from the refined Hilbert series. For brevity only the unrefined Hilbert series is presented:
\begin{equation}
    \hs\left[\mathcal C\left(\text{\Quiver{fig:A5FourU2}}\right)\right]=\frac{\left(\begin{aligned}1 &+ 4 t^2 + 13 t^4 + 39 t^6 + 107 t^8 + 265 t^{10} + 603 t^{12} + 
   1249 t^{14} \\&+ 2385 t^{16} + 4202 t^{18} + 6871 t^{20} + 10449 t^{22} + 
   14843 t^{24} + 19706 t^{26} \\&+ 24519 t^{28} + 28619 t^{30} + 31394 t^{32} + 
   32366 t^{34} + \pal + t^{68}\end{aligned}\right)}{(1 - t^4)^3 (1 - t^8)^3 (1 - t^6)^4 (1 - t^{10})^2}.
\label{eq:A5FourU2HS}
\end{equation}

The same Hilbert series as in \eqref{eq:A5FourU2HS} is obtained by Weyl integration of the Coulomb branch Hilbert series of \Quiver{fig:A5Four} by $\urm(2)$ with the embedding specified in \Figref{fig:CyclicEmbed}.
The conclusion is that:
\begin{equation}
    \mathcal C\left(\text{\Quiver{fig:A5Four}}\right)///\urm(2)=\mathcal C\left(\text{\Quiver{fig:A5FourU2}}\right)\label{eq:A5FourU2}.
\end{equation}

The legs of $(1)-(2)-$ in \Quiver{fig:A5Four} are connected to a gauge node of $\urm(4)$. Importantly, polymerisation does not require that these legs of $(1)-(2)-$ are sub-legs of a "maximal leg". In general, the node being superimposed in polymerisation may be connected to any other node with any multiplicity as long as the monopole formula converges.

\subsection{Cyclic polymerisation with two non-simply laced edges}

\begin{figure}[h!]
    \centering
    \begin{tikzpicture}
    \node[gauge, label=below:$1$] (1l) []{};
    \node[gauge, label=below:$2$] (2l) [right=of 1l]{};
    \node[gauge, label=below:$3$] (3) [right=of 2l]{};
    \node[gauge, label=below:$2$] (2r) [right=of 3]{};
    \node[gauge, label=below:$1$] (1r) [right=of 2r]{};
    \node[flavour,label=right:$2$] (1t) [above=of 3]{};

    \draw[-] (1t)--(3);

    \draw [line width=1pt, double distance=3pt,
             arrows = {-Latex[length=0pt 2 0]}] (2l) -- (3);
    \draw [line width=1pt, double distance=3pt,
             arrows = {-Latex[length=0pt 2 0]}] (2r) -- (3);

    \draw[-] (1l)--(2l) (2r)--(1r);

    \node[flavour, label=right:$2$] (1res)[below=of 3]{};
    \node[gauge, label=below:$3$] (3res) [below=of 1res]{};
    \node[gauge, label=below:$2$] (2res) [right=of 3res]{};

    \draw[-] (1res)--(3res);
    \draw [line width=1pt, double distance=3pt,
             arrows = {-Latex[length=0pt 2 0]}] (2res) -- node[midway, above]{$2\times$} (3res);

    \node[] (topghost) [left=of 1l]{};
    \node[] (bottomghost) [left=of 3res]{};
    \draw [->] (topghost) to [out=-150,in=150,looseness=1] (bottomghost);
    \end{tikzpicture}
    \caption{$\urm(2)$ \hyperref[fig:Cyclic]{cyclic polymerisation} of quiver \Quiver{fig:A5NSL} to produce \Quiver{fig:A5NSLU2}.}
    \label{fig:A5NSLU2}
\end{figure}

The $\urm(2)$ \hyperref[fig:Cyclic]{cyclic polymerisation} of quiver \Quiver{fig:A5NSL} is shown in \Figref{fig:A5NSLU2}. This produces quiver \Quiver{fig:A5NSLU2}, whose Coulomb branch has an $\surm(2)\times \urm(1)$ global symmetry, as can be confirmed from the refined Hilbert series. The label of $2$ above the non-simply laced edge indicates that there are two non-simply laced edges of multiplicity two connecting the $\urm(2)$ and $\urm(3)$ gauge nodes. For brevity only the unrefined Hilbert series is presented:
\begin{equation}
    \hs\left[\mathcal C\left(\text{\Quiver{fig:A5NSLU2}}\right)\right]=\frac{\left(\begin{aligned}1 &+ t^2 + 5 t^4 + 9 t^6 + 27 t^8 + 61 t^{10} + 132 t^{12} + 233 t^{14} \\& +
   424 t^{16} + 670 t^{18} + 1027 t^{20} + 1413 t^{22} + 1882 t^{24} \\&+ 
   2290 t^{26} + 2692 t^{28} + 2903 t^{30} + 3012 t^{32} + \pal + 
   t^{64}\end{aligned}\right)}{(1 - t^2)^3 (1 - t^8) (1 - t^{12})^2 (1 - t^6) (1 - t^{10})^3}.
\label{eq:A5NSLU2HS}
\end{equation}

The same Hilbert series as in \eqref{eq:A5NSLU2HS} is obtained by Weyl integration of the Coulomb branch Hilbert series of \Quiver{fig:A5NSL} by $\urm(2)$ with the embedding specified in \Figref{fig:CyclicEmbed}.
The conclusion is that:
\begin{equation}
    \mathcal C\left(\text{\Quiver{fig:A5NSL}}\right)///\urm(2)=\mathcal C\left(\text{\Quiver{fig:A5NSLU2}}\right)\label{eq:A5NSLU2}.
\end{equation}

In Section \ref{subsec:TwistedAffD} two non-simply laced quivers were \hyperref[fig:Chain]{chain polymerised} together on the side corresponding to long roots. In the example presented above, a quiver with two non-simply laced edges may also be \hyperref[fig:Cyclic]{cyclic polymerised} at nodes corresponding to long roots. This only works since the multiplicities of these non-simply laced edges are the same.

\subsection{Cyclic polymerisation with two edges not matching}

\begin{figure}[h!]
    \centering
    \begin{tikzpicture}
        \node[gauge, label=below:$1$] (1l) []{};
    \node[gauge, label=below:$2$] (2l) [right=of 1l]{};
    \node[gauge, label=below:$3$] (3) [right=of 2l]{};
    \node[gauge, label=below:$2$] (2r) [right=of 3]{};
    \node[gauge, label=below:$1$] (1r) [right=of 2r]{};
    \node[gauge,label=right:$1$] (1t) [above=of 3]{};

    \draw[double,double distance=3pt,line width=0.4pt] (2l)--(3);
    \draw[-] (1l)--(2l) (1t)--(3)--(2r)--(1r);

    \node[gauge, label=right:$1$] (1res)[below=of 3]{};
    \node[gauge, label=below:$3$] (3res) [below=of 1res]{};
    \node[gauge, label=below:$2$] (2res) [right=of 3res]{};

    \draw[-] (1res)--(3res)--(2res);
    \draw[transform canvas={yshift=1.5pt}] (3res)--(2res);
    \draw[transform canvas={yshift=-1.5pt}] (3res)--(2res);

    \node[] (topghost) [left=of 1l]{};
    \node[] (bottomghost) [left=of 3res]{};
    \draw [->] (topghost) to [out=-150,in=150,looseness=1] (bottomghost);
    \end{tikzpicture}
    \caption{$\urm(2)$ \hyperref[fig:Cyclic]{cyclic polymerisation} of quiver \Quiver{fig:A5TwoOneEdge} to produce \Quiver{fig:A5TwoOneEdgeU2}.}
    \label{fig:A5TwoOneEdgeU2}
\end{figure}

The $\urm(2)$ \hyperref[fig:Cyclic]{cyclic polymerisation} of quiver \Quiver{fig:A5TwoOneEdge} is shown in \Figref{fig:A5TwoOneEdgeU2}. This produces quiver \Quiver{fig:A5TwoOneEdgeU2}, whose Coulomb branch has an $\surm(2)\times \urm(1)$ global symmetry, as can be confirmed from the refined Hilbert series. For brevity only the unrefined Hilbert series is presented:
\begin{equation}
    \hs\left[\mathcal C\left(\text{\Quiver{fig:A5TwoOneEdgeU2}}\right)\right]=\frac{\left(\begin{aligned}1 &+ t^3 + 4 t^4 + 2 t^5 + 5 t^6 + 7 t^7 + 11 t^8 + 11 t^9 + 
 15 t^{10}\\& + 15 t^{11} + 22 t^{12} + 19 t^{13} + 21 t^{14} + \pal + t^{28}\end{aligned}\right)}{(1 - t^2)^2 (1 - t^4)^2 (1 - t^3)^3 (1 - 
   t^5)^2 (1 - t^7)}
\label{eq:A5TwoOneEdgeU2HS}
\end{equation}

The same Hilbert series as in \eqref{eq:A5TwoOneEdgeU2HS} is obtained by Weyl integration of the Coulomb branch Hilbert series of \Quiver{fig:A5TwoOneEdge} by $\urm(2)$ with the embedding specified in \Figref{fig:CyclicEmbed}.
The conclusion is that:
\begin{equation}
    \mathcal C\left(\text{\Quiver{fig:A5TwoOneEdge}}\right)///\urm(2)=\mathcal C\left(\text{\Quiver{fig:A5TwoOneEdgeU2}}\right)\label{eq:A5NSLU2}.
\end{equation}

In this example the two nodes of $\urm(2)$ are connected to the same node of $\urm(3)$ with edges of multiplicity two and one. Under \hyperref[fig:Cyclic]{cyclic polymerisation} these $\urm(2)$ gauge nodes are superimposed and still connected to the node of $\urm(3)$ but with edge multiplicity three.
\subsection{Cyclic polymerisation with ``the full Monty"}
\begin{figure}[h!]
    \centering
    \begin{tikzpicture}

    \node[gauge, label=below:$1$] (1l) []{};
    \node[gauge, label=below:$2$] (2l) [right=of 1l]{};
    \node[gauge, label=below:$4$] (3) [right=of 2l]{};
    \node[gauge, label=below:$2$] (2r) [right=of 3]{};
    \node[gauge, label=below:$1$] (1r) [right=of 2r]{};
    \node[gauge, label=left:$1$] (1o) [above=of 2l]{};
    \node[gauge,label=right:$1$] (1t) [above=of 3]{};

    \draw[double,double distance=3pt,line width=0.4pt] (1t)--(3);
    \draw[double,double distance=3pt,line width=0.4pt] (2l)--(3);

    \draw[-] (1l)--(2l) (3)--(2r)--(1r);
    \draw[double,double distance=3pt,line width=0.4pt] (1o)--(2l);
    \draw[-] (2r) to [out=0, in=60,looseness=14] (2r)to [out=60, in=120,looseness=14] (2r)to [out=120, in=180,looseness=14] (2r);

    \node[gauge, label=right:$1$] (1res)[below=of 3]{};
    \node[gauge, label=below:$4$] (3res) [below=of 1res]{};
    \node[gauge, label=below:$2$] (2res) [right=of 3res]{};
    \node[gauge, label=below:$1$] (1ores) [right=of 2res]{};

    \draw[double,double distance=3pt,line width=0.4pt] (3res)--(1res);
    \draw[-] (3res)--(2res);
    \draw[transform canvas={yshift=1.5pt}] (3res)--(2res);
    \draw[transform canvas={yshift=-1.5pt}] (3res)--(2res);
    \draw[double,double distance=3pt,line width=0.4pt] (2res)--(1ores);
    \draw[-] (2res) to [out=15, in=65,looseness=14] (2res)to [out=65, in=115,looseness=14] (2res)to [out=115, in=165,looseness=14] (2res);

    \node[] (topghost) [left=of 1l]{};
    \node[] (bottomghost) [left=of 3res]{};
    \draw [->] (topghost) to [out=-150,in=150,looseness=1] (bottomghost);

    \end{tikzpicture}
    \caption{$\urm(2)$ \hyperref[fig:Cyclic]{cyclic polymerisation} of quiver \Quiver{fig:A5threeloopnode} to produce \Quiver{fig:A5threeloopnodeU2}.}
    \label{fig:A5threeloopnodeU2}
\end{figure}

The $\urm(2)$ \hyperref[fig:Cyclic]{cyclic polymerisation} of the magnetic quiver \Quiver{fig:A5threeloopnode} is shown in \Figref{fig:A5threeloopnodeU2}. This produces quiver \Quiver{fig:A5threeloopnodeU2} whose Coulomb branch has a $\surm(2)\times \urm(1)^2$ global symmetry, as can be confirmed from the refined Hilbert series. For brevity only the unrefined Hilbert series is presented: \begin{equation}
    \hs\left[\mathcal C\left(\text{\Quiver{fig:A5threeloopnodeU2}}\right)\right]=\frac{(1-t^2)^3P_{168}(t)}
   {(1-t^4)^4(1-t^6) (1-t^8)^3  (1-t^{14})^5(1-t^{18})^4},
\label{eq:A5threeloopnodeU2HS}
\end{equation}where $P_{168}(t)$ is the following palindromial of degree 168,
\begin{align}
    P_{168}(t)&=1 + 8 t^2 + 38 t^4 + 140 t^6 + 442 t^8 + 1247 t^{10} + 3222 t^{12} + 
 7742 t^{14} + 17494 t^{16} \nonumber\\&+ 37494 t^{18} + 76726 t^{20} + 150650 t^{22} + 
 284902 t^{24} + 520470 t^{26} + 920601 t^{28} \nonumber\\&+ 1579537 t^{30} + 
 2633008 t^{32} + 4270103 t^{34} + 6745839 t^{36} + 10393109 t^{38} + 
 15632183 t^{40} \nonumber\\&+ 22975327 t^{42} + 33024005 t^{44} + 46456070 t^{46} + 
 64001059 t^{48} + 86402316 t^{50} \nonumber\\&+ 114365948 t^{52} + 148497917 t^{54} + 
 189232573 t^{56} + 236757419 t^{58} + 290941404 t^{60} \nonumber\\&+ 351274874 t^{62} + 
 416829757 t^{64} + 486247365 t^{66} + 557759352 t^{68} + 629243963 t^{70} \nonumber\\&+ 
 698317772 t^{72} + 762458658 t^{74} + 819152424 t^{76} + 866051722 t^{78} + 
 901133393 t^{80} \nonumber\\&+ 922838997 t^{82} + 930186066 t^{84} + \pal + t^{168}.
\end{align}

The same Hilbert series as in \eqref{eq:A5threeloopnodeU2HS} is obtained by Weyl integration of the Coulomb branch Hilbert series of \Quiver{fig:A5threeloopnode} by $\urm(2)$ with the embedding specified in \Figref{fig:CyclicEmbed}.
The conclusion is that:
\begin{equation}
    \mathcal C\left(\text{\Quiver{fig:A5threeloopnode}}\right)///\urm(2)=\mathcal C\left(\text{\Quiver{fig:A5threeloopnodeU2}}\right)
    \label{eq:A5FourU2}.
\end{equation}

This example is "the full Monty" and includes many of the above features of polymerisation at the same time. In particular, both legs are non-maximal, the edge multiplicity of $\urm(2)$ node to the $\urm(4)$ node do not match, one $\urm(2)$ node is attached to a $\urm(1)$ gauge node and the other has three adjoint hypermultiplets.




\bibliographystyle{JHEP}
\bibliography{references}
\end{document}